\documentclass[11pt,a4paper]{amsart}
\usepackage[margin=2.5cm]{geometry}

\usepackage{amsmath,amsfonts,amssymb,amscd,amsthm}
\usepackage[usenames,dvipsnames,svgnames]{xcolor}

\usepackage[utf8]{inputenc}
\usepackage{graphicx}
\usepackage{caption}
\usepackage{subcaption}
\usepackage[numbers]{natbib}
\usepackage{doi}
\usepackage{autonum}
\usepackage{makecell}
\usepackage{hyperref}
\usepackage{acronym}
\usepackage{listings}

\DeclareMathOperator{\grad}{\boldsymbol\nabla}

\def\0{\boldsymbol{0}}

\def\R{\boldsymbol{R}}

\def\x{\boldsymbol{x}}

\def\geophy{K}
\def\georef{{\hat K}}

\def\geomap{\boldsymbol{\Phi}}

\def\jacobian{\boldsymbol{J}_K}

\def\FEMPAR{{\texttt{FEMPAR}}}
\def\R{\mathbb{R}}

\def\georef{{\hat K}}
\def\geophy{K}
\def\geomap{\boldsymbol{\Phi}}
\def\shapetest#1{\phi^{#1}}
\def\shapetrial#1{\psi^{#1}}

\def\boundary{\partial \Omega}

\def\triang{\mathcal{T}_h}

\def\fematrix{\mathbf{A}}
\def\ferhs{\mathbf{f}}
\def\fesol{\mathbf{u}}

\def\blform#1#2{{a}(#1,#2)}
\def\blformh#1#2{{a_h}(#1,#2)}
\def\blformlocal#1#2#3{{a}_{#1}(#2,#3)}
\def\rhsform#1{\ell( #1) }
\def\rhsformh#1{\ell_h( #1) }
\def\rhsformlocal#1#2{{\ell}_{#1}(#2)}

\def\jumpl{\lbrack\!\lbrack}
\def\jumpr{\rbrack\!\rbrack}
\def\jump#1{\jumpl #1 \jumpr}
\def\mean#1{\{\! \!\{ #1\}\! \!\}}

\def\facephy{F}

\def\facejacobian{\boldsymbol{J}_F}

\def\normal{\boldsymbol{n}}
\newcommand{\closure}[2][3]{%
{}\mkern#1mu\overline{\mkern-#1mu#2}}
\def\domain{\Omega}
\def\integrandfe{\boldsymbol{\mathcal{F}}}
\def\ddom{{\rm d}\domain}
\def\gp{\rm gp}
\def\quadrature{\mathrm{Q}}
\def\xh{\hat{\x}}

\newcommand{\pluseq}{\mathrel{+}=}

\acrodef{dof}[DOF]{Degree Of Freedom}
\acrodefplural{dof}[DOFs]{Degrees Of Freedom}
\acrodef{vef}[VEF]{Vertex, Edge, and Face}
\acrodefplural{vef}[VEFs]{Vertices, Edges, and Faces}
\acrodef{sfc}[SFC]{Space-Filling Curve}
\acrodefplural{sfc}[SFCs]{Space-Filling Curves}
\acrodef{tm}[TM]{Tetrahedral Morton}
\acrodefplural{vef}[VEFs]{Vertices, Edges and Faces}
\acrodef{bddc}[BDDC]{Balancing Domain Decomposition by Constraints}
\acrodef{dg}[DG]{Discontinuous Galerkin}
\acrodef{cg}[CG]{Continuous Galerkin}
\acrodef{mpi}[MPI]{Message Passing Interface}
\acrodef{ip}[IP]{Interior Penalty}
\acrodef{bc}[BC]{Boundary Condition}
\acrodef{bcs}[BCs]{Boundary Conditions}
\acrodef{spd}[SPD]{Symmetric Positive Definite}

\acrodef{fe}[FE]{Finite Element}
\acrodefplural{fe}[FEs]{Finite Elements}
\acrodef{pde}[PDE]{Partial Differential Equation}
\acrodefplural{pde}[PDEs]{Partial Differential Equations}
\acrodef{amr}[AMR]{Adaptive Mesh Refinement and coarsening}
\acrodef{sfc}[SFC]{Space-Filling Curve}
\acrodef{amg}[AMG]{Algebraic MultiGrid}
\acrodef{hpc}[HPC]{High Performance Computing}
\acrodef{oo}[OO]{Object-Oriented}
\acrodef{bddc}[BDDC]{Balancing Domain Decomposition by Constraints}
\acrodef{mpi}[MPI]{Message Passing Interface}
\acrodef{cse}[CSE]{Computational Science \& Engineering}
\acrodefplural{ilu}[ILUs]{Incomplete LU decompositions}
\acrodef{tbp}[TBP]{Type-Bound Procedure}
\acrodef{dd}[DD]{Domain Decomposition}
\acrodef{dd}[DD]{Domain Decomposition}
\acrodef{cla}[CLA]{Command-Line-Argument}
\acrodefplural{cla}[CLAs]{Command-Line-Arguments}
\acrodef{cli}[CLI]{Command-Line-Interface}

\def\FEMPAR{{\texttt{FEMPAR}}}
\def\p4est{{\texttt{p4est}}}

\def\tutorial#1{\texttt{tutorial\_#1}}
\def\Tutorial#1{\texttt{Tutorial\_#1}}

\begin{document}

\title[A tutorial-driven introduction to \FEMPAR{} v1.0.0]{A tutorial-driven introduction to the parallel finite element library \FEMPAR{} v1.0.0}

\author[S. Badia]{Santiago Badia$^{1,2}$}

\author[A. F. Mart\'in]{Alberto F. Mart\'in$^{2,3}$}

\thanks{$^1$ School of Mathematics, Monash University, Clayton, Victoria, 3800, Australia. ---$^2$ Centre Internacional de M\`etodes Num\`erics en Enginyeria, Esteve Terrades 5, E-08860 Castelldefels, Spain. $^3$ Universitat Polit\`ecnica de Catalunya, Jordi Girona 1-3, Edifici C1, E-08034 Barcelona. \\ E-mails: {\tt santiago.badia@monash.edu} (SB), {\tt amartin@cimne.upc.edu} (AM)
}

\thanks{SB gratefully acknowledges the support received from the Catalan Government through the ICREA Acad\`emia Research Program. The authors thankfully acknowledge the computer resources at Marenostrum-IV and the technical support provided by the Barcelona Supercomputing Center (RES-ActivityID: FI-2019-1-0007).}
\date{\today}

\begin{abstract}
  This work is a user guide to the \FEMPAR{} scientific software library. \FEMPAR{} is an open-source object-oriented framework for the simulation of partial differential equations (PDEs) using finite element methods on distributed-memory platforms. It provides a rich set of tools for numerical discretization and built-in scalable solvers for the resulting linear systems of equations. An application expert that wants to simulate a PDE-governed problem has to extend the framework with a description of the weak form of the PDE at hand (and additional perturbation terms for non-conforming approximations). We show how to use the library by going through three different tutorials. The first tutorial simulates a linear PDE (Poisson equation) in a serial environment for a structured mesh using both continuous and discontinuous Galerkin finite element methods. The second tutorial extends it with adaptive mesh refinement on octree meshes. The third tutorial is a distributed-memory version of the previous one that combines a scalable octree handler and a scalable domain decomposition solver. The exposition is restricted to linear PDEs and simple geometries to keep it concise. The interested user can dive into more tutorials available in the \FEMPAR{} public repository to learn about further capabilities of the library, e.g., nonlinear PDEs and nonlinear solvers, time integration, multi-field PDEs, block preconditioning, or unstructured mesh handling.
\end{abstract}

\maketitle


\noindent{{\bf {Keywords}}: Mathematical Software, Finite Elements, Object-Oriented Programming, Partial Differential Equations}

\section{Introduction} \label{sec:introduction}

This work is a user-guide introduction to the first public release, i.e., version 1.0.0, of the scientific software \FEMPAR{}. \FEMPAR{} is an \ac{oo} framework for the numerical approximation of \acp{pde} using \acp{fe}. From inception, it has been designed to be scalable on supercomputers and to easily handle multiphysics problems.  \FEMPAR{} v1.0.0 has about 300K lines of code written in (mostly) \ac{oo} Fortran using the features defined in the 2003 and 2008 standards of the language. \FEMPAR{} is publicly available in the Git repository \href{https://github.com/fempar/fempar}{\texttt{https://github.com/fempar/fempar}}.

\FEMPAR{} \ac{fe} technology includes not only arbitrary order Lagrangian \acp{fe}, but also curl- and div-conforming ones. The library supports n-cube and n-simplex  meshes. Continuous and discontinuous spaces can be used, providing all the machinery for the integration of  \ac{dg} terms on facets. It also provides support for dealing with hanging nodes in non-conforming meshes resulting from $h$-adaptivity. In all cases, \FEMPAR{} provides mesh partitioning tools for unstructured meshes and parallel octree handlers for $h$-adaptive simulations (relying on \p4est \cite{burstedde_p4est_2011}), together with algorithms for the parallel generation of \acp{fe} spaces (e.g., global numbering of \acp{dof} across processors).

\FEMPAR{} has been applied to a broad set of applications that includes the simulation of turbulent flows \cite{colomes_assessment_2015,colomes_segregated_2016,colomes_mixed_2016,colomes_segregated_2017}, magnetohydrodynamics \cite{badia_unconditionally_2013,badia_unconditionally_2013-1,planas_approximation_2011,smolentsev_approach_2015,badia_analysis_2015}, monotonic \acp{fe} \cite{badia_discrete_2015,badia_monotonicity-preserving_2014,hierro_shock_2016,badia_monotonicity-preserving_2017,badia_differentiable_2017}, unfitted \acp{fe} and embedded boundary methods \cite{Badia2018-agg-stokes,Verdugo2019},  additive manufacturing simulations \cite{badia_am,Neiva2018,Neiva2019}, and electromagnetics and superconductors \cite{Olm2019,Olm2019a}.

\FEMPAR{} includes a highly scalable built-in numerical linear algebra module based on state-of-the-art domain decomposition solvers; {the multilevel \ac{bddc} solver in \FEMPAR{} has scaled up to 1.75 million MPI tasks in the JUQUEEN Supercomputer \cite{badia_scalability_2015,badia_multilevel_2016}.} This linear algebra framework has been designed to efficiently tackle the linear systems that arise from \ac{fe} discretizations, exploiting the underlying mathematical structure of the \acp{pde}. It is a difference with respect to popular multi-purpose linear algebra packages like PETSc \cite{balay_petsc_2016}, for which \FEMPAR{} also provides wrappers. The library also supplies block preconditioning strategies for multiphysics applications \cite{elman_finite_2005}. The numerical linear algebra suite is very customizable and has already been used for the implementation and scalability analysis of different \ac{dd} solvers \cite{badia_implementation_2013,badia_scalability_2015,badia_enhanced_2013,badia_multilevel_2016,badia_balancing_2016,Badia2019-physics,badia_highly_2014,badia_space-time_2017,badia_robust_2017,Badia2019-subobjects,Badia2019-electromagnetic-solvers} and block preconditioners for multiphysics applications \cite{badia_block_2014}.

The design of such a large numerical library is a tremendous task. A comprehensive presentation of the underlying design of the building blocks of the library can be found in \cite{badia_fempar:_2017}. This reference is more oriented to \FEMPAR{} developers that want to extend or enhance the library capabilities. Fortunately, users whose application requirements are already fulfilled by \FEMPAR{} do not require to know all these details, which can certainly be overwhelming. In any case, \cite[Sect. 3]{badia_fempar:_2017} can be a good introduction to the main abstractions in a \ac{fe} code.

Even though new capabilities are steadily being added to the library, \FEMPAR{} is in a quite mature state. Minor changes on the interfaces relevant to users have been made during the last two years. It has motivated the recent public release of its first stable version and the elaboration of this user guide.

This user guide is tutorial-driven. We have designed three different tutorials covering an important part of \FEMPAR{} capabilities. The first tutorial addresses a Poisson problem with a known analytical solution that exhibits an internal layer. It considers both a continuous \ac{fe} and a \ac{ip} \ac{dg} numerical discretization of the problem. The second tutorial builds on the first one, introducing an adaptive mesh refinement strategy. The third tutorial consists in the parallelization of the second tutorial for distributed-memory machines and the set up of a scalable \ac{bddc} preconditioner at the linear solver step.

In order to keep the presentation concise, we do not aim to be exhaustive. Many features of the library, e.g., the set up of time integrators and nonlinear solvers or multiphysics capabilities, have not been covered here. Instead, we encourage the interested reader to explore the tutorials section in the Git repository \href{https://github.com/fempar/fempar}{\texttt{https://github.com/fempar/fempar}}, where one can find the tutorials presented herein and other (more advanced) tutorials that make use of these additional tools.

\section{Brief overview of \FEMPAR{} main software abstractions} \label{sec:fempar_sw_abstractions}

In this section, we introduce the main mathematical abstractions in \ac{fe} problems that are provided by \FEMPAR{}. In any case, we refer the reader to~\cite[Sect. 3.1]{badia_fempar:_2017} for a more comprehensive exposition. Typically, a \FEMPAR{} user aims to approximate a (system of) \acp{pde} posed in a bounded physical domain $\Omega$  stated in weak form as follows: find $u_g \in V_g$ such that
\begin{align}
\blform{u_g}{v} = \rhsform{v}, \, \qquad \hbox{for any }  v \in V, \label{eq-poisson-weak}
\end{align}
where $\blform{u}{v}$ and $\rhsform{v}$ are the corresponding bilinear and linear forms of the problem. $V_g$ is a Hilbert space supplemented with possibly non-homogeneous Dirichlet \acp{bc} on the Dirichlet boundary $\Gamma_D \subseteq \partial \Omega$, whereas $V$ is the one with homogeneous \acp{bc} on $\Gamma_D$. In any case, the non-homogeneous problem can easily be transformed into a homogeneous one by picking an arbitrary $Eg \in V_g$ and solving \eqref{eq-poisson-weak} with the right-hand side $\ell_g(v) \doteq \rhsform{v} - \blform{Eg}{v}$ for $u = u_g - Eg \in V$.

The numerical discretization of these equations relies on the definition of a finite-dimensional space $V_{h}$ that is a good approximation of $V$, i.e., it satisfies some approximability property. Conforming approximations are such that $V_{h} \subset V$. Instead, non-conforming approximations, e.g., \ac{dg} techniques, violate this inclusion but make use of perturbed version $a_h$ and $\ell_{g,h}$ of $a$ and $\ell_g$, resp. Using \ac{fe} methods, such finite-dimensional spaces can be defined with a mesh $\triang$ covering $\Omega$, a local \ac{fe} space on every cell of the mesh (usually defined as a polynomial space in a reference \ac{fe} combined with a geometrical map), and a global numbering of \acp{dof} to provide trace continuity across cells. In the most general case, the \ac{fe} problem can be stated as: find $u_h \in V_h$ such that
\begin{align}
a_h(u_h,v) = \ell_{g,h}(v_h), \, \qquad \hbox{for any }  v_h \in V_h. \label{eq-poisson-weak-fem}
\end{align}
One can also define the affine operator
\begin{align}\label{problem1h-op} \mathcal{F}_h(u_h) = a_h(u_h,\cdot) - \ell_{g,h}(\cdot) \in V_h',\end{align}
and alternatively state \eqref{eq-poisson-weak-fem} as finding the root of $\mathcal{F}_h$: find $u_h \in V_h$ such that $\mathcal{F}_h(u_h) = 0$.

If we denote as $N_h$ the dimension of  $V_h$, any
discrete function $u_h \in V_h$ can be uniquely represented by
a vector $\fesol \in \mathbb{R}^{N_h}$ as
$u_h = \sum_{b=1}^{N_h} \shapetest{b} \fesol_b$, where $\{\shapetest{b}\}_{b=1}^{N_h}$ is the canonical basis of (global) shape functions of $V_h$ with respect to the \acp{dof} of $V_h$~\cite[Sect. 3]{badia_fempar:_2017}. With these ingredients, \eqref{eq-poisson-weak-fem} can be re-stated as the solution of a linear system of equations $\fematrix \fesol = \ferhs$, with $\fematrix_{ab} \doteq \blformh{\shapetest{b}}{\shapetest{a}}$ and $\ferhs_{a} \doteq \ell_{g,h}({\shapetest{a}})$. The \ac{fe} affine operator in \eqref{problem1h-op} can be represented as $\mathcal{F}_h(u_h) \doteq \fematrix \fesol - \ferhs$, i.e., a matrix and a vector of size ${|\mathcal{N}_h|}$.

The global \ac{fe} space can be defined as cell-wise local \ac{fe} spaces $V_h|_K$ with a basis $\{\shapetest{a}_K\}$ of local shape functions, for $a,b = 1, \ldots, \mathrm{dim}(V_h|_K)$, and an index map $[\cdot]$ that transforms local \ac{dof} identifiers into global \ac{dof} identifiers. Furthermore, it is assumed that the bilinear form can be split into cell-wise contributions of cell-local shape functions for conforming \ac{fe} formulations, i.e.,
\begin{align}
	\blform{u}{v} = \sum_{K \in \triang} \blformlocal{K}{u|_K}{v|_K}.\label{eq:cellwise-blf}
\end{align}
In practice, the computation of $\fematrix$ makes use of this cell-wise expression. For every cell $K \in \triang$,
one builds a cell matrix $(\fematrix_K)_{a b}\doteq\blformlocal{K}{\shapetest{a}_{K}}{\shapetest{b}_{K}}$ and cell vector $(\ferhs_K)_{a}\doteq\rhsformlocal{K}{\shapetest{a}_{K}}$. Then, these are assembled into $\fematrix$ and $\ferhs$, resp., as $\fematrix_{[a][b]} \pluseq (\fematrix_K)_{a b}$ and $\ferhs_{[a]} \pluseq (\ferhs_K)_{a}$, where $[\cdot]$ is an index map that transforms local \ac{dof} identifiers into global \ac{dof} identifiers.
The cell-local
  bilinear form can be has the form:
  $$\blformlocal{\geophy}{\shapetest{b}_K}{\shapetrial{a}_K} = \int_\geophy \integrandfe(\x) \ddom,$$
where the evaluation of $\integrandfe(\x)$ involves the evaluation of shape function derivatives. The integration is never performed on the cell in the physical space. Instead, the cell (in the physical space) is usually expressed as a geometrical map $\geomap_K$ over a reference cell (e.g., the $[-1,1]^3$ cube for hexahedral meshes), and integration is performed at the reference space. Let us represent the Jacobian of the geometrical mapping with $ \jacobian \doteq \frac{\partial \geomap_K}{\partial \x}$. We can rewrite the cell integration in the reference cell, and next consider a quadrature rule ${\rm Q}$ defined by a set of points/weights $(\xh_{\rm gp}, {\rm w}_{\rm gp})$, as follows:
\begin{align}\label{fematrix}
  \int_\geophy \integrandfe(\x) {\rm d} \Omega
  = \int_\georef \integrandfe \circ \geomap(\x) |\jacobian| {\rm d}\Omega
  = \sum_{{\xh_{\rm gp} \in \quadrature}} \integrandfe \circ \geomap(\xh_{\rm gp}) {\rm w}(\xh_{\gp}) |\jacobian(\xh_{\gp})|.
  \end{align}

\def\jacobf{\boldsymbol{J}_F}
For \ac{dg} methods,the additional stabilization terms should also be written as the sum of cell or facet-wise contributions. In this case, the computation of the matrix entries involves numerical integration in cells and facets (see Sect.~\ref{sec:dg_discrete_int} for an example). A facet is shared by two cells that we represent with $K^+$ and $K^-$. All the facet terms in a \ac{dg} method can be written as the facet integral of an operator over the trial shape functions of $K^+$ or $K^-$ times an operator over the test shape functions
of $K^+$ or $K^-$, i.e.,
$$\blformlocal{F}{{\shapetest{b}_{K^{\alpha}}}}{{\shapetrial{a}_{K^{\beta}}}} = \int_F \integrandfe^{\alpha,\beta}(\x) {\rm d}F, \quad \hbox{for } \alpha, \, \beta \in \{+,-\}.$$
Thus, we have four possible combinations of local facet matrices. As for cells, we can consider a reference facet $\hat{F}$, and a mapping $\geomap_F : \hat{F} \rightarrow F$ from the reference facet to the every facet of the triangulation (in the physical
space). Let us represent the Jacobian of the geometrical mapping with
$ \jacobf \doteq \frac{\partial {\geomap}_F}{\partial \x}$, which has values  in $\mathbb{R}^{(d-1) \times d}$. We can
rewrite the facet integral in the reference facet, and next consider
a quadrature rule ${\rm Q}$ on $\hat{F}$ defined by a set of points/weights
$(\xh_{\rm gp}, {\rm w}_{\rm gp})$, as follows:
\begin{align}\label{facematrix}
  \int_F \integrandfe^{\alpha,\beta}(\x) {\rm d} \Omega
  = \int_{\hat{F}} \integrandfe^{\alpha,\beta} \circ \geomap_F(\x) |\jacobf| {\rm d}F
  = \sum_{{\xh_{\rm gp} \in \quadrature}} \integrandfe^{\alpha,\beta} \circ \geomap_F(\xh_{\rm gp}) {\rm w}(\xh_{\gp}) |\jacobf(\xh_{\gp})|.
\end{align} $|\facejacobian|$ is defined as:
\begin{equation} \label{eq:face_jacobian_measure}
|\facejacobian|=\left \| \frac{{\rm d}\geomap_\facephy}{{\rm d} x} \right \|_2 \
\mbox{ and } \ |\facejacobian|=\left \| \frac{\partial\geomap^1_\facephy}{\partial \xh} \times \frac{\partial\geomap_\facephy^2}{\partial \xh} \right \|_2,
\end{equation}
for $d=2,3$, respectively.

A \FEMPAR{} user must explicitly handle a set of data types that represent some of the previous mathematical abstractions. In particular,
 the main software abstractions in \FEMPAR{} and their roles in the solution of the problem are:
\begin{itemize}
	\item \texttt{triangulation\_t}: The triangulation $\triang$, which represents a partition of the physical domain $\Omega$ into polytopes (e.g., tetrahedra or hexahedra).
	\item \texttt{fe\_space\_t}: The \ac{fe} space, which represents both the test space $V_h$ (with homogeneous \acp{bc}) and the non-homogeneous \ac{fe} space (by combining $V_h$ and $Eg$). It requires as an input the triangulation $\triang$ and (possibly) the Dirichlet data $g$, together with other additional parameters like the order of the approximation.
	\item \texttt{fe\_function\_t}: A \ac{fe} function $u_h \in V_h$, represented with the corresponding \ac{fe} space (where, e.g., the Dirichlet boundary data is stored) and the free \ac{dof} values.
	\item \texttt{discrete\_integration\_t}: The discrete integration is an abstract class to be extended by the user, which computes the cell-wise matrices by integrating  $a_K(\shapetest{b}_K,\shapetest{a}_K)$ and $\ell_K(\shapetest{a}_K)$ (analogously for facet terms). At this level, \FEMPAR{} provides a set of tools required to perform numerical integration (e.g., quadratures and geometrical maps) described in Sect.~\ref{sec:disc_int_conf} for cell integrals and in Sect.~\ref{sec:dg_discrete_int} for facet integrals.
	\item \texttt{fe\_affine\_operator}: The linear (affine) operator $\mathcal{F}_h$, defined in terms of a \ac{fe} space and the discrete forms $a_h$ and $\ell_{h}$. It provides $\fematrix$ and $\ferhs$.
  \item \texttt{quadrature\_t}: A simple type that contains the integration point coordinates and weights in the reference cell/facet.
\end{itemize}
The user also interacts with a set of data types for the solution of the resulting linear system of equations, providing interfaces with different direct and iterative Krylov subspace solvers and preconditioners (either provided by \FEMPAR{} or an external library). \FEMPAR{} also provides some visualization tools for postprocessing the computed results.

\section{Downloading and installing \FEMPAR{} and its tutorial programs} \label{sec:installing_fempar}

The quickest and easiest way to start with \FEMPAR{} is using Docker. Docker is a tool designed to easily create, deploy, and run applications by using containers. \FEMPAR{} provides a set of Docker containers with the required environment (serial or parallel, debug or release) to compile the project source code and to run tutorials and tests. A detailed and very simple installation guide can be found in  \href{https://github.com/fempar/fempar}{\texttt{https://github.com/fempar/fempar}}, together with instructions for the compilation of the tutorial programs explained below.

\section{Common structure and usage instructions of \FEMPAR{} tutorials} \label{sec:common_tutorial_structure}

A \FEMPAR{} tutorial is a \ac{fe} application program that uses the tools (i.e., Fortran200X derived data types and their \acp{tbp}) provided by \FEMPAR{} in order to approximate the solution of a \ac{pde} (or, more generally, a system of such equations).
We strive to follow a common structure for all \FEMPAR{} tutorials in the seek of uniformity and ease of presentation of the different features of the library. Such structure is sketched in Listing~\ref{lst:tutorial_structure}.
Most of the code of tutorial programs is encompassed within a single program unit.
Such unit is in turn composed of four parts:~1)~import of external module symbols (Lines~\ref{loc:tutorial_0X_fempar_names}-\ref{loc:tutorial_0X_rest_support_modules}); 2)~declaration of tutorial parameter constants and variables (Lines~\ref{loc:tutorial_0X_constants}-\ref{loc:tutorial_0X_cla_values}); 3)~the main executable code of the tutorial (Lines~\ref{loc:tutorial_0X_fempar_init}-\ref{loc:tutorial_0X_fempar_finalize}); and 4)~implementation of helper procedures within the \texttt{contains} section of the program unit (Line~\ref{loc:tutorial_0X_helper_procedures}).

\begin{lstlisting}[float=htbp,language={[03]Fortran},escapechar=@,caption=Structure of a prototypical \FEMPAR{} tutorial. The symbol \texttt{\#} denotes the tutorial identifier; e.g. \texttt{tutorial\_01\_...},label={lst:tutorial_structure}]
program tutorial_#_...
  use fempar_names @\label{loc:tutorial_0X_fempar_names}@
  use tutorial_#_support_module1_names @\label{loc:tutorial_0X_support_modules}@
  use tutorial_#_support_module2_names
  ... ! Usage of the rest of support modules @\label{loc:tutorial_0X_rest_support_modules}@
  implicit none
  ... ! Declaration of tutorial_#_... parameter constants @\label{loc:tutorial_0X_constants}@
  ... ! Declaration of data type instances required by tutorial_#_... @\label{loc:tutorial_0X_instances}@
  ... ! Declaration of variables storing CLA values particular to tutorial_#_... @\label{loc:tutorial_0X_cla_values}@
  call fempar_init() ! (i.e., construct system-wide objects) @\label{loc:tutorial_0X_fempar_init}@
  call setup_parameter_handler() @\label{loc:tutorial_0X_setup_parameter_handler}@
  call get_tutorial_cla_values() @\label{loc:tutorial_0X_get_clas}@
  ... ! Calls to the rest of helper procedures within the contains section  @\label{loc:tutorial_0X_call_helpers}@
      ! in order to drive all the necessary steps in the FE simulation
  call fempar_finalize() ! (i.e., destroy system-wide objects)  @\label{loc:tutorial_0X_fempar_finalize}@
contains
  ... ! Implementation of helper procedures @\label{loc:tutorial_0X_helper_procedures}@
end program tutorial_#_...
\end{lstlisting}

In part 1), the tutorial uses the \texttt{fempar\_names} module to import all \FEMPAR{} library symbols (i.e., derived types, parameter constants, system-wide variables, etc.), and a set of tutorial-specific module units, which are not part of the \FEMPAR{} library, but developed specifically for the problem at hand. Each of these modules defines a tutorial-specific data type and its \acp{tbp}. Although not necessarily, these are typically type extensions (i.e., subclasses) of parent classes defined within \FEMPAR{}. These data type extensions let the user define problem-specific ingredients such as, e.g., the source term of the \ac{pde}, the function to be imposed on the Dirichlet and/or Neumann boundaries, or the definition of a discrete weak form suitable for the problem at hand, while leveraging (re-using) the code within \FEMPAR{} by means of Fortran200X native support of run-time polymorphism.

In part 2), the tutorial declares a set of parameter constants, typically the tutorial name, authors, problem description, etc. (to be output on screen on demand by the user), the tutorial data type instances in charge of the \ac{fe} simulation, such as the triangulation (mesh) of the computational domain or the \ac{fe} space from which the approximate solution of the \ac{pde} is sought (see Sect.~\ref{sec:fempar_sw_abstractions}), and a set of variables to hold the values of the \acp{cla} which are specific to the tutorial. As covered in the sequel in more detail, tutorial users are provided with a \ac{cli}. Such interface constitutes the main communication mechanism to provide the input required by tutorial programs apart from, e.g., mesh data files generated from the GiD \cite{_gid_2016} unstructured mesh generator (if the application problem requires such kind of meshes).

Part 3) contains the main tutorial executable code, which is in charge of driving all the necessary \ac{fe} simulation steps. This code in turn relies on part 4), i.e., a set of helper procedures implemented within the \texttt{contains} section of the program unit. The main tasks of a \ac{fe} program (and thus, a \FEMPAR{} tutorial), even for transient, non-linear \ac{pde} problems, typically encompass:
a)~to set up a mesh and a \ac{fe} space;
b)~to assemble a discrete linear or linearized algebraic system of equations; c)~to solve the system built in b); d)~to numerically post-process and/or visualize the solution. As will be seen along the paper, there is an almost one-to-one correspondence among these tasks and the helper procedures within the \texttt{contains} section.

The main executable code of the prototypical \FEMPAR{} tutorial in Listing~\ref{lst:tutorial_structure} is (and {\em must be}) encompassed within calls to \texttt{fempar\_init()} (Line~\ref{loc:tutorial_0X_fempar_init}) and \texttt{fempar\_finalize()} (Line~\ref{loc:tutorial_0X_fempar_finalize}). The former constructs/initializes all system-wide objects, while the latter performs the reverse operation.
For example, in the call to \texttt{fempar\_init()}, a system-wide dictionary of creational methods for iterative linear solver instances is set up. Such dictionary lays at the kernel of a {\em Creational} \ac{oo} design pattern \cite{gamma_e._design_1995,freeman_head_2004} that lets \FEMPAR{} users to add new iterative linear solver implementations {\em without the need to recompile the library at all}. Apart from these two calls, the tutorial main executable code also calls the
\texttt{setup\_parameter\_handler()} and \texttt{get\_tutorial\_cla\_values()} helper procedures in
Lines~\ref{loc:tutorial_0X_setup_parameter_handler} and~\ref{loc:tutorial_0X_get_clas}, resp., which are related to \ac{cli} processing, and covered in the sequel.

The code of the \texttt{setup\_parameter\_handler()} helper procedure is shown in Listing~\ref{lst:setup_parameter_handler}. It sets up the so-called \texttt{parameter\_handler} system-wide object, which is  directly connected with the tutorial \ac{cli}. The \texttt{process\_parameters} \ac{tbp} registers/defines a set of \acp{cla} to be parsed, parses the \acp{cla} provided by the tutorial user through the \ac{cli}, and internally stores their values into a parameter dictionary of \textless{\em key},{\em value}\textgreater pairs;
 a pointer to such dictionary can be obtained by calling \texttt{parameter\_handler\%get\_values()} later on.\footnote{This parameter dictionary, with type name \texttt{parameterlist\_t}, is provided by a stand-alone external software library called \texttt{FPL} \cite{FPL}.}

\begin{lstlisting}[float=htbp,language={[03]Fortran},escapechar=@,caption=The \texttt{setup\_parameter\_handler()} tutorial helper procedure.,label={lst:setup_parameter_handler}]
subroutine setup_parameter_handler()
  call parameter_handler%process_parameters(& 
       define_user_parameters_procedure=define_tutorial_clas,& 
       progname=tutorial_name,&
       version=tutorial_version,&
       description=tutorial_description,&
       authors=tutorial_authors)
end subroutine setup_parameter_handler
\end{lstlisting}

 There are essentially two kind of \acp{cla} registered by \texttt{process\_parameters}. {\em On the one hand}, \FEMPAR{} itself registers a large bunch of \acp{cla}. Each of these \acp{cla} corresponds one-to-one to a particular \FEMPAR{} derived type.
 The data type a \ac{cla} is linked with can be easily inferred from the convention followed for \FEMPAR{}  \ac{cla} names, which prefixes the name of the data type (or an abbreviation of it) to the \ac{cla} name.
 Many of the \FEMPAR{} data types require a set of parameter values in order to customize their behaviour and/or the way they are set up.
 These data types are designed such that these parameter values may be provided by an instance of the aforementioned parameter dictionary. Thus, by extracting the parameter dictionary stored within \texttt{parameter\_handler}, and passing it to the \FEMPAR{} data type instances, one directly connects the \ac{cli} with the instances managed by the \ac{fe} program. This is indeed the mechanism followed by all tutorial programs. In any case, \FEMPAR{} users are not forced to use this mechanism in their \ac{fe} application programs. They can always build and pass an ad-hoc parameter dictionary to the corresponding instance, thus by-passing the parameter values provided to the \ac{cli}.

 {\em On the other hand}, the tutorial program itself (or, in general, any \ac{fe} application program) may optionally register tutorial-specific \acp{cla}.
 This is achieved by providing a user-declared procedure to the optional
 \texttt{define\_user\_parameters\_procedure} dummy argument of \texttt{process\_parameters}. In Listing~\ref{lst:setup_parameter_handler}, the particular procedure passed is called \texttt{define\_tutorial\_clas}. An excerpt of the code of such procedure is shown in Listing~\ref{lst:define_tutorial_clas}. In this listing, the reader may observe that registering a \ac{cla} involves defining a parameter dictionary {\em key} ({\texttt{"FE\_FORMULATION"}), a \ac{cla} name (\texttt{"--FE\_FORMULATION"}), a default value for the \ac{cla} in case it is not passed (\texttt{"CG"}), a help message, and (optionally) a set of admissible choices for the \ac{cla}.

\begin{lstlisting}[float=htbp,language={[03]Fortran},escapechar=@,caption=An excerpt of a tutorial helper procedure that is used to register tutorial-specific \acp{cla}.,label={lst:define_tutorial_clas}]
subroutine define_tutorial_clas()
  call parameter_handler%add("FE_FORMULATION", "--FE_FORMULATION", "CG", &
          help="Select Finite Element formulation for the problem at hand; &
                either Continuous (CG) or Discontinuous Galerkin (DG)", &
          choices="CG,DG")
  ! ... Register the rest of tutorial-specific CLAs
end subroutine define_tutorial_clas
\end{lstlisting}

 The parameter dictionary key passed when registering a \ac{cla} can be used later on in order to get the value of the corresponding \ac{cla} or to override it with a fixed value, thus ignoring the value provided to the \ac{cla}. This is achieved by means of the \texttt{get...()} and \texttt{update()} \acp{tbp} of \texttt{parameter\_handler}.
 Listing~\ref{lst:get_tutorial_cla_values} shows an excerpt of the helper subroutine called in Line~\ref{loc:tutorial_0X_get_clas} of Listing~\ref{lst:tutorial_structure}. This subroutine uses the \texttt{getasstring()} \ac{tbp} of \texttt{parameter\_handler} in order to obtain the {\em string} passed by the tutorial user to the \texttt{"--FE\_FORMULATION"} tutorial-specific \ac{cla}. Examples on the usage of \texttt{update()} can be found, e.g., in Sect.~\ref{sec:tutorial01}.

\begin{lstlisting}[float=htbp,language={[03]Fortran},escapechar=@,caption=An excerpt of a tutorial helper procedure that is used to obtain tutorial-specific \ac{cla} values.,label={lst:get_tutorial_cla_values}]
subroutine get_tutorial_cla_values()
  call parameter_handler%getasstring("FE_FORMULATION", fe_formulation)
  call parameter_handler%get("ALPHA", alpha)
  ... ! Obtain the rest of tutorial-specific CLA values 
end subroutine get_tutorial_cla_values 
\end{lstlisting}

The full set of tutorial \acp{cla}, along with rich help messages, can be output on screen by calling the tutorial program with the \texttt{"--help"} \ac{cla},
while the full list of parameter dictionary of \textless{\em key},{\em value}\textgreater pairs {\em after parsing}, with the \texttt{"--PARAMETER\_HANDLER\_PRINT\_VALUES"} one.
This latter \ac{cla} may be useful to confirm that the tutorial program invocation from command-line produces the desired effect on the values actually handled by the program.

\section{\Tutorial{01}: Steady-state Poisson with a circular wave front} \label{sec:tutorial01}

\subsection{Model problem} \label{sec:tutorial01_model_problem}
\Tutorial{01} tackles the Poisson problem. In strong form this problem reads: find $u$ such that
\begin{align}
-\Delta u = f \qquad \hbox{in } \, \Omega,   \label{eq-poisson}
\end{align}
where $f : \Omega \rightarrow \R$ is a given source term, and $\Omega:=[0,1]^d$ is the unit box domain, with $d:=2,3$ being the number of space dimensions.
Prob. \eqref{eq-poisson} is supplied with inhomogeneous Dirichlet\footnote{Other \acp{bc}, e.g., Neumann or Robin (mixed)
conditions can also be considered for the Poisson problem. While these sort of \acp{bc} are supported by \FEMPAR{} as well, we do not consider them in \tutorial{01} for simplicity.} \acp{bc} $u=g$ on $\partial \Omega$, with $g : \partial \Omega \rightarrow \R$ a given function defined on the domain boundary.
We in particular consider the standard benchmark problem in \cite{Mitchell2011}. The source term $f$ and Dirichlet function $g$ are chosen such that the exact (manufactured) solution of
\eqref{eq-poisson} is:
\begin{align}
u(\x) := \mathrm{arctan}(\alpha(\sqrt{(\x-\x_c)\cdot(\x-\x_c)}-r)). \label{eq-poisson-solution}
\end{align}
This solution has a sharp circular/spherical wave front of radius $r$ centered at $\x_c$.
Fig.\ref{fig-2D_sol} and Fig.~\ref{fig-3D_sol} illustrate the solution for $d=2,3$, resp., and parameter values
$\alpha=200$, $r=0.7$, and $\x_c = (-0.05, -0.05)$, $\x_c = (-0.05, -0.05, -0.05)$ for $d=2,3$, resp.

\begin{figure}[t!]
    \centering
    \begin{subfigure}[t]{0.45\textwidth}
        \includegraphics[width=\textwidth]{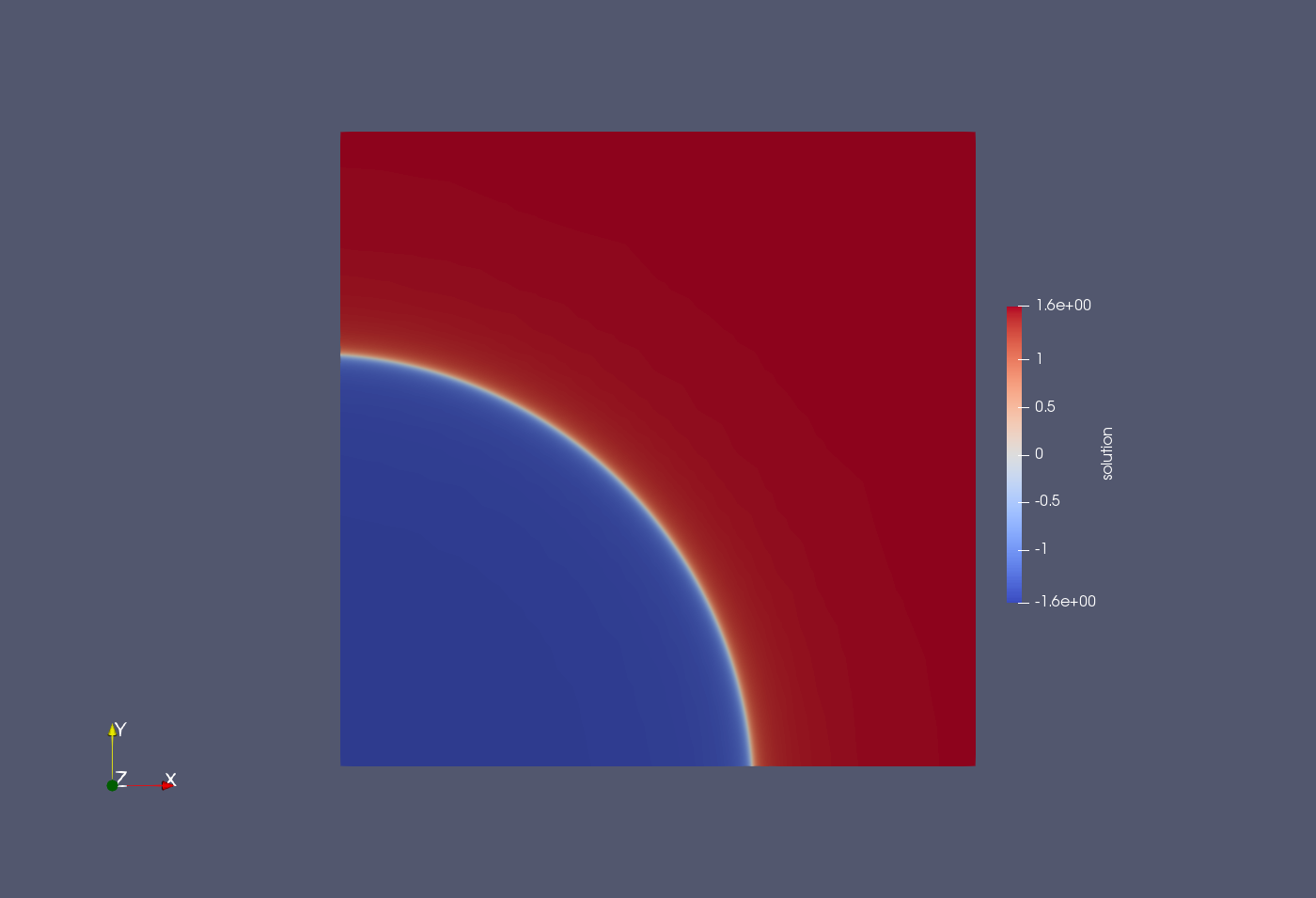}
        \caption{2D benchmark problem.}
        \label{fig-2D_sol}
    \end{subfigure}
    \begin{subfigure}[t]{0.45\textwidth}
        \includegraphics[width=\textwidth]{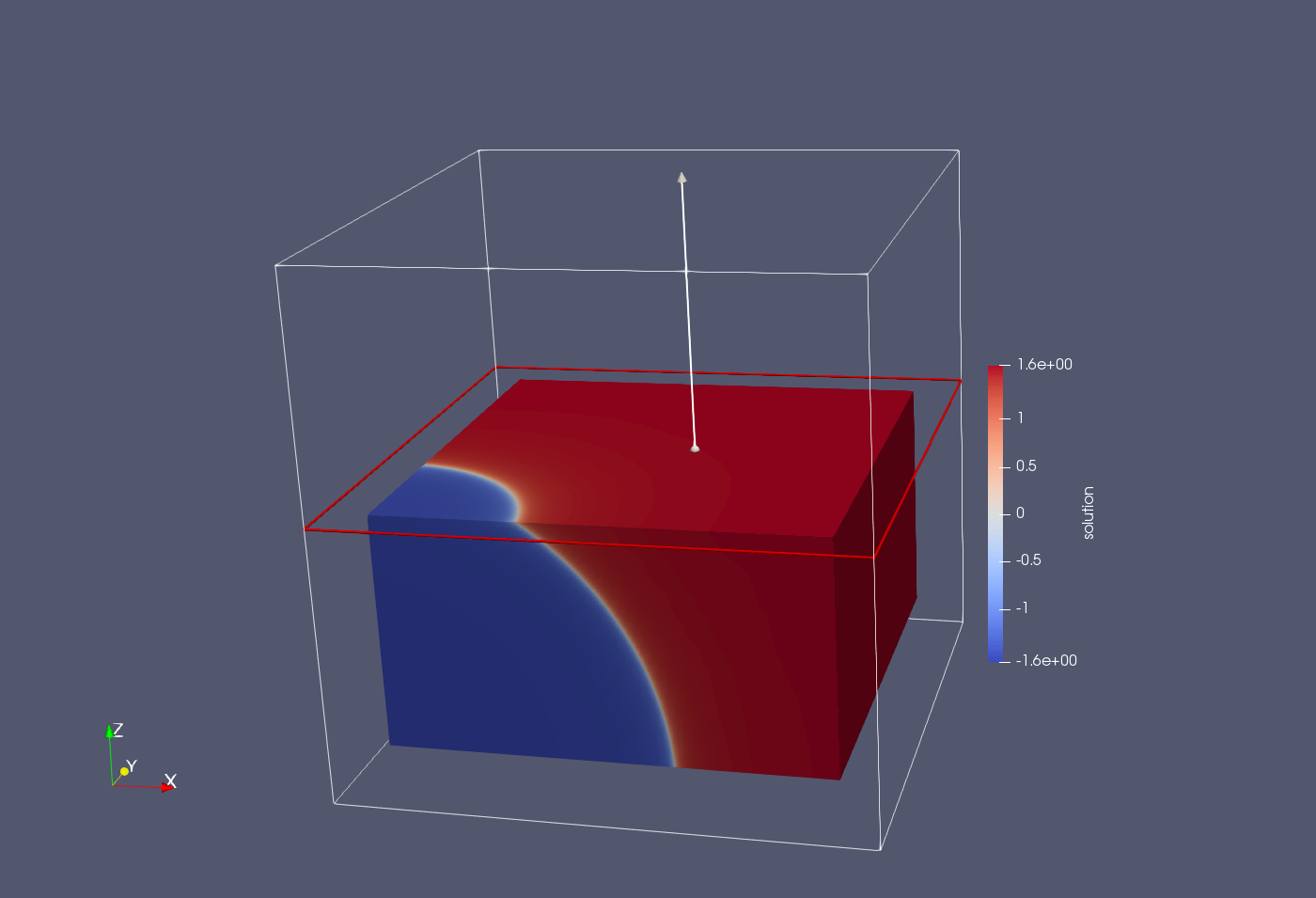}
        \caption{3D benchmark problem.}
        \label{fig-3D_sol}
    \end{subfigure}
    \caption{Manufactured solution \eqref{eq-poisson-solution} for Prob. \eqref{eq-poisson}.}\label{fig:solution_poisson_problem}
\end{figure}

\subsection{\ac{fe} discretization} \label{sec:tutorial01_fe_discretization}

\Tutorial{01} implements two different \ac{fe} formulations for the Poisson problem. On the one hand, a conforming \ac{cg} formulation, which is covered in Sect.~\ref{sec:cg_formulation_poisson}, and a non-conforming \ac{dg} one, covered in Sect.~\ref{sec:dg_formulation_poisson}. In this tutorial, both formulations are used in combination with a uniform (thus conforming) triangulation $\triang$ of $\Omega$ made of quadrilateral/hexahedral cells. Apart from solving \eqref{eq-poisson}, \tutorial{01} also evaluates the \ac{fe} discretization error. In particular, for each cell $K$, it computes the square of the error energy norm, which for the Poisson problem is defined as $e_K^2 := \int_K \grad (u-u_h)\cdot \grad (u-u_h)$, with $u$ and $u_h$ being the exact and \ac{fe} solution, resp. It also records and prints on screen the total error $e:=(\sum_K e_K^2)^{1/2}$. On user-demand, the cell quantities $e_K^2$ can be written to post-processing data files for later visualization; see Sect.~\ref{sec:tutorial01_commented_code} for more details.

\subsubsection{\ac{cg} \ac{fe} formulation} \label{sec:cg_formulation_poisson}

In order to derive a \ac{cg} \ac{fe} formulation, one starts with the weak form of the problem at hand, the Poisson problem in the case of \tutorial{01}. This problem can be stated as in \eqref{eq-poisson-weak} with $\blform{u}{v}:=\int_\Omega \grad u \cdot \grad v$ and $\ell(v) :=\int_\Omega f v$. The weak form \eqref{eq-poisson-weak} is discretized by
replacing $H_0^1(\Omega)$ by a finite-dimensional space $V_h$, without any kind of perturbation. As we aim at building a conforming \ac{fe} formulation, i.e.,  $V_h \subset H_0^1(\Omega)$, we must ensure that the conformity requirements of $H_0^1(\Omega)$ are fulfilled. In particular, we must ensure that the functions in $V_h$ have continuous trace across cell interfaces. We refer to ~\cite[Sect. 3]{badia_fempar:_2017} for a detailed exposition of how $H_0^1(\Omega)$-conforming \ac{fe} spaces can be built using the so-called Lagrangian (a.k.a. nodal) \acp{fe}. When such \ac{fe} is combined with hexahedral cells, $V_h|_K := \mathcal{Q}_q(K)$, i.e., the space of multivariate polynomials that are of degree less or equal to $q$ with respect to each variable in $(x_1, \ldots, x_d) \in K$.  The system matrix can be computed as described in \eqref{eq:cellwise-blf} with $\blformlocal{K}{u}{v}:=\int_K \grad u \cdot \grad v$, $\rhsformlocal{K}{v}:=\int_K f v$ (assuming $f \in L^2(\Omega)$).

\subsubsection{\ac{dg} \ac{fe} formulation} \label{sec:dg_formulation_poisson}

\Tutorial{01} also implements a {\em non-conforming} \ac{dg} formulation for the Poisson problem.
In particular, an \ac{ip} \ac{dg} formulation described in \cite{quarteroni_numerical_2014}.
This formulation, as any other non-conforming discretization method, extracts the discrete solution  $u_h$ from
a global \ac{fe} space $V_h$ which does not conform to $H^1(\Omega)$, i.e., $V_h \not\subset H^1(\Omega)$.
In particular, $V_h$ is composed of functions that
are continuous within each cell, but discontinuous across cells, i.e.,
$V_h = \{ u_h \in L_2(\Omega): u_h|_K \in \mathcal{Q}_q(K) \subset
H^1(K),\ K \in \triang\}$, with $\mathcal{Q}_q(K)$ as defined in Sect.~\ref{sec:cg_formulation_poisson}.
While this extremely simplifies the construction of $V_h$, as one does not have to take care
of the inter-cell continuity constraints required for $H^1(\Omega)$-conformity (see ~\cite[Sect. 3]{badia_fempar:_2017} for more details), one cannot plug $V_h$ directly into \eqref{eq-poisson-weak}, since the original bilinear form has no sense
for a non-conforming \ac{fe} space. Instead, one requires judiciously numerical perturbations of the continuous
bilinear and linear forms in \eqref{eq-poisson-weak} in order to \emph{weakly} enforce conformity.

In the \ac{ip} \ac{dg} \ac{fe} formulation at hand, and in contrast to the one presented in
Sect.~\ref{sec:cg_formulation_poisson} (that imposes essential Dirichlet \acp{bc} strongly)
the condition $u=g$ on $\Omega$ is weakly imposed,
as it is natural in this kind of formulations. If we denote $\mathcal{F}^{\Omega}_{h}$ and
$\mathcal{F}^{\boundary}_{h}$ as the set of interior and boundary
facets of $\triang$, resp., the discrete weak form of \ac{ip} \ac{dg} method implemented by \tutorial{01} reads:
find $u_h \in V_h$ such that
\begin{align} \label{eq-poisson-weak-dg}
\blformh{u_h}{v_h} = \rhsformh{v_h}, \, \qquad \hbox{for any }  v_h \in V_h,
\end{align}
with
\begin{equation} \label{eq:ip_dg_bilinear}
\begin{split}
&  \blformh{u_h}{v_h}  = \sum_{K\in \triang} \int_K \grad u_h \cdot \grad v_h -  \\
& \sum_{\facephy \in \mathcal{F}^{\Omega}_{h}} \int_\facephy \jump{v_h} \cdot \mean{\grad u_h}
 - \tau \sum_{\facephy \in \mathcal{F}^{\Omega}_{h}} \int_\facephy \jump{u_h} \cdot \mean{\grad v_h}
 + \sum_{\facephy \in \mathcal{F}^{\Omega}_{h}} \gamma |\facephy|^{-1} \int_\facephy  \jump{u_h} \cdot \jump{v_h} - \\
&   \sum_{\facephy \in \mathcal{F}^{\boundary}_{h}} \int_\facephy v_h \grad{u_h} \cdot \normal
 -  \tau  \sum_{\facephy \in \mathcal{F}^{\boundary}_{h}} \int_\facephy u_h \grad{v_h} \cdot \normal
 +  \sum_{\facephy \in \mathcal{F}^{\boundary}_{h}} \gamma |\facephy|^{-1} \int_\facephy  u_h v_h,
\end{split}
\end{equation}
and
\begin{equation} \label{eq:ip_dg_linear}
\begin{split}
\rhsformh{v_h} = \sum_{K\in \triang} \int_K f v_h - \tau \sum_{\facephy \in \mathcal{F}^{\boundary}_{h}} \int_\facephy g \grad{v_h} \cdot \normal
 + \sum_{\facephy \in \mathcal{F}^{\boundary}_{h}} \gamma |\facephy|^{-1} \int_\facephy g v_h.
\end{split}
\end{equation}
In \eqref{eq:ip_dg_bilinear} and \eqref{eq:ip_dg_linear}, $\tau$ is a fixed constant that characterizes the particular method at hand, $\gamma$
is a facet-wise positive constant referred to as penalty parameter, and $|\facephy|$ denotes the surface
of the facet; $\tau$ and $\gamma$ should be suitably chosen such that $a_h(\cdot,\cdot)$ is well-posed (stable and continuous) in the discrete setting, and the \ac{fe} formulation enjoys optimal rates of
convergence~\cite{quarteroni_numerical_2014}. Finally, if we denote as $K^+$ and $K^-$ the
two cells that share a given facet, then $\mean{w_h}$ and $\jump{w_h}$ denote mean values and
jumps of $w_h$ across cells facets:
\begin{equation} \label{eq:mean_and_jump_ops}
\mean{w_h} = \frac{w_h^++w_h^-}{2}, \quad  \quad \jump{w_h} = w_h^+ \normal^+ + w_h^- \normal^-,
\end{equation}
with $\normal^+$, $\normal^-$ being the facet outward unit normals, and
$w_h^+$, $w_h^-$ the restrictions of $w_h$ to the facet, both from either the
perspective of $K^+$ and $K^-$, resp.

The  assembly of the cell integrals in \eqref{eq:ip_dg_bilinear} and \eqref{eq:ip_dg_linear} is performed as described in Sect.~\ref{sec:cg_formulation_poisson}. With regard to the facet integrals, assuming that we are sitting on an interior facet $\facephy \in \mathcal{F}^{\Omega}_{h}$, four facet-wise matrices, namely $\fematrix^{\facephy}_{K^+ K^+}$, $\fematrix^{\facephy}_{K^+ K^-}$, $\fematrix^{\facephy}_{K^- K^+}$, and  $\fematrix^{\facephy}_{K^- K^-}$, are computed.\footnote{The case of boundary facets $\facephy \in \mathcal{F}^{\boundary}_{h}$ is just a degenerated case of the one corresponding to interior facets where only a single facet-wise matrix $\fematrix^{\facephy}_{K^+ K^+}$ has to be computed; we omit these facets from the discussion in order to keep the presentation short.} The entries of, e.g., $\fematrix^{\facephy}_{K^+ K^-}$, are defined as:
\begin{equation} \label{eq:local_to_face_matrix}
\left(\fematrix^{\facephy}_{K^+ K^-}\right)_{ab} = -\int_\facephy \jump{\shapetest{b}_{K^-}} \cdot \mean{\grad \shapetest{a}_{K^+}}  -
                                        \tau \int_\facephy \jump{\shapetest{a}_{K^+}} \cdot \mean{\grad \shapetest{b}_{K^-}} +
                                         \gamma |\facephy|^{-1} \int_\facephy  \jump{\shapetest{a}_{K^+}} \cdot \jump{\shapetest{b}_{K^-}},
\end{equation}
for $a,b = 1, \ldots, \mathrm{dim}(\mathcal{Q}_q(K))$. This matrix is assembled
into $\fematrix$ as $\fematrix_{[a][b]} \pluseq (\fematrix^{\facephy}_{K^+ K^-})_{ab}$.

\subsection{The commented code} \label{sec:tutorial01_commented_code}

The main program unit of \tutorial{01} is shown in Listing~\ref{lst:tutorial_01_program_unit}. Apart from \texttt{fempar\_names}, it also uses three tutorial-specific support modules in Lines~\ref{loc:tutorial_01_di_names}-\ref{loc:tutorial_01_error_names}. The one used in Line~\ref{loc:tutorial_01_di_names} implements the data type instances declared in Line~\ref{loc:tutorial_01_cg_di} and~\ref{loc:tutorial_01_dg_di}, the one in Line~\ref{loc:tutorial_01_func_names}, the ones declared in Lines~\ref{loc:tutorial_01_source_term} and~\ref{loc:tutorial_01_exact_solution}, and the one in Line~\ref{loc:tutorial_01_error_names}, the data type instance declared in Line~\ref{loc:tutorial_01_error_estimator}. The rest of data type instances declared in part 2) of the tutorial program (Lines~\ref{loc:tutorial_01_part2_start}-\ref{loc:tutorial_01_part2_stop}) are implemented within \FEMPAR{}.
\Tutorial{01}'s main executable code spans Lines~\ref{loc:tutorial_01_part3_start}-\ref{loc:tutorial_01_part3_stop}.
As there is almost a one-to-one mapping among the data type instances
and the helper procedures called by \tutorial{01}, we will introduce them in the sequel step-by-step along with code snippets of the corresponding helper procedures.
The \texttt{setup\_parameter\_handler} and \texttt{get\_tutorial\_cla\_values} helper procedures have been already introduced in Sect.~\ref{sec:common_tutorial_structure}.
\Tutorial{01} registers \acp{cla} to select the values of $\alpha$, $r$, and $x_c$
(see Sect.~\ref{sec:tutorial01_model_problem}), and the \ac{fe} formulation to be used (see Sect.~\ref{sec:tutorial01_fe_discretization}).

\begin{lstlisting}[float=htbp,language={[03]Fortran},escapechar=@,caption=\Tutorial{01} program unit.,label={lst:tutorial_01_program_unit}]
program tutorial_01_poisson_sharp_circular_wave
  use fempar_names
  use tutorial_01_discrete_integration_names @\label{loc:tutorial_01_di_names}@
  use tutorial_01_functions_names            @\label{loc:tutorial_01_func_names}@
  use tutorial_01_error_estimator_names      @\label{loc:tutorial_01_error_names}@
  ... ! Declaration of tutorial_01_... parameter constants  @\label{loc:tutorial_01_part2_start}@
  type(serial_context_t)                  :: world_context
  type(environment_t)                     :: environment
  type(serial_triangulation_t)            :: triangulation
  type(serial_fe_space_t)                 :: fe_space
  type(strong_boundary_conditions_t)      :: strong_boundary_conditions 
  type(sharp_circular_wave_source_term_t) :: source_term       @\label{loc:tutorial_01_source_term}@
  type(sharp_circular_wave_solution_t)    :: exact_solution    @\label{loc:tutorial_01_exact_solution}@
  type(fe_affine_operator_t)              :: fe_affine_operator
  type(fe_function_t)                     :: discrete_solution
  type(cg_discrete_integration_t), target :: cg_discrete_integration @\label{loc:tutorial_01_cg_di}@
  type(dg_discrete_integration_t), target :: dg_discrete_integration @\label{loc:tutorial_01_dg_di}@
  type(direct_solver_t)                   :: direct_solver
  type(output_handler_t)                  :: output_handler
  type(poisson_error_estimator_t)         :: error_estimator @\label{loc:tutorial_01_error_estimator}@
  ... ! Declaration of variables storing CLA values particular to tutorial_01_... @\label{loc:tutorial_01_part2_stop}@
  call fempar_init()                                              @\label{loc:tutorial_01_part3_start}@
  call setup_parameter_handler()
  call get_tutorial_cla_values()
  call setup_context_and_environment()
  call setup_triangulation()   @\label{loc:tutorial_01_setup_triangulation}@
  call setup_problem_functions()
  call setup_strong_boundary_conditions() @\label{loc:tutorial_01_setup_strong_bcs}@
  call setup_fe_space()
  call setup_discrete_solution() @\label{loc:tutorial_01_setup_discrete_solution}@
  call setup_and_assemble_fe_affine_operator() @\label{loc:tutorial_01_assemble_operator}@
  call solve_system()  
  call compute_error() @\label{loc:tutorial_01_compute_error}@
  call write_postprocess_data_files()
  call free_all_objects()
  call fempar_finalize()                                          @\label{loc:tutorial_01_part3_stop}@
contains
  ... ! Implementation of helper procedures
end program tutorial_01_poisson_sharp_circular_wave
\end{lstlisting}

The \texttt{setup\_context\_and\_environment} helper procedure is shown in Listing~\ref{lst:tutorial_01_environment}. Any \FEMPAR{} program requires to set up (at least) one {\em context} and one {\em environment}. In a nutshell, a {\em context} is a software abstraction for a group of parallel tasks (a.k.a. processes) and the communication layer that orchestrates their concurrent execution. There are several context implementations in \FEMPAR{}, depending on the computing environment targeted by the program. As \tutorial{01} is designed to work in serial computing environments, \texttt{world\_context} is declared of type \texttt{serial\_context\_t}. This latter data type represents a degenerated case in which the group of tasks is just composed by a single task. On the other hand, the {\em environment} organizes the tasks of the context from which it is set up into subgroups of tasks, referred to as levels, and builds up additional communication mechanisms to communicate tasks belonging to different levels.
As \texttt{world\_context} represents a single-task group,
we force the environment to handle just a single level, in turn composed of a single task. This is achieved using the \texttt{parameter\_handler\%update(...)} calls in Lines~\ref{loc:tutorial_01_env_force1} and \ref{loc:tutorial_01_env_force2} of Listing~\ref{lst:tutorial_01_environment}, resp; see Sect.~\ref{sec:common_tutorial_structure}. 
The rationale behind the environment will become clearer in \tutorial{03}, which is designed to work in distributed-memory computing environments using \ac{mpi} as the communication layer.

\begin{lstlisting}[float=htbp,language={[03]Fortran},escapechar=@,caption=The \texttt{setup\_context\_and\_environment} procedure.,label={lst:tutorial_01_environment}]
subroutine setup_context_and_environment()
  call world_context%create()
  call parameter_handler%update(environment_num_levels_key, 1) @\label{loc:tutorial_01_env_force1}@
  call parameter_handler%update(environment_num_tasks_x_level_key, [1]) @\label{loc:tutorial_01_env_force2}@
  call environment%create(world_context, parameter_handler%get_values())
end subroutine setup_context_and_environment
  
\end{lstlisting}

The triangulation $\triang$ of $\Omega$ is set up in Listing~\ref{lst:tutorial_01_triangulation}.
\FEMPAR{} provides a data type hierarchy of triangulations rooted at the so-called \texttt{triangulation\_t} abstract data type~\cite[Sect. 7]{badia_fempar:_2017}. In the most general case, \texttt{triangulation\_t} represents a {\em non-conforming} mesh partitioned into a set of subdomains (i.e., distributed among a set of parallel tasks) that can be $h$-adapted and/or re-partitioned (i.e., re-distributed among the tasks) in the course of the simulation.
\Tutorial{01}, however, uses a particular type extension of \texttt{triangulation\_t}, of type \texttt{serial\_triangulation\_t}, which represents a {\em conforming} mesh centralized on a single task that remains static in the course of the simulation. For this triangulation type, the user may select to automatically generate a uniform mesh for simple domains (e.g., a unit cube),
currently of brick (quadrilateral or hexahedral) cells, or, for more complex domains, import it from mesh data files, e.g., generated by the GiD unstructured mesh generator \cite{_gid_2016}. Listing~\ref{lst:tutorial_01_triangulation} follows the first itinerary.
In particular, in Line~\ref{loc:tutorial_01_triang_force1}, it forces the \texttt{serial\_triangulation\_t} instance to be generated by a uniform mesh generator of brick cells, and in Lines~\ref{loc:tutorial_01_triang_force2_start}-\ref{loc:tutorial_01_triang_force2_stop}, that $\Omega=[0,1]^d$ is the domain to be meshed, as required by our model problem. The rest of parameter values of this mesh generator, such as, e.g., the number of cells per space dimension, are not forced, so that the user may specify them as usual via the corresponding \acp{cla}.
The actual set up of the triangulation occurs in the call at Line~\ref{loc:tutorial_01_triang_force2_set_up}.

\begin{lstlisting}[float=htbp,language={[03]Fortran},escapechar=@,caption=The \texttt{setup\_triangulation} procedure.,label={lst:tutorial_01_triangulation}]
subroutine setup_triangulation()
  integer(ip) :: num_dims
  call parameter_handler%update(static_triang_generate_from_key, & @\label{loc:tutorial_01_triang_force1}@
       static_triang_generate_from_struct_hex_mesh_generator) 
  call parameter_handler%get(struct_hex_mesh_generator_num_dims_key, num_dims) @\label{loc:tutorial_01_triang_force2_start}@
  if ( num_dims == 2 ) then
     call parameter_handler%update(struct_hex_mesh_generator_domain_limits_key, &
          [0.0_rp,1.0_rp,0.0_rp,1.0_rp])
  else
     call parameter_handler%update(struct_hex_mesh_generator_domain_limits_key, &
          [0.0_rp,1.0_rp,0.0_rp,1.0_rp,0.0,1.0_rp])
  end if @\label{loc:tutorial_01_triang_force2_stop}@
  call triangulation%create(environment, parameter_handler%get_values())@\label{loc:tutorial_01_triang_force2_set_up}@
end subroutine setup_triangulation
\end{lstlisting}

Listing~\ref{lst:tutorial_01_functions} sets up the \texttt{exact\_solution} and \texttt{source\_term} objects. These represent the exact (analytical) solution $u$ and source term $f$ of our problem. The program does not need to implement the Dirichlet function $g$, as it is equivalent to $u$ in our model problem.
While we do not cover their implementation here, the reader is encouraged to inspect the \texttt{tutorial\_01\_functions\_names} module. Essentially, this module implements a pair of program-specific data types extending the so-called \texttt{scalar\_function\_t} \FEMPAR{} data type. This latter data type represents an scalar-valued function $h$, and provides interfaces for the evaluation of $h(\x)$, $\grad h(\x)$, etc., with
$\x \in \closure[1]{\Omega}$, to be implemented by type extensions. In particular, \tutorial{01} requires $u(\x)$ and $\grad u(\x)$ for the imposition of Dirichlet \acp{bc}, and the evaluation of the energy norm, resp., and $f(\x)$ for the evaluation of the source term.

\begin{lstlisting}[float=htbp,language={[03]Fortran},escapechar=@,caption=The \texttt{setup\_problem\_functions} subroutine.,label={lst:tutorial_01_functions}]
subroutine setup_problem_functions()
  call source_term%create(triangulation%get_num_dims(),& 
       alpha,circle_radius,circle_center)
  call exact_solution%create(triangulation%get_num_dims(),&
       alpha,circle_radius,circle_center)
end subroutine setup_problem_functions
\end{lstlisting}

Listing~\ref{lst:tutorial_01_bcs} sets up
the \texttt{strong\_boundary\_conditions} object.
With this object one can define the regions of the domain boundary on which to impose strong \acp{bc}, along with the function to be imposed on each of these regions.
It is required for the \ac{fe} formulation in Sect.~\ref{sec:cg_formulation_poisson} as, in this method, Dirichlet \acp{bc} are imposed strongly.
It is not required by the \ac{ip} \ac{dg} formulation, and thus not set up by Listing~\ref{lst:tutorial_01_bcs} for such formulation.
The rationale behind Listing~\ref{lst:tutorial_01_bcs} is as follows.
Any \FEMPAR{} triangulation handles internally (and provides on client demand) a set identifier (i.e., an integer number) per each \ac{vef} of the mesh. On the other hand, it is assumed that the mesh generator from which the triangulation is imported classifies
the boundary of the domain into geometrical regions, and that, when the mesh is generated, it assigns the same set identifier to all \acp{vef} which belong to the same geometrical region.\footnote{We stress, however, that the \ac{vef} set identifiers can be modified by the user if the classification provided by the mesh generator it is not suitable for their needs.} For example, for $d=2$, the uniform mesh generator classifies the domain into 9 regions, namely the four corners of the box, which are assigned  identifiers $1, \ldots, 4$, the four faces of the box, which are assigned identifiers $1,\ldots,8$ , and the interior of the box, which is assigned identifier 9. For $d=3$, we have 27 regions, i.e., 8 corners, 12 edges, 6 faces, and the interior of the domain. (At this point, the reader should be able to grasp where the numbers \texttt{8} and \texttt{26} in Listing~\ref{lst:tutorial_01_bcs} come from.)
With the aforementioned in mind, Listing~\ref{lst:tutorial_01_bcs} sets up the \texttt{strong\_boundary\_conditions} instance
conformally with how the \acp{vef} of the triangulation laying on the boundary are flagged with set identifiers. In the loop spanning Lines~\ref{loc:tutorial_01_bcs_loop_start}-\ref{loc:tutorial_01_bcs_loop_stop}, it defines a strong boundary condition to be imposed for each of the regions that span the boundary of the unit box domain, and the same function, i.e., \texttt{exact\_solution}, to be imposed on all these regions, as required by the model problem of \tutorial{01}.

\begin{lstlisting}[float=htbp,language={[03]Fortran},escapechar=@,caption=The \texttt{setup\_strong\_boundary\_conditions} subroutine.,label={lst:tutorial_01_bcs}]
subroutine setup_strong_boundary_conditions()
  integer(ip) :: i, boundary_ids
  if ( fe_formulation == "CG" ) then
     boundary_ids = merge(8, 26, triangulation%get_num_dims() == 2) 
     call strong_boundary_conditions%create()
     do i = 1, boundary_ids @\label{loc:tutorial_01_bcs_loop_start}@
        call strong_boundary_conditions%insert_boundary_condition(boundary_id=i, &
             field_id=1, &
             cond_type=component_1, &
             boundary_function=exact_solution)
     end do@\label{loc:tutorial_01_bcs_loop_stop}@
  end if   
end subroutine setup_strong_boundary_conditions
\end{lstlisting}

In Listing~\ref{lst:tutorial_01_fe_space}, \tutorial{01} sets up the \texttt{fe\_space} instance, i.e.,  the computer representation of $V_h$. In Line~\ref{loc:tutorial_01_fes_num_fields}, the subroutine forces \texttt{fe\_space} to build a single-field \ac{fe} space, and in Line~\ref{loc:tutorial_01_fes_same}, that it is built from the same local \ac{fe} for all cells $K \in \triang$. In particular, from a scalar-valued Lagrangian-type \ac{fe} (Lines~\ref{loc:tutorial_01_fes_ftype1}-\ref{loc:tutorial_01_fes_ftype2} and~\ref{loc:tutorial_01_fes_reftype1}-\ref{loc:tutorial_01_fes_reftype2}, resp.), as required by our model problem. The parameter value corresponding to the polynomial order of $V_h|_K$ is not forced, and thus can be selected from the \ac{cli}.
Besides, Listing~\ref{lst:tutorial_01_fe_space} also forces the construction of either a conforming or non-conforming $V_h$, depending on the \ac{fe} formulation selected by the user (see Lines~\ref{loc:tutorial_01_fes_conf} and~\ref{loc:tutorial_01_fes_nonconf}, resp.).
The actual set up of \texttt{fe\_space} occurs in Line~\ref{loc:tutorial_01_fes_create_conf} and~\ref{loc:tutorial_01_fes_create_nonconf} for the conforming and non-conforming variants of $V_h$, resp. In the latter case, \texttt{fe\_space} does not require \texttt{strong\_boundary\_conditions}, as there are not \acp{bc} to be strongly enforced in this case.\footnote{In any case, passing it would not result in an error condition, but in unnecessary overhead to be paid.} The call in these lines generates a global numbering of the \acp{dof} in $V_h$, and (if it applies) identifies the \acp{dof} sitting on the regions of the boundary of the domain which are subject to strong \acp{bc} (combining the data provided by the triangulation and \texttt{strong\_boundary\_conditions}).
Finally, in Lines~\ref{loc:tutorial_01_fes_cell_integ} and~\ref{loc:tutorial_01_face_integ}, Listing~\ref{lst:tutorial_01_fe_space} activates the internal data structures that \texttt{fe\_space} provides for the numerical evaluation of cell and facet integrals, resp. These are required later on to evaluate the discrete bilinear and linear forms of the \ac{fe} formulations implemented by \tutorial{01}.
We note that the call in Line~\ref{loc:tutorial_01_face_integ} is only required for the \ac{ip} \ac{dg} formulation,  as in the \ac{cg} formulation there are not facet integrals to be evaluated.

\begin{lstlisting}[float=htbp,language={[03]Fortran},escapechar=@,caption=The \texttt{setup\_fe\_space} subroutine.,label={lst:tutorial_01_fe_space}]
subroutine setup_fe_space()
  type(string) :: fes_field_types(1), fes_ref_fe_types(1)
  call parameter_handler%update(fes_num_fields_key,1) @\label{loc:tutorial_01_fes_num_fields}@
  call parameter_handler%update(fes_same_ref_fes_all_cells_key,.true.) @\label{loc:tutorial_01_fes_same}@
  fes_field_types(1) = String(field_type_scalar) @\label{loc:tutorial_01_fes_ftype1}@
  call parameter_handler%update(fes_field_types_key,fes_field_types)  @\label{loc:tutorial_01_fes_ftype2}@
  fes_ref_fe_types(1) = String(fe_type_lagrangian) @\label{loc:tutorial_01_fes_reftype1}@
  call parameter_handler%update(fes_ref_fe_types_key, fes_ref_fe_types ) @\label{loc:tutorial_01_fes_reftype2}@
  if ( fe_formulation == "CG" ) then
     call parameter_handler%update(fes_ref_fe_conformities_key, [.true.]) @\label{loc:tutorial_01_fes_conf}@
     call fe_space%create( triangulation  = triangulation, &               @\label{loc:tutorial_01_fes_create_conf}@
                           conditions     = strong_boundary_conditions, &
                           parameters     = parameter_handler%get_values() )
  else if ( fe_formulation == "DG" ) then
     call parameter_handler%update(fes_ref_fe_conformities_key, [.false.]) @\label{loc:tutorial_01_fes_nonconf}@
     call fe_space%create( triangulation  = triangulation, &                @\label{loc:tutorial_01_fes_create_nonconf}@
                           parameters     = parameter_handler%get_values() )
  end if
  call fe_space%set_up_cell_integration()  @\label{loc:tutorial_01_fes_cell_integ}@ 
  if ( fe_formulation == "DG" ) then
     call fe_space%set_up_facet_integration()  @\label{loc:tutorial_01_face_integ}@
  end if
end subroutine setup_fe_space 
\end{lstlisting}

Listing~\ref{lst:tutorial_01_discrete_solution} sets up the \texttt{discrete\_solution} object, of type \texttt{fe\_function\_t}. This \FEMPAR{} data type represents an element of $V_h$, the \ac{fe} solution $u_h $ in the case of \tutorial{01}.
In Line~\ref{loc:tutorial_01_ds_create}, this subroutine allocates room for storing the \ac{dof} values of $u_h$, and in Line~\ref{loc:tutorial_01_ds_interp}, it computes the values of those \acp{dof} of $u_h$ which lay on a region of the domain boundary which is subject to strong \acp{bc}, the whole boundary of the domain in the case of \tutorial{01}.
The  \texttt{interpolate\_dirichlet\_values} \ac{tbp} of \texttt{fe\_space} interpolates \texttt{exact\_solution}, i.e., $u$, which is extracted from \texttt{strong\_boundary\_conditions}, using a suitable \ac{fe} interpolator for the \ac{fe} space at hand, i.e., the Lagrangian interpolator in the case of \tutorial{01}. \FEMPAR{} supports interpolators for curl- and div-conforming \ac{fe} spaces as well. Interpolation in such \ac{fe} spaces involves the numerical evaluation of the functionals (moments) associated to the \acp{dof} of $V_h$ \cite{Olm2019a}.

\begin{lstlisting}[float=htbp,language={[03]Fortran},escapechar=@,caption=The \texttt{setup\_discrete\_solution} subroutine.,label={lst:tutorial_01_discrete_solution}]
subroutine setup_discrete_solution()
  call discrete_solution%create(fe_space) @\label{loc:tutorial_01_ds_create}@
  if ( fe_formulation == "CG" ) then
     call fe_space%interpolate_dirichlet_values(discrete_solution) @\label{loc:tutorial_01_ds_interp}@
  end if
end subroutine setup_discrete_solution
\end{lstlisting}

In Listing~\ref{lst:tutorial_01_fe_operator}, the tutorial program builds
the \texttt{fe\_affine\_operator} instance, i.e.,  the computer representation of the operator defined in \eqref{problem1h-op}. Building this instance is a two-step process. First,
we have to call the \texttt{create} \ac{tbp} (see Line~\ref{loc:tutorial_01_setup_op_create}). Apart from specifying
the  data storage format and properties of the stiffness matrix $\fematrix$\footnote{In particular, Listing~\ref{lst:tutorial_01_fe_operator} specifies sparse matrix CSR format and symmetric storage, i.e., that only the upper triangle is stored, and that the matrix is \ac{spd}. These hints are used by some linear solver implementations in order to choose the most appropriate solver driver for the particular structure of $\fematrix$, assuming that the user does not force a particular solver driver, e.g., from the \ac{cli}. }, this call equips \texttt{fe\_affine\_operator\_t}
with all that it requires in order to evaluate the entries of the operator. Second, we have to call the \texttt{compute} TBP (Line~\ref{loc:tutorial_01_setup_op_compute}), that triggers the actual numerical evaluation and  assembly of the discrete weak form. With extensibility and flexibility in mind, this latter responsibility does not fall on \texttt{fe\_affine\_operator\_t}, but actually on the data type extensions of a key  abstract class defined within \FEMPAR{}, referred to as \texttt{discrete\_integration\_t}~\cite[Sect. 11.2]{badia_fempar:_2017}.  \Tutorial{01} implements two type extensions of \texttt{discrete\_integration\_t}, namely \texttt{cg\_discrete\_integration\_t}, and \texttt{dg\_...\_t}. The former implements (the discrete variant of) \eqref{eq-poisson-weak}, while the second, \eqref{eq-poisson-weak-dg}.
In Lines~\ref{loc:tutorial_01_setup_op_di_start}-\ref{loc:tutorial_01_setup_op_di_stop}, Listing~\ref{lst:tutorial_01_fe_operator} passes to these instances the source term and boundary function to be imposed on the Dirichlet boundary. We note that the \ac{ip} \ac{dg} formulation works directly with the analytical expression of the boundary function (i.e., \texttt{exact\_solution}), while the \ac{cg} formulation with its interpolation (i.e., \texttt{discrete\_solution}). Recall that the \ac{cg} formulation
imposes Dirichlet \acp{bc} strongly. In the approach followed by \FEMPAR{}, the strong imposition of \acp{bc} requires the values of the \acp{dof} of $u_h(=g_h)$ sitting on the Dirichlet boundary when assembling the contribution of the cells that touch the Dirichlet boundary; see~\cite[10.4]{badia_fempar:_2017}. On the other hand, the \ac{ip} \ac{dg} formulation needs to evaluate the last two facet integrals in \eqref{eq:ip_dg_linear} with the boundary function $g$ (i.e., $u$ in our case) as integrand. It is preferable that this formulation works with the analytical expression of the function, instead of its \ac{fe} interpolation, to avoid an extra source of approximation error.

\begin{lstlisting}[float=htbp,language={[03]Fortran},escapechar=@,caption=The \texttt{setup\_and\_assemble\_fe\_affine\_operator}  subroutine.,label={lst:tutorial_01_fe_operator}]
subroutine setup_and_assemble_fe_affine_operator()
  class(discrete_integration_t), pointer :: discrete_integration
  if ( fe_formulation == "CG" ) then @\label{loc:tutorial_01_setup_op_di_start}@
     call cg_discrete_integration%set_source_term(source_term)
     call cg_discrete_integration%set_boundary_function(discrete_solution)
     discrete_integration => cg_discrete_integration
  else if ( fe_formulation == "DG" ) then 
     call dg_discrete_integration%set_source_term(source_term)
     call dg_discrete_integration%set_boundary_function(exact_solution) 
     discrete_integration => dg_discrete_integration
  end if @\label{loc:tutorial_01_setup_op_di_stop}@
  call fe_affine_operator%create ( sparse_matrix_storage_format = csr_format, &  @\label{loc:tutorial_01_setup_op_create}@
       diagonal_blocks_symmetric_storage = [ .true. ], &
       diagonal_blocks_symmetric = [ .true. ], &
       diagonal_blocks_sign = [ SPARSE_MATRIX_SIGN_POSITIVE_DEFINITE ], &
       fe_space = fe_space, &
       discrete_integration = discrete_integration )
  call fe_affine_operator%compute() @\label{loc:tutorial_01_setup_op_compute}@
end subroutine setup_and_assemble_fe_affine_operator 
\end{lstlisting}

Listing~\ref{lst:tutorial_01_solve_system} finds the root of the \ac{fe} operator, i.e., $u_h \in V_h$ such that $\mathcal{F}_h(u_h) = 0$. For such purpose, it first sets up the \texttt{direct\_solver} instance (Lines~\ref{loc:tutorial_01_solve_system_setup_start}-\ref{loc:tutorial_01_solve_system_setup_stop})  and later uses its \texttt{solve} \ac{tbp} (Line~\ref{loc:tutorial_01_solve_system_solve}). This latter method is fed with the ``translation'' of $\mathcal{F}_h$, i.e., the right hand side $\ferhs$ of the linear system, and the free \ac{dof} values of $u_h$ as the unknown to be found. The free \acp{dof} of $u_h$ are those whose values are not constrained, e.g., by strong \acp{bc}.
\texttt{direct\_solver\_t} is a \FEMPAR{} data type that offers interfaces to (non-distributed) sparse direct solvers provided by external libraries. In its first public release, \FEMPAR{} provides interfaces to PARDISO~\cite{_intel_????} (the version available in the Intel MKL library) and UMFPACK~\cite{Davis2004}, although it is designed such that additional sparse direct solver implementations can be easily added. We note that Listing~\ref{lst:tutorial_01_solve_system} does not force any parameter value related to \texttt{direct\_solver\_t}; the default \ac{cla} values are appropriate for the Poisson problem. In any case, at this point the reader is encouraged to inspect the \acp{cla} linked to \texttt{direct\_solver\_t} and play around them.

\begin{lstlisting}[float=htbp,language={[03]Fortran},escapechar=@,caption=The \texttt{solve\_system} subroutine.,label={lst:tutorial_01_solve_system}]
subroutine solve_system()
  class(vector_t), pointer :: dof_values
  call direct_solver%set_type_from_pl(parameter_handler%get_values()) @\label{loc:tutorial_01_solve_system_setup_start}@
  call direct_solver%set_parameters_from_pl(parameter_handler%get_values())
  call direct_solver%set_matrix(fe_affine_operator%get_matrix())      @\label{loc:tutorial_01_solve_system_setup_stop}@
  dof_values => discrete_solution%get_free_dof_values()                      @\label{loc:tutorial_01_solve_system_get_dofs}@
  call direct_solver%solve(fe_affine_operator%get_translation(),dof_values)  @\label{loc:tutorial_01_solve_system_solve}@
end subroutine solve_system
\end{lstlisting}

As any other \ac{fe} program, \tutorial{01} post-processes the computed \ac{fe} solution. In particular, in the \texttt{compute\_error} subroutine (Listing~\ref{lst:tutorial_01_compute_error}), it computes $e_K^2$ for each $K \in \triang$, and the global error $e$; see Sect.~\ref{sec:tutorial01_fe_discretization}. These quantities along with the \ac{fe} solution $u_h$ are written to output data files for later visualization in the \texttt{write\_postprocess\_data\_files} helper subroutine (Listing~\ref{lst:tutorial_01_oh}). The first subroutine relies on the \texttt{error\_estimator} instance, implemented in one of the support modules of \tutorial{01}. In particular, the actual computation of $e_K^2$ occurs at the call to the \texttt{compute\_local\_true\_errors} \ac{tbp} of this data type (Line~\ref{loc:tutorial_01_compute_error_lte} of Listing~\ref{lst:tutorial_01_compute_error}). At this point, the user is encouraged to inspect the implementation of this data type in order to grasp how the numerical evaluation of the integrals required for the computation of $e_K^2$ is carried out using \FEMPAR{}; see also \cite[Sect. 8, 10.5]{badia_fempar:_2017}.

\begin{lstlisting}[float=htbp,language={[03]Fortran},escapechar=@,caption=The \texttt{compute\_error} subroutine.,label={lst:tutorial_01_compute_error}]
subroutine compute_error()
  real(rp) :: global_error_energy_norm
  call error_estimator%create(fe_space,parameter_handler%get_values())
  call error_estimator%set_exact_solution(exact_solution)
  call error_estimator%set_discrete_solution(discrete_solution)
  call error_estimator%compute_local_true_errors() @\label{loc:tutorial_01_compute_error_lte}@
  global_error_energy_norm = &
         sqrt(sum(error_estimator%get_sq_local_true_error_entries()))
  ... ! Write error on screen along with the number of cells and DOFs 
end subroutine compute_error
\end{lstlisting}

\begin{lstlisting}[float=htbp,language={[03]Fortran},escapechar=@,caption=The \texttt{write\_postprocess\_data\_files} subroutine.,label={lst:tutorial_01_oh}]
subroutine write_postprocess_data_files()
 if ( write_postprocess_data ) then
   call output_handler%create(parameter_handler%get_values())
   call output_handler%attach_fe_space(fe_space)
   call output_handler%add_fe_function(discrete_solution, 1, 'solution')
   call output_handler%add_cell_vector(&
                       error_estimator%get_sq_local_true_errors(),&
                       'cell_error_energy_norm_squared')
   call output_handler%open()
   call output_handler%write()
   call output_handler%close()
 end if
end subroutine write_postprocess_data_files
\end{lstlisting}

The generation of output data files in Listing~\ref{lst:tutorial_01_oh} is
in charge of  the \texttt{output\_handler} instance. This instance lets the user to register an arbitrary number of FE functions (together with the corresponding FE space these functions were generated from) and cell data arrays (e.g., material properties or error estimator indicators), to be output in the appropriate format for later visualization. The user may also select to apply a differential operator to the \ac{fe} function, such as divergence, gradient or curl, which involve further calculations to be performed on each cell.
The first public release of \FEMPAR{} supports two different data output formats,  the standard-open model VTK~\cite{Schroeder:1998:VTO:272980}, and the XDMF+HDF5~\cite{xdmf} model. The first format is the recommended option for serial computations (or parallel computations on a moderate number of tasks).
The second model is designed with the parallel I/O data challenge in mind. It is therefore the recommended option for large-scale simulations in high-end computing environments. The data format to be used relies on parameter values passed to \texttt{output\_handler}.
As Listing~\ref{lst:tutorial_01_oh} does not force any parameter value related to \texttt{output\_handler\_t}, \tutorial{01} users may select the output data format readily from the \ac{cli}. At this point, the reader may inspect the \acp{cla} linked to \texttt{output\_handler\_t} and play around them to see the differences in the output data files generated by Listing~\ref{lst:tutorial_01_oh}.

\subsubsection{Discrete integration for a conforming method}\label{sec:disc_int_conf}

As commented in Sect.~\ref{sec:fempar_sw_abstractions} and one can observe in Listing~\ref{lst:tutorial_01_fe_operator}, the definition of the \ac{fe} operator requires a method that conceptually traverses all the cells in the triangulation, computes the element matrix and right-hand side at every cell, and assembles it in a global array. The abstract type in charge of this is the \texttt{discrete\_integration\_t}, which must be extended by the user to integrate their \ac{pde} system. In Listing~\ref{lst:tutorial_01_cg_discrete_integration}, we consider a concrete version of this type for the Poisson equation using conforming Lagrangian \acp{fe}.

\begin{lstlisting}[float=htbp,language={[03]Fortran},escapechar=@,caption=The {\tt integrate\_galerkin}
\ac{tbp} of the \tutorial{01}-specific {\tt cg\_discrete\_integration\_t} data type..,label={lst:tutorial_01_cg_discrete_integration}]
subroutine cg_discrete_integration_integrate_galerkin (this, fe_space, assembler)
  implicit none
  class(cg_discrete_integration_t), intent(in)    :: this
  class(serial_fe_space_t)        , intent(inout) :: fe_space
  class(assembler_t)              , intent(inout) :: assembler
  class(fe_cell_iterator_t), allocatable  :: fe
  type(quadrature_t)       , pointer      :: quad
  type(point_t)            , pointer      :: quad_coords(:) 
  type(vector_field_t)     , allocatable  :: shape_gradients(:,:) @\label{loc:cg_di_fe_iterator_decgrad}@
  real(rp)                 , allocatable  :: shape_values(:,:) @\label{loc:cg_di_fe_iterator_decval}@
  real(rp), allocatable :: elmat(:,:), elvec(:)
  integer(ip)  :: qpoint, idof, jdof
  real(rp)     :: dV, source_term_value

  call fe_space%create_fe_cell_iterator(fe) @\label{loc:cg_di_fe_iterator}@
  ... ! Allocate elmat, elvec
  do while ( .not. fe%has_finished() ) @\label{loc:cg_di_fe_iterator_fin}@
     ! Update cell-integration related data structures
     call fe%update_integration() @\label{loc:cg_di_fe_iterator_update_integration}@
     ! Get quadrature and quadrature coordinates mapped to physical space
     quad        => fe%get_quadrature() @\label{loc:cg_di_fe_iterator_quadrature}@
     quad_coords => fe%get_quadrature_points_coordinates() @\label{loc:cg_di_fe_iterator_coords}@
     ! Compute cell matrix and vector
     elmat = 0.0_rp; elvec = 0.0_rp
     call fe%get_values(shape_values) @\label{loc:cg_di_fe_iterator_vals}@
     call fe%get_gradients(shape_gradients) @\label{loc:cg_di_fe_iterator_grads}@
     do qpoint = 1, quad%get_num_quadrature_points()@\label{loc:cg_di_fe_iterator_numgp}@
        dV = fe%get_det_jacobian(qpoint) * quad%get_weight(qpoint) @\label{loc:cg_di_fe_iterator_jac}@
        ! Compute cell matrix (bilinear form)
        do idof = 1, fe%get_num_dofs()@\label{loc:cg_di_fe_iterator_numshf}@
           do jdof = 1, fe%get_num_dofs()
              elmat(idof,jdof) = elmat(idof,jdof) + & @\label{loc:cg_di_fe_iterator_elmat}@
                   dV * shape_gradients(jdof,qpoint) * shape_gradients(idof,qpoint)@\label{loc:cg_di_fe_iterator_elmat2}@
           end do
        end do
        ! Compute cell vector (linear form)
        call this%source_term%get_value(quad_coords(qpoint),source_term_value) @\label{loc:cg_di_fe_iterator_force}@
        do idof = 1, fe%get_num_dofs()
           elvec(idof) = elvec(idof) + &
                   dV * source_term_value * shape_values(idof,qpoint)
        end do
     end do
     ! Assemble elmat/elvec into assembler while taking care of strong
     ! Dirichlet BCs (in turn extracted from this%discrete_boundary_function)
     call fe%assembly( this%discrete_boundary_function, elmat, elvec, assembler )@\label{loc:cg_di_fe_iterator_assemble}@
     call fe%next() @\label{loc:cg_di_fe_iterator_next}@
     end do
  call fe_space%free_fe_cell_iterator(fe)
  ... ! Free shape_* arrays, elmat, elvec
end subroutine cg_discrete_integration_integrate_galerkin
\end{lstlisting}
The integration of the (bi)linear forms requires cell integration machinery, which is provided by \texttt{fe\_space\_t} through the creation of the \texttt{fe\_cell\_iterator\_t} in Line \ref{loc:cg_di_fe_iterator}. Conceptually, the computed cell iterator is an object that provides mechanisms to iterate over all the cells of the triangulation (see Line~\ref{loc:cg_di_fe_iterator_fin} and Line~\ref{loc:cg_di_fe_iterator_next}). Positioned in a cell, it provides a set of cell-wise queries. All the integration machinery of a new cell is computed in Line~\ref{loc:cg_di_fe_iterator_update_integration}. After this update of integration data, one can extract from the iterator the number of local shape functions (see Line~\ref{loc:cg_di_fe_iterator_numshf}) and an array with their values (resp., gradients) at the integration points in Line~\ref{loc:cg_di_fe_iterator_vals} (resp., Line~\ref{loc:cg_di_fe_iterator_grads}). We note that the data types of the entries of these arrays can be scalars (see Line~\ref{loc:cg_di_fe_iterator_decval}), vectors (of type \texttt{vector\_field\_t}, see Line~\ref{loc:cg_di_fe_iterator_decgrad}), or tensors (of type \texttt{tensor\_field\_t}). It can also return information about the Jacobian of the geometrical transformation (see, e.g., the query that provides the determinant of the Jacobian of the cell map in Line~\ref{loc:cg_di_fe_iterator_jac}). The iterator also provides a method to fetch the cell  quadrature (see Line~\ref{loc:cg_di_fe_iterator_quadrature}), which in turn has procedures to get the number of integration points (Line \ref{loc:cg_di_fe_iterator_numgp}) and their associated weights (Line \ref{loc:cg_di_fe_iterator_jac}).

With all these ingredients, the implementation of the (bi)linear forms is close to its blackboard expression, making it compact, simple, and intuitive (see Line~\ref{loc:cg_di_fe_iterator_elmat}). It is achieved using operator overloading for different vector and tensor operations, e.g., the contraction and scaling operations (see, e.g., the inner product of vectors in Line~\ref{loc:cg_di_fe_iterator_elmat2}). The computation of the right-hand side is similar. The only peculiarity is the consumption of the expression for $f$ in Line~\ref{loc:cg_di_fe_iterator_force},  {which has to receive the quadrature points coordinates in physical space, i.e., with the entries of the {\tt quad\_coords(:)} array. We recall that {\tt cg\_discrete\_integration} was supplied with the expression for $f$ in Listing~\ref{lst:tutorial_01_fe_operator}.}

The \texttt{fe\_cell\_iterator\_t} data type also offers methods to assemble the element matrix and vector into \texttt{assembler}, which is the object that ultimately holds the global system matrix and right-hand side within the \ac{fe} affine operator. The assembly is carried out in Line~\ref{loc:cg_di_fe_iterator_assemble}. {The call in this line is provided with the \ac{dof} values of the function to be imposed strongly on the Dirichlet boundary, i.e., {\tt discrete\_boundary\_function}; see discussion accompanying Listing~\ref{lst:tutorial_01_fe_operator} in Sect.~\ref{sec:tutorial01_commented_code}.}

\subsubsection{Discrete integration for a non-conforming method}\label{sec:dg_discrete_int}

Let us consider the numerical integration of the \ac{ip} \ac{dg} method in \eqref{eq:ip_dg_bilinear}. For non-conforming \ac{fe} spaces, the formulation requires also a loop over the facets to integrate the perturbation \ac{dg} terms. It can be written in a similar fashion to Sect.~\ref{sec:disc_int_conf}, but considering also facet-wise structures. For such purposes, \FEMPAR{} provides \ac{fe} facet iterators (see Line~\ref{loc:dg_di_l2}), which are similar to the \ac{fe} cell iterators but traversing all the facets in the triangulation. This iterator provides a method to distinguish between boundary facets and interior facets (see Line~\ref{loc:dg_di_l3}), since different terms have to be integrated in each case. As above, at every facet, one can compute all the required numerical integration tools (see Line~\ref{loc:dg_di_l4}). After this step, one can also extract a facet quadrature (Line~\ref{loc:dg_di_l5}), which provides the number of quadrature points (see Line~\ref{loc:dg_di_l6}) and its weights (see Line~\ref{loc:dg_di_l9}). The \ac{fe} facet iterator also provides methods that return $\normal^+$, $\normal^-$ (facet outward unit normals) (see Line~\ref{loc:dg_di_l7}), a characteristic facet size (see Line~\ref{loc:dg_di_l8}), or the determinant of the Jacobian of the geometrical transformation of the facet from the reference to the physical space (see Line~\ref{loc:dg_di_l9}).

An interior facet is shared between two and only two cells. After some algebraic manipulation, the \ac{dg} terms in \eqref{eq:ip_dg_bilinear} can be decomposed into a set of terms that involve test functions (and gradients) of both cells, as shown in \eqref{eq:local_to_face_matrix}. The loop over the four facet matrices is performed in Lines~\ref{loc:dg_di_l10}-\ref{loc:dg_di_l13}. For each cell sharing the facet, one can also compute the shape functions and its gradients (see Lines~\ref{loc:dg_di_l11}-\ref{loc:dg_di_l12} and \ref{loc:dg_di_l14}-\ref{loc:dg_di_l15}). With all these ingredients, we compute the facet matrices in Lines~\ref{loc:dg_di_l18}-\ref{loc:dg_di_l22}.
We note that the constant $\gamma$ in \eqref{eq:ip_dg_bilinear} has been defined in Line~\ref{loc:dg_di_l1} as $10 p^2$, where $p$ is the order of the \ac{fe} space. The \ac{fe} facet iterator also provides a method for the assembly of the facet matrices into global structures (see Line~\ref{loc:dg_di_l23}).

\begin{lstlisting}[float=htbp,language={[03]Fortran},escapechar=@,caption=The {\tt integrate\_galerkin}
\ac{tbp} of the \tutorial{01}-specific {\tt dg\_discrete\_integration\_t} data type..,label={lst:tutorial_01_dg_discrete_integration}]
subroutine dg_discrete_integration_integrate_galerkin (this, fe_space, assembler)
 implicit none
 class(dg_discrete_integration_t), intent(in)    :: this
 class(serial_fe_space_t)        , intent(inout) :: fe_space
 class(assembler_t)              , intent(inout) :: assembler
 ... ! Declare local variables in Listing @\ref{lst:tutorial_01_cg_discrete_integration}@
 class(fe_facet_iterator_t), allocatable :: fe_face
 real(rp), allocatable :: shape_values_Kminus(:,:), &
                          shape_values_Kplus(:,:)
 type(vector_field_t), allocatable :: shape_gradients_Kminus(:,:), &
                                      shape_gradients_Kplus(:,:)
 type(vector_field_t) :: normals(2)
 real(rp) :: h_length, C_IP ! Interior Penalty constant
 real(rp), allocatable :: facemat(:,:,:,:), facevec(:,:)
 integer(ip) :: Kminus, Kplus

 ... ! Integrate and assemble cell integrals (see Listing @\ref{lst:tutorial_01_cg_discrete_integration}@)
 C_IP = 10.0_rp * real(fe%get_order()**2, rp) @\label{loc:dg_di_l1}@
 call fe_space%create_fe_facet_iterator(fe_face)@\label{loc:dg_di_l2}@
 ... ! Allocate facemat, facevec
 do while ( .not. fe_face%has_finished() ) ! Loop over all triangulation facets
   if ( .not. fe_face%is_at_boundary() ) then ! Interior facet @\label{loc:dg_di_l3}@
    ! Update facet-integration related data structures
    call fe_face%update_integration()@\label{loc:dg_di_l4}@
    quad => fe_face%get_quadrature() ! Get facet quadrature @\label{loc:dg_di_l5}@
    facemat = 0.0_rp
    do qpoint = 1, quad%get_num_quadrature_points()   @\label{loc:dg_di_l6}@
      call fe_face%get_normal(qpoint,normals) @\label{loc:dg_di_l7}@
      h_length = fe_face%compute_characteristic_length(qpoint) @\label{loc:dg_di_l8}@
      dV = fe_face%get_det_jacobian(qpoint) * quad%get_weight(qpoint) @\label{loc:dg_di_l9}@
      do Kminus = 1, fe_face%get_num_cells_around() @\label{loc:dg_di_l10}@
       call fe_face%get_values(Kminus,shape_values_Kminus) @\label{loc:dg_di_l11}@
       call fe_face%get_gradients(Kminus,shape_gradients_Kminus) @\label{loc:dg_di_l12}@
       do Kplus = 1, fe_face%get_num_cells_around() @\label{loc:dg_di_l13}@
         call fe_face%get_values(Kplus,shape_values_Kplus) @\label{loc:dg_di_l14}@
         call fe_face%get_gradients(Kplus,shape_gradients_Kplus) @\label{loc:dg_di_l15}@
         do idof = 1, fe_face%get_num_dofs(Kminus) @\label{loc:dg_di_l16}@
          do jdof = 1, fe_face%get_num_dofs(Kplus) @\label{loc:dg_di_l17}@
            !- ({{grad u}}[[v]] + (1-xi)*[[u]]{{grad v}}) + C*p^2/h * [[u]] [[v]]
            facemat(idof,jdof,Kminus,Kplus) = facemat(idof,jdof,Kminus,Kplus) + & @\label{loc:dg_di_l18}@
               dV *  & @\label{loc:dg_di_l19}@
               (-0.5_rp*shape_gradients_Kplus(jdof,qpoint)*normals(Kminus)*shape_values_Kminus(idof,qpoint) - & @\label{loc:dg_di_l20}@
               0.5_rp*shape_gradients_Kminus(idof,qpoint)*normals(Kplus)*shape_values_Kplus(jdof,qpoint) + & @\label{loc:dg_di_l21}@
               c_IP / h_length * shape_values_Kplus(jdof,qpoint)*shape_values_Kminus(idof,qpoint)*normals(Kminus)*normals(Kplus)) @\label{loc:dg_di_l22}@
          end do
         end do
       end do
      end do
    end do
    call fe_face%assembly( facemat, assembler ) @\label{loc:dg_di_l23}@
   else if ( fe_face%is_at_boundary() ) then
    ... ! Integrate and assemble boundary facet terms
   end if
   call fe_face%next()
 end do
 call fe_space%free_fe_facet_iterator(fe_face)
 ... ! Free shape_* arrays, facemat, facevec
end subroutine dg_discrete_integration_integrate_galerkin
\end{lstlisting}

\section{\Tutorial{02}: \tutorial{01} problem tackled with AMR}

\subsection{Model problem} \label{sec:tutorial02_model_problem}
See Sect.~\ref{sec:tutorial01_model_problem}.

\subsection{\ac{fe} discretization} \label{sec:tutorial02_fe_discretization}

	While \tutorial{01} uses a uniform (thus conforming) mesh $\triang$
	of $\Omega$, \tutorial{02} combines the two \ac{fe} formulations presented in Sect.~\ref{sec:tutorial01_fe_discretization} with a more clever/efficient domain discretization approach. Given that the solution of Prob.~\eqref{eq-poisson} exhibits highly localized features, in particular an internal sharp circular/spherical wave front layer, \tutorial{02} exploits a \FEMPAR{} triangulation data structure that efficiently supports dynamic $h$-adaptivity techniques (a.k.a. \ac{amr}), i.e., the ability of the mesh to be refined {\em in the course of the simulation} in those regions of the
	domain that present a complex behaviour (e.g., the internal layer in the case of Prob.~\eqref{eq-poisson}), and to be coarsened  in those areas where essentially nothing relevant happens (e.g., those areas away from the internal layer).
	\Tutorial{02} restricts itself to $h$-adaptivity techniques with a fixed polynomial order. This is in contrast to $hp$-adaptivity techniques, in which the local \ac{fe} space polynomial order $p$ also varies among cells.
	In its first public release, the support of $hp$-adaptivity
	techniques in \FEMPAR{} is restricted to non-conforming \ac{fe} formulations.

	In order to support \ac{amr} techniques, \FEMPAR{} relies on the so-called forest-of-trees approach for efficient mesh generation and adaptation.
	Forest-of-trees can be seen as a two-level decomposition of
	$\Omega$, referred to as macro and micro level, resp.
	In the macro level, we have the so-called coarse mesh,
	i.e., a {\em conforming} partition $\mathcal{C}_h$ of $\Omega$  into cells $K \in \mathcal{C}_h$. For efficiency reasons,
	$\mathcal{C}_h$ should be as coarse as possible, but it should also keep the geometrical discretization error within tolerable margins.
	For complex domains, $\mathcal{C}_h$ is usually generated by an unstructured mesh generator, and then imported into the program. For simple domains, such as boxes in the case of Prob.~\eqref{eq-poisson}, a single coarse cell is
	sufficient to resolve the geometry of $\Omega$. On the other hand, in the micro level, each of the cells of $\mathcal{C}_h$ becomes the root of an adaptive tree that can be subdivided arbitrarily (i.e., recursively refined) into finer cells. The locally refined mesh $\triang$ to be used for \ac{fe} discretization is defined as the union of the leaves of all adaptive trees.

	In the case of quadrilateral (2D) or hexahedral (3D) adaptive meshes, the recursive application of the standard isotropic 1:4 (2D) and 1:8 (3D) refinement rule to the coarse mesh cells (i.e., to the adaptive tree roots) leads to adaptive trees that are referred to as quadtrees and octrees, resp.,
	and the data structure resulting from patching them together is called {\em forest-of-quadtrees} and {\em -octrees}, resp., although the latter term is typically employed in either case.
	\FEMPAR{} provides a triangulation data structure suitable for the construction of {\em generic} \ac{fe} spaces $V_h$ (grad-, div-, and curl-conforming \ac{fe} spaces) which exploits the \p4est{}~\cite{burstedde_p4est_2011} library as its {\em forest-of-octrees} specialized meshing engine.
	We refer to \cite[Sect. 3]{Badia2019a} for a detailed exposition of the design criteria underlying the $h$-adaptive triangulation  in \FEMPAR{}, and the approach followed in order to reconstruct it from the light-weight, memory-efficient representation of the forest-of-octrees that \p4est{} handles internally.

	Tree-based meshes provide multi-resolution capability by local adaptation. The cells in $\triang$ (i.e., the leaves of the adaptive trees) might be located at different refinement level.  However, these meshes are (potentially) {\em non-conforming}, i.e., they contain the so-called {\em hanging \acp{vef}}.  These occur at the interface of neighboring cells with different refinement levels.
	Mesh non-conformity introduces additional complexity in the implementation of both conforming and non-conforming \ac{fe} formulations.
	In the former case, \acp{dof} sitting on \emph{hanging} \acp{vef} cannot have an arbitrary value, as this would result in $V_h$ violating the trace continuity requirements for conformity across interfaces shared by a coarse cell and its finer cell neighbours.
	In order to restore conformity, the space $V_h$ has to be supplied with suitably defined algebraic constraints that express the value of hanging \acp{dof} (i.e., \acp{dof} sitting on hanging \acp{vef}) as linear combinations of {\em true} \acp{dof} (i.e., \acp{dof} sitting on regular \acp{vef}); see \cite[Sect. 4]{Badia2019a}.
	These constraints, in most approaches available in the literature, are either applied to the local cell matrix and vector right before assembling them into their global counterparts,
	or, alternatively, eliminated from the global linear system later on (i.e.,  after \ac{fe} assembly}). In particular, \FEMPAR{} follows the first approach. Therefore, hanging \acp{dof} are not associated to an equation/unknown in the global linear system; see \cite[Sect. 5]{Badia2019a}. On the other hand, in the case of non-conforming \ac{fe} formulations, the facet integration machinery has to support the evaluation of flux terms across neighbouring cells at different refinement levels, i.e., facet integrals on each of the finer subfacets hanging on a coarser facet. \FEMPAR{} supports such kind of facet integrals as well.

	Despite the aforementioned, we note the following. First, the degree of implementation complexity is significantly reduced by
	enforcing the so-called \emph{2:1 balance} constraint, i.e., adjacent cells may
	differ at most by a single level of refinement; the $h$-adaptive triangulation in \FEMPAR{} always satisfies this constraint \cite{Badia2019a}. Second, the library is entirely responsible for handling such complexity. Library users are not aware of mesh non-conformity when evaluating and assembling the discrete weak form of the \ac{fe} formulation at hand. Indeed, as will be seen in Sect.~\ref{sec:tutorial02_commented_code}, \tutorial{02} re-uses ``as-is'' the \texttt{cg\_discrete\_integration\_t}, and \texttt{dg\_...\_t} objects of \tutorial{01}; see Sect.~\ref{sec:tutorial01_commented_code}.

	\subsection{The commented code} \label{sec:tutorial02_commented_code}

	In order to illustrate the \ac{amr} capabilities in \FEMPAR{}, while generating a suitable mesh for Prob.~\eqref{eq-poisson}, \tutorial{01} performs an \ac{amr} loop comprising the steps shown in Fig.~\ref{fig:tutorial_02_amr_loop}.

	\begin{figure}[ht!]
	\begin{enumerate}
	\setlength{\parskip}{0pt}
	\setlength{\itemsep}{0pt plus 1pt}
	 \item Generate a conforming mesh $\triang$ by uniformly refining, a number user-defined steps, a single-cell coarse mesh $\mathcal{C}_h$ representing the unit box domain (i.e., $\Omega$ in Prob.~\eqref{eq-poisson}).
	 \item Compute an approximate \ac{fe} solution $u_h$ using the current mesh $\triang$.
	 \item Compute $e_K^2$ for all cells $K \in \triang$; see Sect.~\ref{sec:tutorial01_fe_discretization}.
	 \item Given user-defined refinement and coarsening fractions, denoted by $\alpha_r$ and $\alpha_c$, resp., find thresholds $\theta_r$ and $\theta_c$ such that the number of cells with $e_K >\theta_r$  (resp., $e_K < \theta_c$)  is (approximately) a fraction $\alpha_r$  (resp., $\alpha_c$) of the number of cells in $\triang$.
	 \item Refine and coarsen the mesh cells, i.e., generate a new mesh $\triang$,  accordingly to the input provided by the previous step.
	 \item Repeat steps (2)-(5) a number of user-defined steps.
	\end{enumerate}
	\caption{\Tutorial{02} \ac{amr} loop.} \label{fig:tutorial_02_amr_loop}
	\end{figure}

	The main program unit of \tutorial{02} is shown in Listing~\ref{lst:tutorial_02_program_unit}. \Tutorial{02} re-uses ``as-is'' the tutorial-specific modules of \tutorial{01}: no adaptations of these modules are required despite the higher complexity underlying \ac{fe} discretization on non-conforming meshes; see Sect.~\ref{sec:tutorial02_fe_discretization}.
	Compared to \tutorial{01}, \tutorial{02} declares the {\tt triangulation} instance to be of type {\tt p4est\_serial\_triangulation\_t} (instead of {\tt serial\_triangulation\_t}). This data type extension of {\tt triangulation\_t} supports dynamic $h$-adaptivity (see Sect.~\ref{sec:tutorial02_fe_discretization}) on serial computing environments (i.e., the triangulation is not actually distributed, but centralized on a single task) using \p4est{} under the hood. On the other hand, it declares the {\tt refinement\_strategy} instance, of type {\tt fixed\_fraction\_refinement\_strategy\_t}. The role of this \FEMPAR{} data type in the loop of Fig.~\ref{fig:tutorial_02_amr_loop} will be clarified along the section.

\begin{lstlisting}[float=htbp,language={[03]Fortran},escapechar=@,caption=\Tutorial{02} program unit.,label={lst:tutorial_02_program_unit}]
program tutorial_02_poisson_sharp_circular_wave_amr
  use fempar_names
  ... ! Use tutorial_01_... support modules
  implicit none
  ... ! Declaration of tutorial_02_... parameter constants  @\label{loc:tutorial_02_part2_start}@
  type(p4est_serial_triangulation_t)         :: triangulation @\label{loc:tutorial_02_triangulation}@
  type(fixed_fraction_refinement_strategy_t) :: refinement_strategy
  ... ! Declaration of remaining objects of tutorial_02_... (@see Listing~\ref{lst:tutorial_01_program_unit}@) @\label{loc:tutorial_02_rest}@
  ... ! Declaration of variables storing tutorial_02_... CLA values @\label{loc:tutorial_02_part2_stop}@
  call fempar_init() @\label{loc:tutorial_02_part3_start}@
  call setup_parameter_handler()
  call get_tutorial_cla_values()
  call setup_context_and_environment()
  current_amr_step = 0                    @\label{loc:tutorial_02_part3_startup_loop_start}@
  ... @\label{loc:tutorial_02_multiline}@ ! See Lines @\ref{loc:tutorial_01_setup_triangulation}@-@\ref{loc:tutorial_01_compute_error}@ of Listing @\ref{lst:tutorial_02_program_unit}@ 
  call setup_refinement_strategy()
  call output_handler_initialize() @\label{loc:tutorial_02_oh_initialize}@
  do @\label{loc:tutorial_02_loop_start}@
    call output_handler_write_current_amr_step() @\label{loc:tutorial_02_oh_write_current_step}@
    if ( current_amr_step == num_amr_steps ) exit @\label{loc:tutorial_02_finish_loop}@
    current_amr_step = current_amr_step + 1  @\label{loc:tutorial_02_increase_step}@
    call refinement_strategy%update_refinement_flags(triangulation) @\label{loc:tutorial_02_refinement_update}@
    call setup_triangulation()  @\label{loc:tutorial_02_setup_triangulation}@
    call setup_fe_space()                              @\label{loc:tutorial_02_solve_start}@
    call setup_discrete_solution()                     @\label{loc:tutorial_02_set_di}@
    call setup_and_assemble_fe_affine_operator()
    call solve_system()                               @\label{loc:tutorial_02_solve_stop}@
    call compute_error() @\label{loc:tutorial_02_compute_error}@
  end do @\label{loc:tutorial_02_loop_end}@
  call output_handler_finalize() @\label{loc:tutorial_02_oh_finalize}@
  call free_all_objects()
  call fempar_finalize() @\label{loc:tutorial_02_part3_stop}@
contains
  ... ! Implementation of helper procedures
end program tutorial_02_poisson_sharp_circular_wave_amr
\end{lstlisting}

	\Tutorial{02}'s main executable code spans Lines~\ref{loc:tutorial_02_part3_start}-\ref{loc:tutorial_02_part3_stop}. Apart from those \acp{cla} registered by \tutorial{01}, in the call to  {\tt setup\_parameter\_handler}, \tutorial{02} registers two additional \acp{cla} to specify the number of user-defined uniform refinement and \ac{amr} steps in Steps~(1) and (6), resp., of Fig.~\ref{fig:tutorial_02_amr_loop}. The main executable code of \Tutorial{02} is mapped to the steps in
	Fig.~\ref{fig:tutorial_02_amr_loop} as follows. Line~\ref{loc:tutorial_02_multiline}, which actually triggers a sequence of calls equivalent to  Lines~\ref{loc:tutorial_01_setup_triangulation}-\ref{loc:tutorial_01_compute_error} of Listing~\ref{lst:tutorial_02_program_unit}, corresponds to Step~(1), (2) and (3) with $\triang$ being the initial conforming mesh generated by means of a user-defined number of uniform refinement steps. In particular, the call to {\tt setup\_triangulation} implements Step~(1), and the call to {\tt compute\_error},  Step~(3). The rest of calls within Lines~\ref{loc:tutorial_01_setup_triangulation}-\ref{loc:tutorial_01_compute_error} of Listing~\ref{lst:tutorial_02_program_unit} are required to implement Step~(2). The loop in Listing~\ref{lst:tutorial_02_program_unit} spanning Lines~\ref{loc:tutorial_02_loop_start}-\ref{loc:tutorial_02_loop_end}, implements the successive generation of a sequence of hierarchically refined meshes. In particular, Line~\ref{loc:tutorial_02_refinement_update} implements Step~(4) of Fig.~\ref{fig:tutorial_02_amr_loop}, Line~\ref{loc:tutorial_02_setup_triangulation} Step~(5), and Lines~\ref{loc:tutorial_02_solve_start}-\ref{loc:tutorial_02_solve_stop} and Line~\ref{loc:tutorial_02_compute_error}
	Step~(2) and Step~(3), resp., with $\triang$ being the newly generated mesh.

	The \tutorial{01} helper subroutines in Listings \ref{lst:tutorial_01_environment}, \ref{lst:tutorial_01_functions}, \ref{lst:tutorial_01_bcs}, and \ref{lst:tutorial_01_discrete_solution}, are re-used ``as-is'' for \tutorial{02}.
	In the particular case of {\tt setup\_strong\_boundary\_conditions}, this is possible  as
	the $h$-adaptive triangulation assigns the \ac{vef} set identifiers equivalently to the triangulation that was used by \tutorial{01} (when both are configured to mesh box domains).
	The rest of \tutorial{02} helper subroutines have to be implemented only slightly differently to the ones of \tutorial{01}. In particular, \tutorial{02} handles the {\tt current\_amr\_step} counter; see Listing~\ref{lst:tutorial_02_program_unit}.
	Right at the beginning, the counter is initialized to zero (see Line~\ref{loc:tutorial_02_part3_startup_loop_start}), and incremented at each iteration of the loop spanning Lines~\ref{loc:tutorial_02_loop_start}-\ref{loc:tutorial_02_loop_end} (Line~\ref{loc:tutorial_02_increase_step}). Thus, this counter can be used by the helper subroutines to distinguish among two possible scenarios. If {\tt current\_amr\_step == 0}, the program is located on the initialization section right before the loop, or within this loop otherwise.
	The usage of {\tt current\_amr\_step} by {\tt setup\_triangulation} and {\tt setup\_fe\_space} is illustrated in Listings~\ref{lst:tutorial_02_triangulation} and \ref{lst:tutorial_02_fe_space}, resp.

\begin{lstlisting}[float=htbp,language={[03]Fortran},escapechar=@,caption=The \texttt{setup\_triangulation} procedure.,label={lst:tutorial_02_triangulation}]
subroutine setup_triangulation()
  integer(ip) :: i, num_dims
  if ( current_amr_step == 0 ) then
     ... ! Force the triangulation to mesh the unit box domain              @\label{loc:tutorial_02_triang_force_unix_box}@
     call triangulation%create(environment, parameter_handler%get_values()) @\label{loc:tutorial_02_triang_create}@
     do i=1, num_uniform_refinement_steps                                   @\label{loc:tutorial_02_triang_loop_start}@
        call flag_all_cells_for_refinement()                               
        call triangulation%refine_and_coarsen()                             @\label{loc:tutorial_02_triang_refine_and_coarsen_1}@
     end do                                                                 @\label{loc:tutorial_02_triang_loop_stop}@
  else
     call triangulation%refine_and_coarsen()                                @\label{loc:tutorial_02_triang_refine_and_coarsen_2}@
  end if
end subroutine setup_triangulation

subroutine flag_all_cells_for_refinement()                                  @\label{loc:tutorial_02_triang_flag_start}@
  class(cell_iterator_t), allocatable :: cell
  call triangulation%create_cell_iterator(cell)
  do while ( .not. cell%has_finished() )  
     call cell%set_for_refinement()
     call cell%next()
  end do
  call triangulation%free_cell_iterator(cell) 
end subroutine flag_all_cells_for_refinement                               @\label{loc:tutorial_02_triang_flag_stop}@
\end{lstlisting}

\begin{lstlisting}[float=htbp,language={[03]Fortran},escapechar=@,caption=The \texttt{setup\_fe\_space} subroutine.,label={lst:tutorial_02_fe_space}]
subroutine setup_fe_space()
  if ( current_amr_step == 0 ) then
     ... ! See Listing @~\ref{lst:tutorial_01_fe_space}@
  else
     call fe_space%refine_and_coarsen()
  end if
end subroutine setup_fe_space
\end{lstlisting}

	In Listing~\ref{lst:tutorial_02_triangulation} one may readily observe that the {\tt setup\_triangulation} helper subroutine behaves differently depending on the value of {\tt current\_amr\_step}. When its value is zero (Lines~\ref{loc:tutorial_02_triang_force_unix_box}-\ref{loc:tutorial_02_triang_loop_stop}), the subroutine first generates a triangulation of the unit box domain which is composed of a single brick cell, i.e., a forest-of-octrees with a single adaptive octree (Lines~\ref{loc:tutorial_02_triang_force_unix_box}~\ref{loc:tutorial_02_triang_create}). Then, it enters a loop in which the octree root is refined uniformly {\tt num\_uniform\_refinement\_steps} times (Lines~\ref{loc:tutorial_02_triang_loop_start}-\ref{loc:tutorial_02_triang_loop_stop}) in order to generate $\triang$ in Step~(1) of Fig.~\ref{fig:tutorial_02_amr_loop}; this program variable holds the value provided by the user through the corresponding \ac{cla}. The loop relies on the {\tt flag\_all\_cells\_for\_refinement} helper procedure, which is shown in Lines~\ref{loc:tutorial_02_triang_flag_start}-\ref{loc:tutorial_02_triang_flag_stop} of Listing~\ref{lst:tutorial_02_triangulation}, and the {\tt refine\_and\_coarsen} \ac{tbp} of {\tt triangulation} (Line~\ref{loc:tutorial_02_triang_refine_and_coarsen_1}). The first procedure walks through over the mesh cells, and for each cell, sets a per-cell flag that tells the triangulation to refine the cell, i.e., we flag all cells for refinement, in order to obtain a uniformly refined mesh.\footnote{Iteration over the mesh cells is performed by means of the polymorphic variable {\tt cell} of declared type {\tt cell\_iterator\_t}. Iterators are data types that provide sequential traversals over the {\em full sets of objects} that all together (conceptually) comprise \texttt{triangulation\_t} as a mesh-like container {\em without exposing its internal organization}. Besides, by virtue of Fortran200X native support of run-time polymorphism, the code of {\tt flag\_all\_cells\_for\_refinement} can be leveraged (re-used) for any triangulation that extends the {\tt triangulation\_t} abstract data type, e.g., the one that will be used by \tutorial{03} in Sect.~\ref{sec:tutorial03}. We refer to \cite[Sect. 7]{badia_fempar:_2017} for the rationale underlying the design of iterators in \FEMPAR{}.} On the other hand, the {\tt refine\_and\_coarsen} \ac{tbp}
	adapts the triangulation based on the cell flags set by the user, while transferring data that the user might have attached to the mesh objects (e.g., cells and \acp{vef} set identifiers) to the new mesh objects generated after mesh adaptation.

	When {\tt current\_amr\_step} is larger than zero, Listing~\ref{lst:tutorial_02_triangulation} invokes the {\tt refine\_and\_coarsen} \ac{tbp} of the triangulation (Line~\ref{loc:tutorial_02_triang_refine_and_coarsen_2}). As commented in the previous paragraph, this procedure adapts the mesh accordingly to how the mesh cells have been marked for refinement, coarsening, or to be left as they were prior to adaptation. This latter responsibility falls on the {\tt refinement\_strategy} instance, and in particular in its {\tt update\_refinement\_flags} \ac{tbp}. This \ac{tbp}, which is called in Line~\ref{loc:tutorial_02_refinement_update} of Listing~\ref{lst:tutorial_02_program_unit}, right before the call to {\tt setup\_triangulation} in Line~\ref{loc:tutorial_02_setup_triangulation} of Listing~\ref{lst:tutorial_02_program_unit},
	follows the strategy in Step~(4) of Fig.~\ref{fig:tutorial_02_amr_loop}.

	Listing~\ref{lst:tutorial_02_fe_space} follows the same pattern as Listing~\ref{lst:tutorial_02_triangulation}. When the value of {\tt current\_amr\_step} is zero, it sets up the {\tt fe\_space} instance using exactly the same sequence of calls as in Listing~\ref{lst:tutorial_01_fe_space}. When the value of {\tt current\_amr\_step} is larger than zero, it calls the {\tt refine\_and\_coarsen} \ac{tbp} of {\tt fe\_space} to build a new global \ac{fe} space $V_h$ from the newly generated triangulation, while trying to re-use its already  allocated internal data buffers as much as possible. Optionally, this \ac{tbp} can by supplied with a \ac{fe} function $u_h$ (or, more generally, an arbitrary number of them). In such a case, {\tt refine\_and\_coarsen} injects the \ac{fe} function provided into the newly generated $V_h$ by using a suitable \ac{fe} projector for the \ac{fe} technology being used, the Lagrangian interpolator in the case of \tutorial{02}. This feature is required by numerical solvers of transient and/or non-linear \acp{pde}.

	Listing~\ref{lst:tutorial_02_solve_system} shows those lines of code of the {\tt solve\_system} procedure of \tutorial{02} which are different to its \tutorial{01} counterpart in Listing~\ref{lst:tutorial_01_solve_system}.\footnote{In order to keep the presentation concise, the adaptation of the rest of \tutorial{01} helper subroutines to support the \ac{amr} loop in Fig.~\ref{fig:tutorial_02_amr_loop} is not covered in this section. At this point, the reader might inspect how the rest of \tutorial{02} subroutines are implemented by looking at the source code Git repository.}
	 Apart from the fact that Listing~\ref{lst:tutorial_02_solve_system} follows the same pattern as Listings~\ref{lst:tutorial_02_triangulation} and \ref{lst:tutorial_02_fe_space}, the reader should note the call to the {\tt update\_hanging\_dof\_values} \ac{tbp} of {\tt fe\_space} in Line~\ref{loc:tutorial_02_solve_system_update_hanging}. This \ac{tbp} computes the values of $u_h$ corresponding to hanging \acp{dof} using the algebraic constraints required to enforce the conformity of $V_h$; see Sect.~\ref{sec:tutorial02_fe_discretization}. This is required each time that the values of $u_h$ corresponding to {\em true} \acp{dof} are updated (e.g., after linear system solution). Not calling it results in unpredictable errors during post-processing, or even worse, during a non-linear and/or time-stepping iterative solution loop, in which the true \ac{dof} values of $u_h$ are updated at each iteration.

\begin{lstlisting}[float=htbp,language={[03]Fortran},escapechar=@,caption=The \texttt{solve\_system} subroutine.,label={lst:tutorial_02_solve_system}]
subroutine solve_system()
  if ( current_amr_step == 0 ) then
     ... ! Lines @\ref{loc:tutorial_01_solve_system_setup_start}@-@\ref{loc:tutorial_01_solve_system_setup_stop}@ of Listing@~\ref{lst:tutorial_01_solve_system}@
  else 
     call direct_solver%reallocate_after_remesh()
  end if
  ... ! Lines @\ref{loc:tutorial_01_solve_system_get_dofs}@-@\ref{loc:tutorial_01_solve_system_solve}@ of Listing@~\ref{lst:tutorial_01_solve_system}@
  if ( fe_formulation == "CG" ) then
     call fe_space%update_hanging_dof_values(discrete_solution)  @\label{loc:tutorial_02_solve_system_update_hanging}@
  end if
end subroutine solve_system
\end{lstlisting}

	Listing~\ref{lst:tutorial_02_ref_strat} shows the helper subroutine that sets up the {\tt refinement\_strategy} object. As mentioned above, this object implements the strategy in Step~(2) of Fig.~\ref{fig:tutorial_02_amr_loop}.\footnote{\FEMPAR{} v1.0.0 also provides an implementation of the Li and Bettess refinement strategy \cite{Li1997}. We stress, nevertheless, that the library is designed such that new refinement strategies can be easily added by developing type extensions (subclasses) of the {\tt refinement\_strategy\_t} abstract data type.} In Lines~\ref{loc:tutorial_02_ref_strat_r}-\ref{loc:tutorial_02_ref_strat_c}, the procedure forces {\tt refinement\_strategy} to use parameter values $\alpha_r=0.1$ and $\alpha_c=0.05$; see Step~(2) of Fig.~\ref{fig:tutorial_02_amr_loop}. The actual set up of {\tt refinement\_strategy} occurs at Line~\ref{loc:tutorial_02_ref_strat_create}. We note that, in this line,   {\tt refinement\_strategy} is supplied with the {\tt error\_estimator} instance, of type \texttt{poisson\_error\_estimator\_t}. This tutorial-specific data type, which is also used by \tutorial{01}, extends the {\tt error\_estimator\_t} \FEMPAR{} abstract data type.
	This latter data type is designed to be a
	place-holder of a pair of cell data arrays storing the per-cell true and estimated errors, resp. These arrays are computed by the data type extensions of {\tt error\_estimator\_t} in the {\tt compute\_local\_true\_errors} and {\tt compute\_local\_estimates} (deferred) \acp{tbp}. Neither \tutorial{02} nor \tutorial{01} actually uses a-posteriori error estimators. Thus, the  {\tt compute\_local\_estimates} \ac{tbp} of  \texttt{poisson\_error\_estimator\_t} just copies the local true errors into the estimated errors array. While the {\tt compute\_error} helper subroutine of \tutorial{01} (Listing~\ref{lst:tutorial_01_compute_error}) does not call the {\tt compute\_local\_estimates} \ac{tbp}, its counterpart in \tutorial{02} {\em must call it}, as the {\tt refinement\_strategy} object extracts the data to work with from the error estimators array of {\tt error\_estimator}.

\begin{lstlisting}[float=htbp,language={[03]Fortran},escapechar=@,caption=The \texttt{setup\_refinement\_strategy} subroutine.,label={lst:tutorial_02_ref_strat}]
subroutine setup_refinement_strategy()
  call parameter_handler%update(ffrs_refinement_fraction_key, 0.10_rp) @\label{loc:tutorial_02_ref_strat_r}@
  call parameter_handler%update(ffrs_coarsening_fraction_key, 0.05_rp) @\label{loc:tutorial_02_ref_strat_c}@
  call parameter_handler%update(ffrs_max_num_mesh_iterations_key, num_amr_steps)
  call refinement_strategy%create(error_estimator,parameter_handler%get_values()) @\label{loc:tutorial_02_ref_strat_create}@
end subroutine setup_refinement_strategy
\end{lstlisting}

	Listing~\ref{lst:tutorial_02_program_unit} handles the output of post-processing data files rather differently from Listing~\ref{lst:tutorial_01_program_unit}. It in particular generates simulation results for the full set of triangulations generated in the course of the simulation. In order to do so, it relies on the ability of {\tt output\_handler\_t} to manage the time steps in transient simulations on a sequence of triangulations which might be different at each single time step.
	The {\tt output\_...\_initialize} procedure resembles Listing~\ref{lst:tutorial_01_oh}, except that it does not call neither the {\tt write} nor the {\tt close} \acp{tbp} of {\tt output\_handler\_t}.
	Besides, it forces a parameter value of {\tt output\_handler\_t} to inform it that the triangulation might be different at each single step. If the mesh does not evolve dynamically, then it is only written once, instead of at every step, saving disk storage (as far as such feature is supported by the output data format).
	The {\tt output\_handler\_write\_...\_step} procedure (see Listing~\ref{lst:tutorial_02_program_unit})
	first calls the {\tt append\_time\_step} and, then, the {\tt write} \acp{tbp} of {\tt output\_handler\_t}. The combination of these two calls outputs a new step into the output data files. Finally, the {\tt output\_handler\_finalize} procedure (see Listing~\ref{lst:tutorial_02_program_unit}) just calls the {\tt close} \ac{tbp} of {\tt output\_handler\_t}. This call closes all file units handled by {\tt output\_handler}, thus flushing into disk all pending write operations.

	\subsection{Numerical results} \label{sec:tutorial02_results}

\begin{figure}[ht!]
	  \centering
	  \begin{subfigure}{\textwidth}
	    \centering
	    \includegraphics[width=0.45\textwidth]{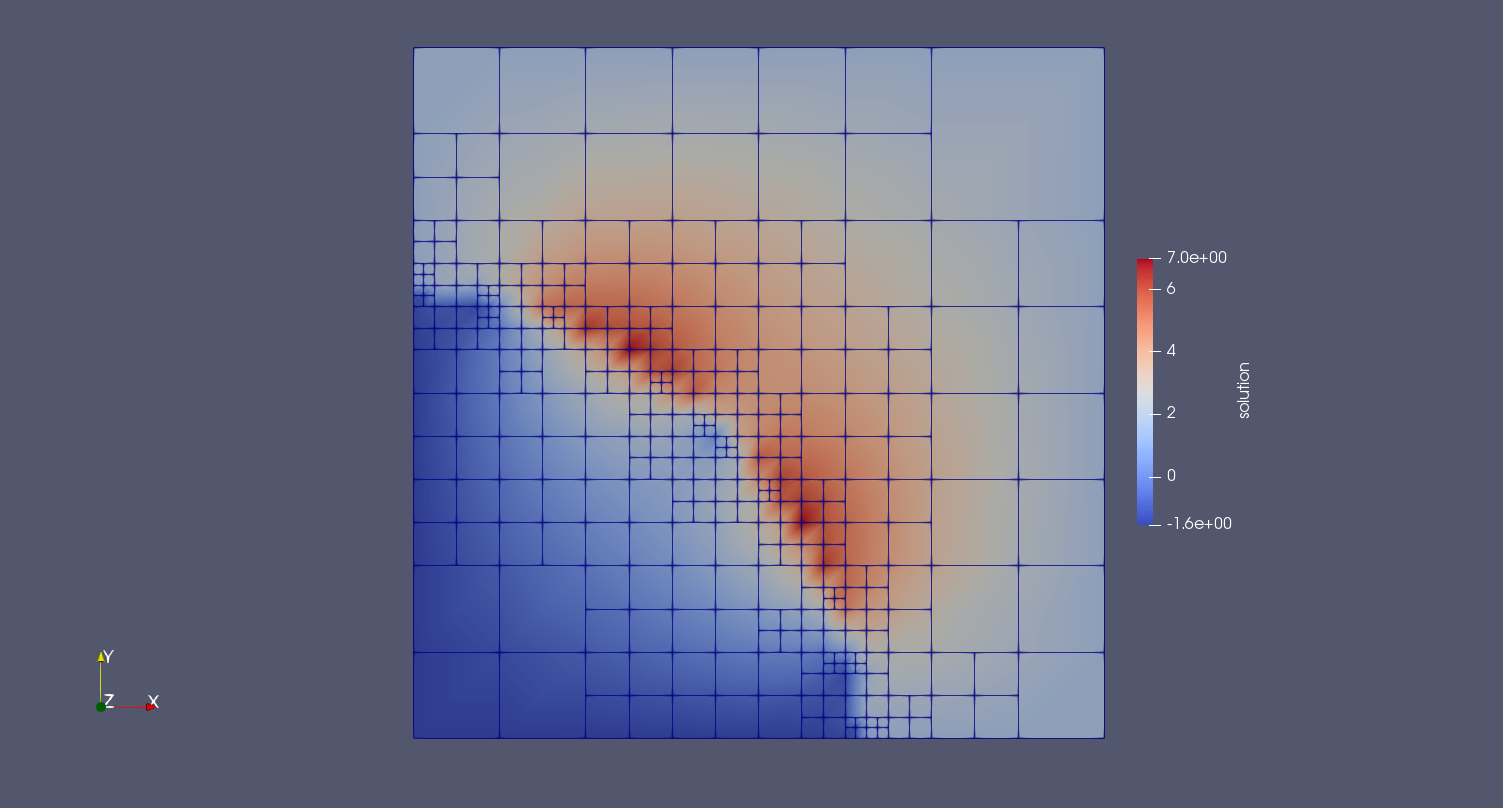}
	    \includegraphics[width=0.45\textwidth]{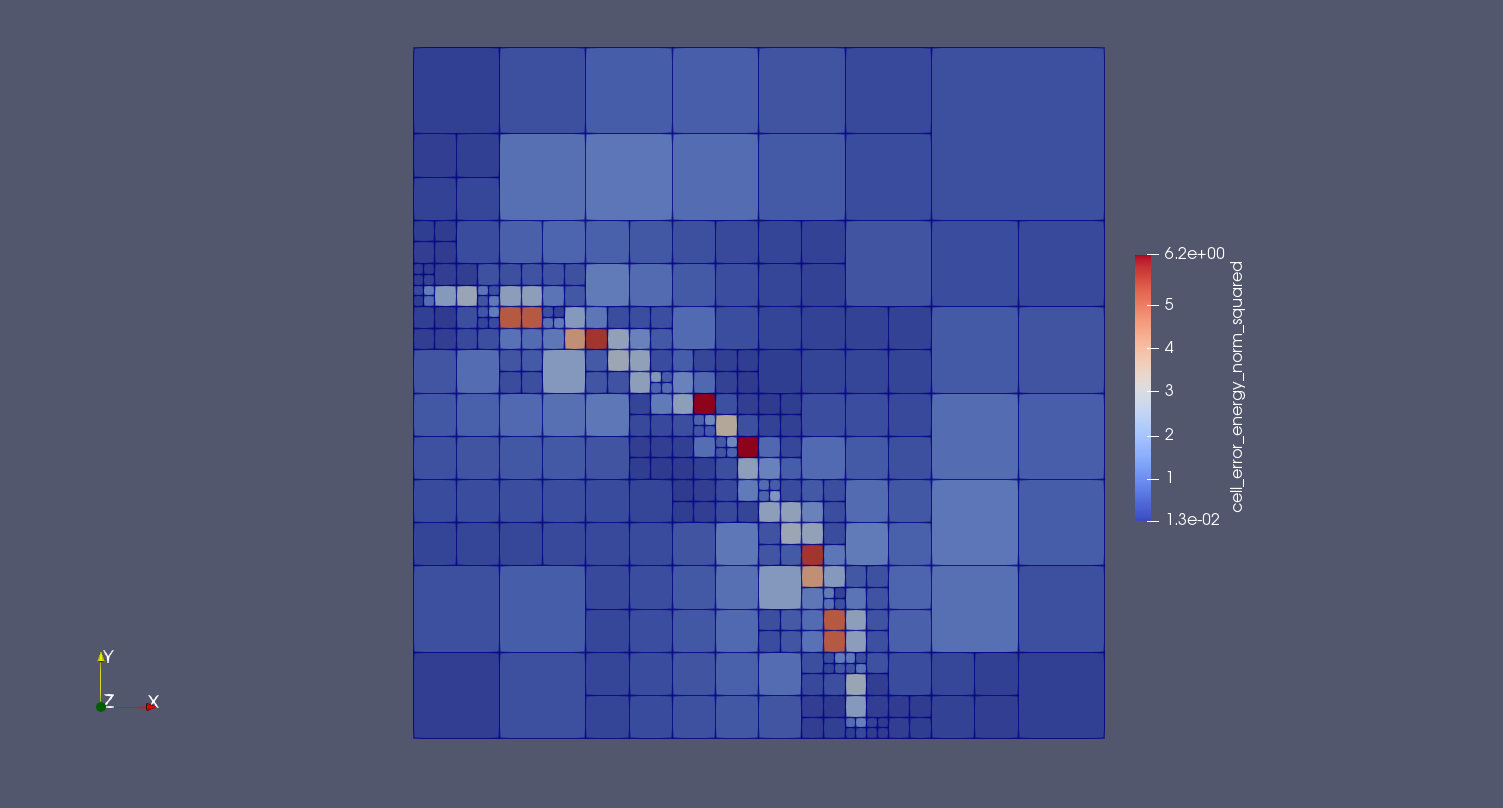}
	    \caption{8 \ac{amr} steps; 322 cells; 243 (true) \acp{dof}.}
	    \label{fig:s1_q1}
	  \end{subfigure}
	  \begin{subfigure}{\textwidth}
	    \centering
	    \includegraphics[width=0.45\textwidth]{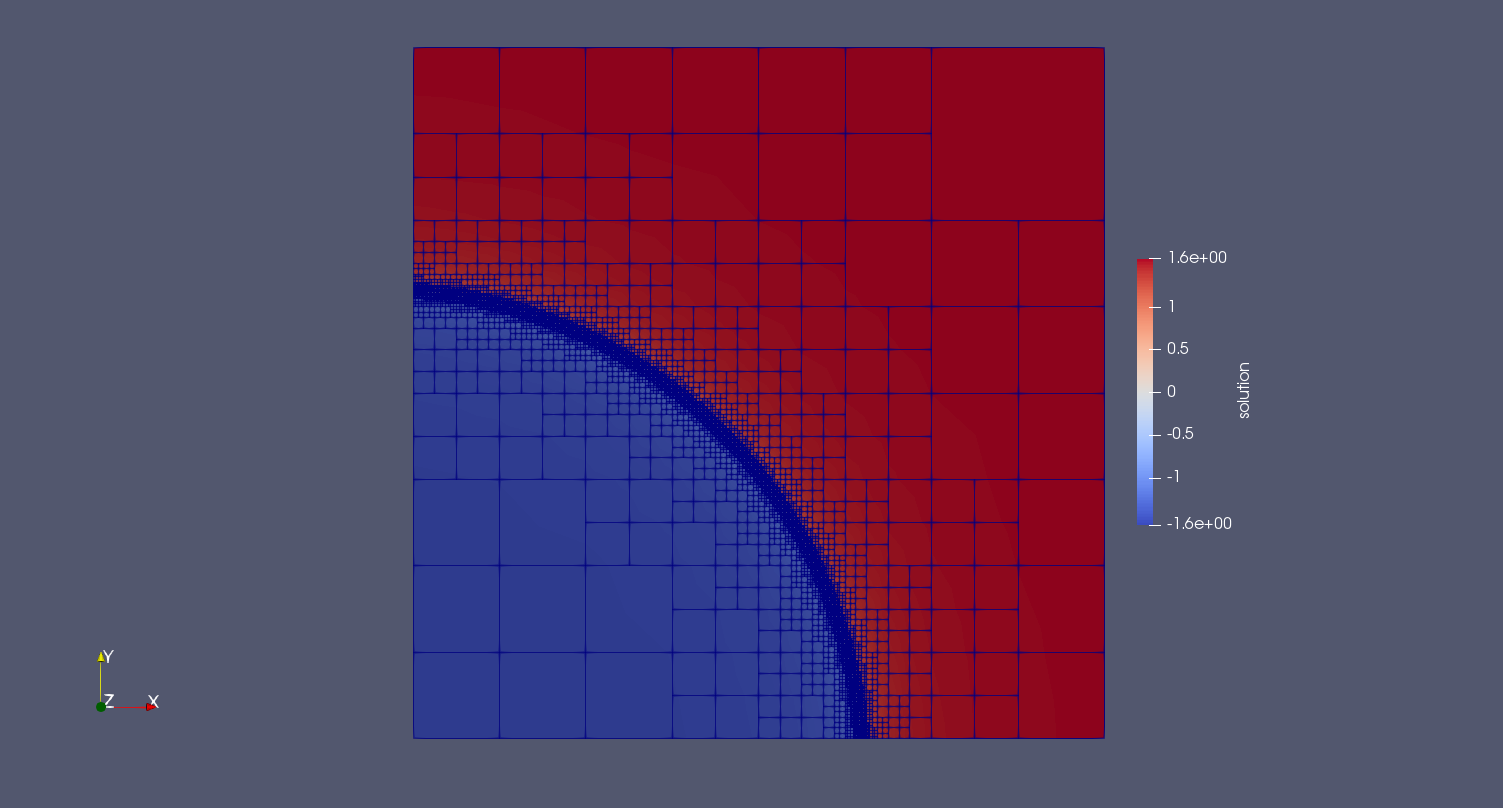}
	    \includegraphics[width=0.45\textwidth]{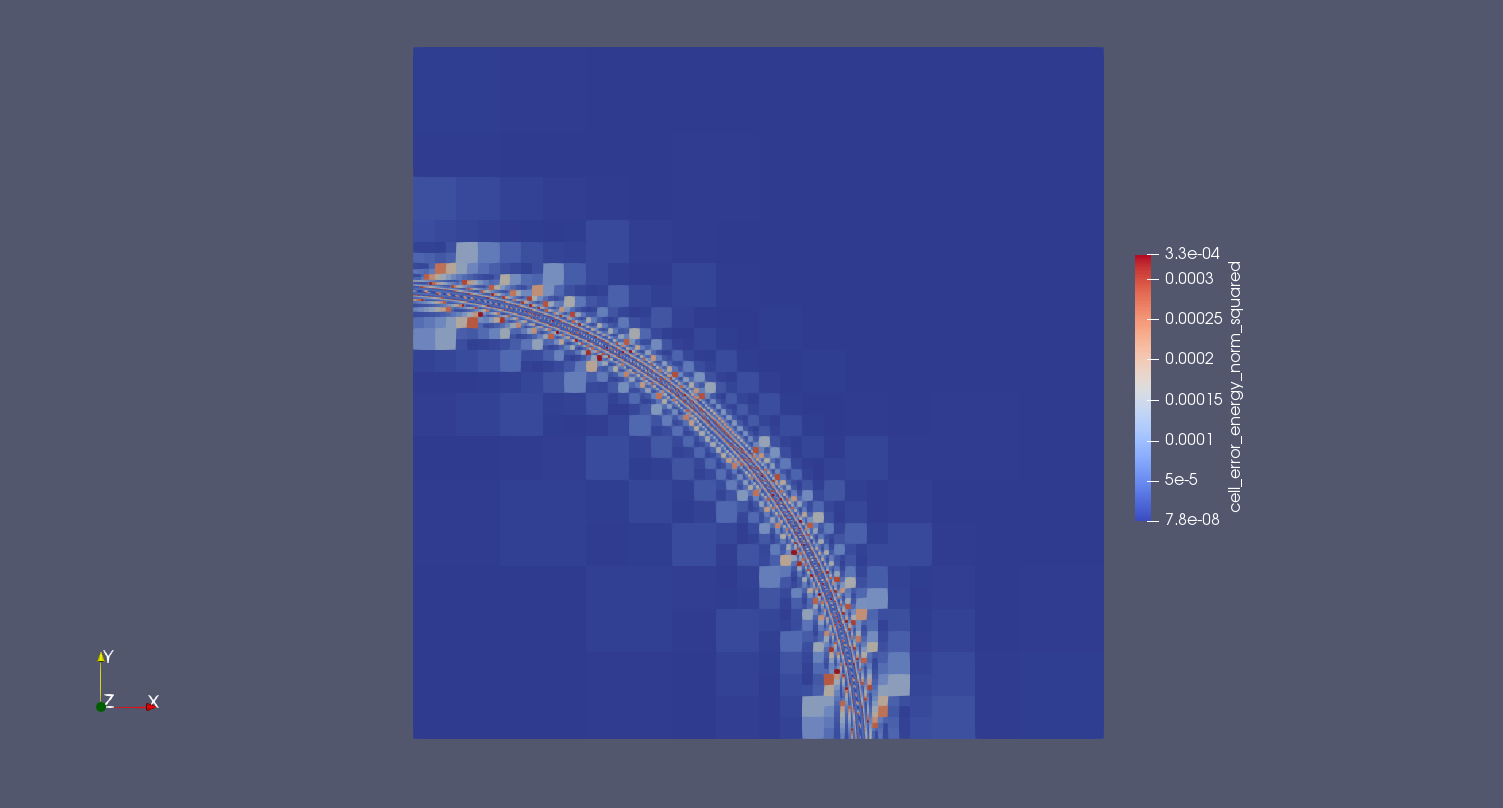}
	    \caption{20 \ac{amr} steps; 13,564 cells; 11,535 (true) \acp{dof}.}
	    \label{fig:s2_q1}
	  \end{subfigure}
	  \caption{Mesh and \ac{fe} solution (left) and $e_K^2$ for all $K \in \triang$ (right).}
	  \label{fig:tutorial_02_results}
	\end{figure}

	In Fig.~\ref{fig:s1_q1} and~\ref{fig:s2_q1} we show the \ac{fe} solution computed by \tutorial{02}, along with $e_K^2$, for all $K \in \triang$, for the 2D version of Problem~\eqref{sec:tutorial01_model_problem} discretized with an adapted mesh resulting from 8 and 20  \ac{amr} steps, resp., and bilinear Lagrangian \acp{fe}. The number of initial uniform refinement steps was set to 2, resulting in an initial conforming triangulation made of 16 quadrilateral cells. As expected, the mesh tends to be locally refined close to the internal layer.

\begin{figure}[t!]
	    \centering
	    \begin{subfigure}[t]{0.45\textwidth}
		\includegraphics[width=\textwidth]{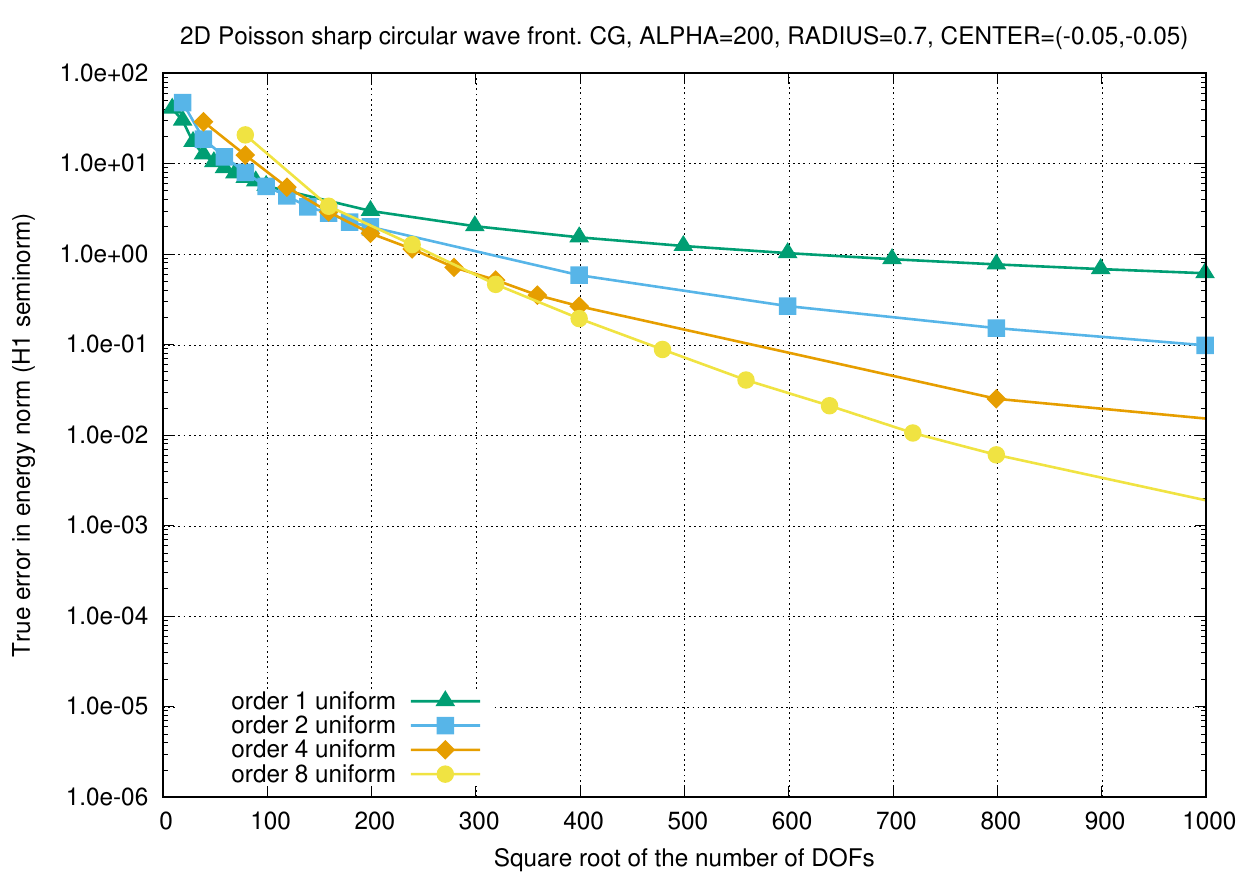}
		\caption{Uniform meshing.}
		\label{fig:conv_uniform}
	    \end{subfigure}
	    \begin{subfigure}[t]{0.45\textwidth}
		\includegraphics[width=\textwidth]{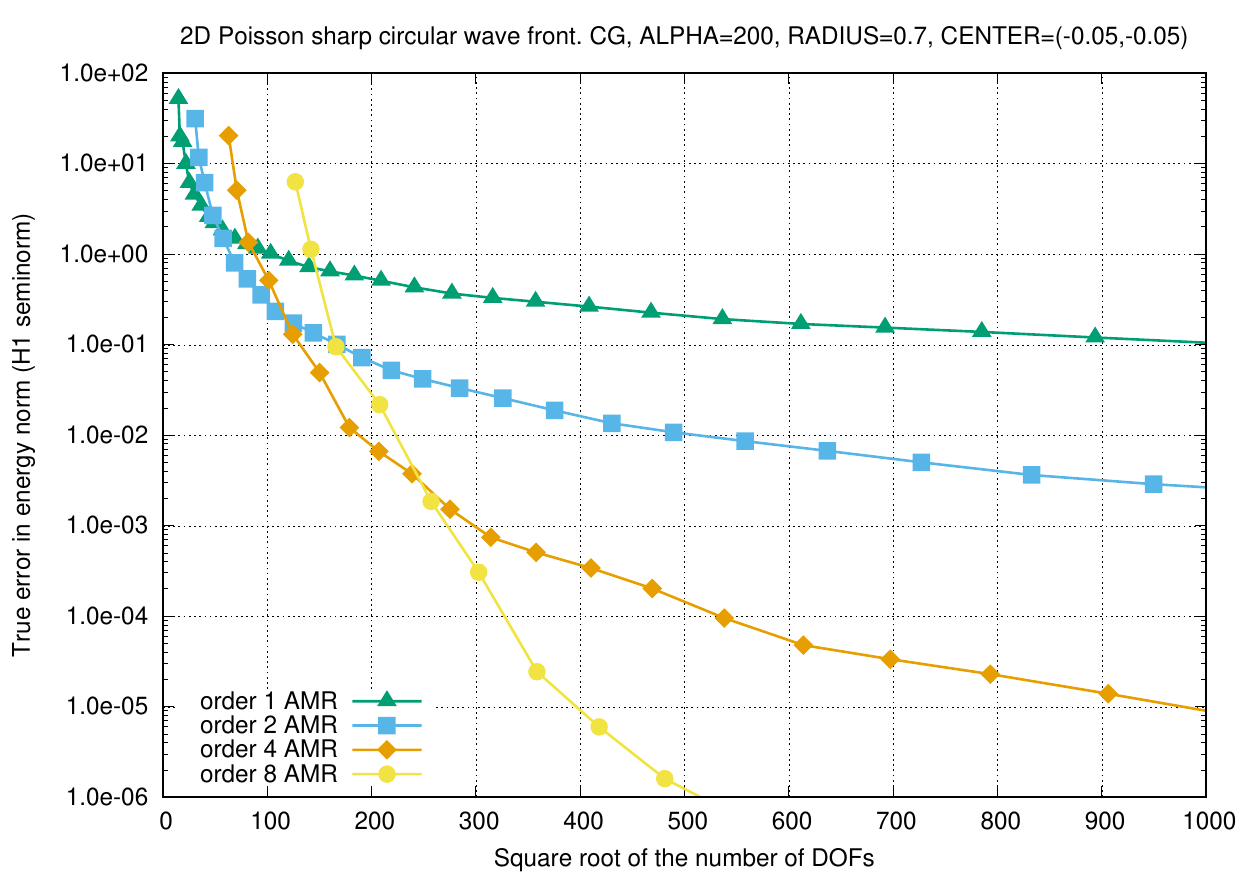}
		\caption{AMR.}
		\label{fig:conv_amr}
	    \end{subfigure}
	    \caption{Convergence history for the 2D benchmark problem using the \ac{cg} \ac{fe} formulation with different polynomial order.}\label{fig:convergence_history}
	\end{figure}

	On the other hand, in Fig.~\ref{fig:convergence_history}, we show error convergence history plots for the 2D benchmark problem.  The results in Fig.~\ref{fig:conv_uniform} were obtained with \tutorial{01}, while those in Fig.\ref{fig:conv_amr}, with \tutorial{02}. As expected, the benefit of using local refinement is substantial for the problem at hand.\footnote{We note that the plots in Fig~\ref{fig:convergence_history} can be automatically generated using the Unix bash shell scripts located at the {\tt convergence\_plot} subfolder accompanying the source code of \tutorial{01} and \tutorial{02}.}

	\section{\Tutorial{03}: Distributed-memory parallelization of \tutorial{02}} \label{sec:tutorial03}

	\subsection{Model problem} \label{sec:tutorial03_model_problem}
	See Sect.~\ref{sec:tutorial01_model_problem}.

	\subsection{Parallel \ac{fe} discretization} \label{sec:tutorial_03_fe_discretization}

	\Tutorial{03} exploits a set of {\em fully-distributed} data structures and associated algorithms available in \FEMPAR{} for the scalable solution of \ac{pde} problems in high-end distributed-memory computers~\cite{Badia2019a}. Such data structures are driven by \tutorial{03} in order to efficiently  parallelize the \ac{amr} loop of \tutorial{02} (Fig.~\ref{fig:tutorial_02_amr_loop}). In order to find
	$u_h$ at each adaptation step (Step~(2), Fig.~\ref{fig:tutorial_02_amr_loop}), \tutorial{03} combines the \ac{cg} \ac{fe} formulation for Prob.~\eqref{eq-poisson} (Sect.~\ref{sec:cg_formulation_poisson}) with a scalable domain decomposition preconditioner for the fast iterative solution of the linear system resulting from \ac{fe} discretization.\footnote{\FEMPAR{} v1.0.0 also supports {\em parallel} \ac{dg}-like non-conforming \ac{fe} formulations for the Poisson problem. However, a scalable domain decomposition preconditioner suitable for this family of \ac{fe} formulations is not yet available in its first public release.
	This justifies why \tutorial{03} restricts itself to the \ac{cg} \ac{fe} formulation.
	In any case, we stress that \FEMPAR{} is designed such that this preconditioner can be easily added in future releases of the library.} In this section, we briefly introduce some key ideas underlying the extension of the approach presented in Sect.~\ref{sec:tutorial02_fe_discretization} to distributed computing environments. On the other hand, Sect.~\ref{sec:tutorial04_preconditioning} overviews the preconditioning approach used by \tutorial{03}, and its parallel implementation in \FEMPAR{}.

\begin{figure}[ht!]
	\centering
	\begin{subfigure}{0.24\textwidth}
	  \centering
	  \includegraphics[width=0.6\textwidth]{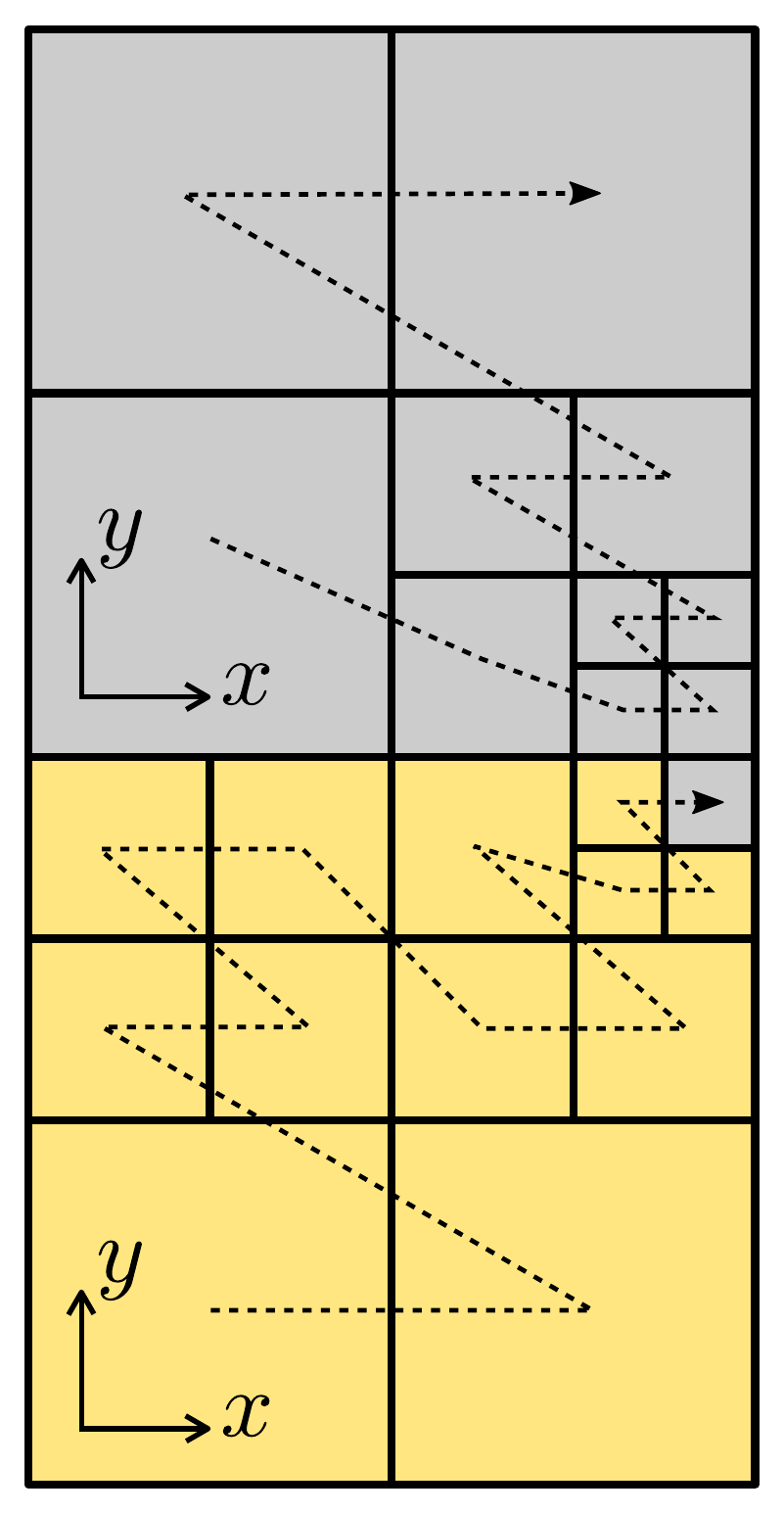}
	  \caption{$\triang$.}
	  \label{fig:mesh-distribution-a}
	\end{subfigure}
	\begin{subfigure}{0.24\textwidth}
	  \centering
	  \includegraphics[width=0.6\textwidth]{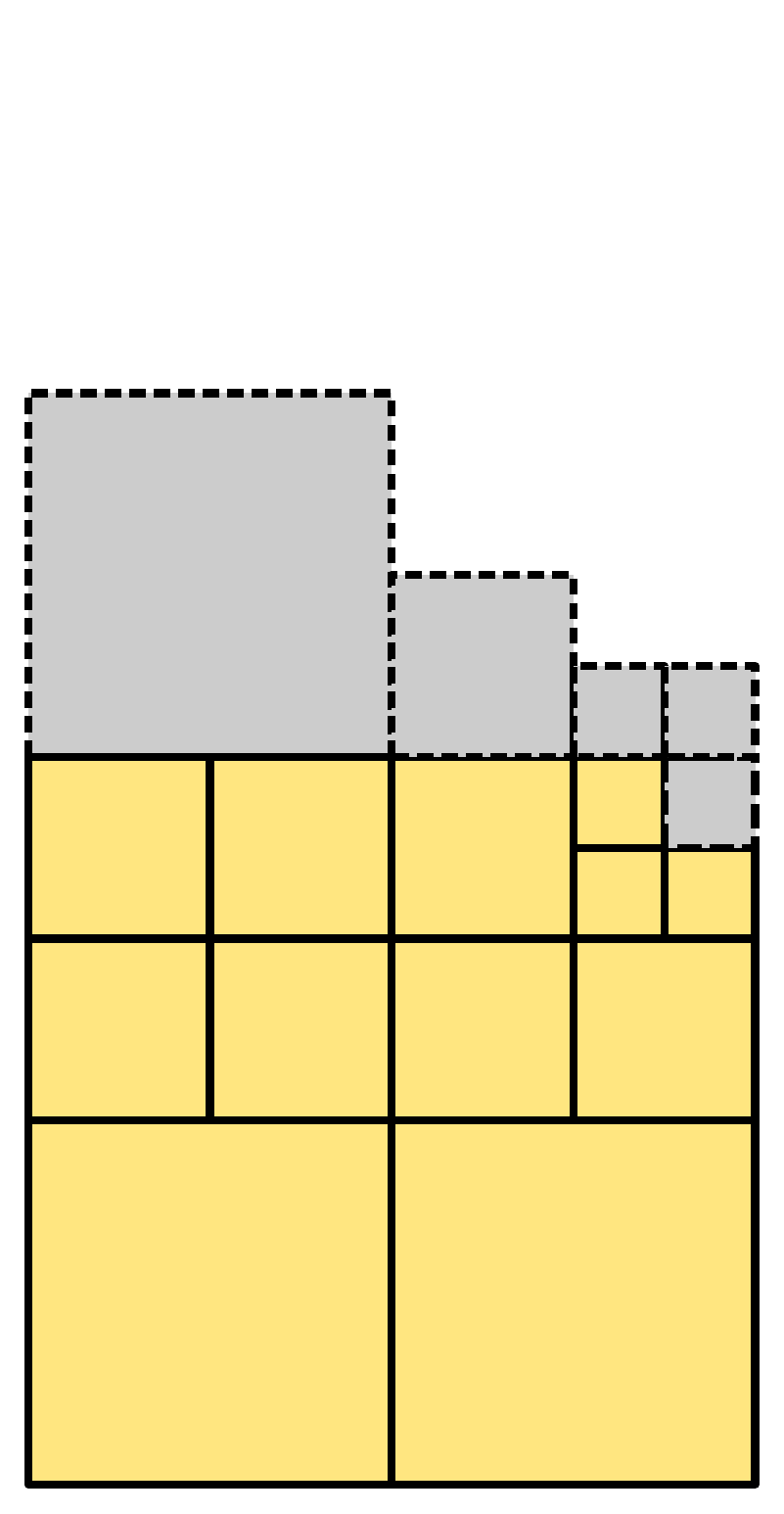}
	  \caption{$\mathcal{T}_h^1$.}
	  \label{fig:mesh-distribution-b}
	\end{subfigure}
	\begin{subfigure}{0.24\textwidth}
	  \centering
	  \includegraphics[width=0.6\textwidth]{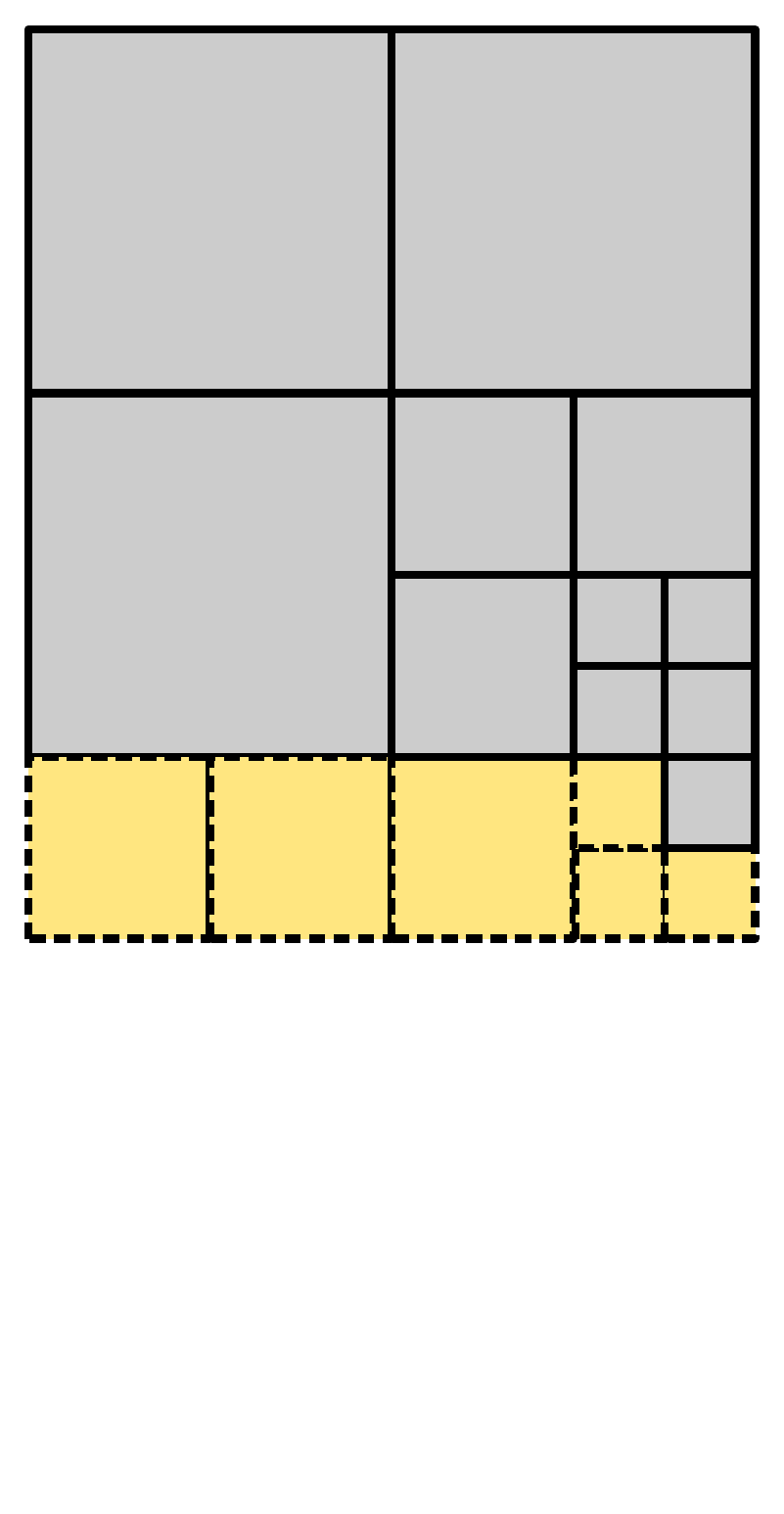}
	  \caption{$\mathcal{T}_h^2$.}
	  \label{fig:mesh-distribution-c}
	\end{subfigure}
	\caption{2:1 balanced forest-of-quadtrees mesh with two quadtrees (i.e., $|\mathcal{C}_h|=2$) distributed among two processors,
		 1:4 refinement and
		 the Morton \ac{sfc}~\cite{burstedde_p4est_2011}.
		 Local cells are depicted
		 with continuous boundary lines, while those in the ghost layer with dashed ones.}
	\label{fig:mesh-distribution}
	\end{figure}

	It order to scale \ac{fe} simulations to large core counts, the adaptive mesh must be partitioned (distributed) among the parallel tasks such that each of these only holds a local portion of the global mesh. (The same requirement applies to the rest of data structures in the \ac{fe} simulation pipeline, i.e., \ac{fe} space, linear system, solver, etc.) Besides, as the solution might exhibit highly localized features, dynamic mesh adaptation can result in an unacceptable amount of load imbalance. Thus, it urges that the adaptive mesh data structure supports {\em dynamic load-balancing}, i.e., that it can be re-distributed among the parallel processes in the course of the simulation.
	As mentioned in Sect.~\ref{sec:tutorial02_fe_discretization}, dynamic $h$-adaptivity in \FEMPAR{}
	relies on forest-of-trees meshes. Modern forest-of-trees manipulation engines provide a scalable, linear runtime solution to the mesh (re-)partitioning problem based on the exploitation of \acp{sfc}. \acp{sfc} provide a natural means to assign an ordering of the forest-of-trees leaves,
	which is exploited for the parallel arrangement of data. For example,
	in the \p4est{} library, the forest-of-octrees leaves are arranged in a global one-dimensional data array in increasing Morton index ordering~\cite{burstedde_p4est_2011}. This ordering corresponds geometrically with the traversal of a $z$-shaped \ac{sfc} (a.k.a. Morton \ac{sfc}) of $\mathcal{T}_h$; see Fig.~\ref{fig:mesh-distribution-a}. This approach allows for fast dynamic repartitioning. A partition of $\triang$ is simply generated by dividing the leaves in the linear ordering induced by the \ac{sfc} into as many equally-sized segments as parallel tasks involved in the computation.

	The parallel $h$-adaptive triangulation in \FEMPAR{} reconstructs the local portion of $\triang$ corresponding to each task from the distributed forest-of-octrees that \p4est{} handles internally~\cite{Badia2019a}. These local portions are illustrated in Fig.~\ref{fig:mesh-distribution-b} and \ref{fig:mesh-distribution-c} when the forest-of-octrees in Fig~\ref{fig:mesh-distribution-a} is distributed among two processors. The local portion of each task is composed by a set of cells that it owns, i.e., the {\em local cells} of the task, and a set of off-processor cells (owned by remote processors) which are in touch with its local cells, i.e., the {\em ghost cells} of the task. This overlapped mesh partition is used by the library to exchange data among nearest neighbours, and to glue together the global \acp{dof} of $V_h$ which are sitting on the interface among subdomains, as required in order to construct \ac{fe} spaces for conforming \ac{fe} formulations in a distributed setting~\cite{Badia2019a}.

	The user of the library, however, should also be aware to some extent of the distributed data layout of the triangulation.
	Depending on the numerical method at hand, it might be required to perform computations that involve the ghost cells, or to completely avoid them.
	For example, the computation of facet integrals on the interface among subdomains requires access to the ghost cells data (e.g., local shape functions values and gradients). On the other hand,
	cell integrals are typically assembled into global data structures distributed across processors (e.g., the linear system or the global energy norm of the error). While it is practically possible to evaluate a cell integral over a ghost cell in \FEMPAR{}, this would result in excess computation, and even worse, to over-assembly due to the overlapped mesh partition (i.e., to wrong results). To this end, cell iterators of the parallel $h$-adaptive triangulation provide \acp{tbp} that let the user to distinguish among local and ghost cells, e.g., in an iteration over all cells of the mesh portion of a parallel task.

	\subsection{Fast and scalable parallel linear system solution} \label{sec:tutorial04_preconditioning}

	\Tutorial{03} solves the linear system resulting from discretization {\em iteratively} via (preconditioned) Krylov subspace solvers~\cite{saad_iterative_2003}. To this end, \FEMPAR{} provides {\em abstract implementations} (i.e., that can be leveraged either in serial or distributed computing environments, and/or for scalar or blocked layouts of the linear(ized) system) of a rich suite of solvers of this kind, such as, e.g., the Conjugate Gradients and GMRES solvers. Iterative solvers are much better suited than sparse direct solvers for the efficient exploitation of distributed-memory computers.
	However, they have to be equipped with an efficient preconditioner, a cornerstone ingredient for convergence acceleration, robustness and scalability.

	Preconditioners based on the \ac{dd} approach~\cite{Toselli2005} are an appealing solution for the fast and scalable parallel iterative solution of linear systems arising from \ac{pde} discretization~\cite{art003,badia_multilevel_2016,zampini_PETSc_scalability}. \ac{dd} preconditioners make explicit use of the partition of the global mesh into sub-meshes (see Fig.~\ref{fig:mesh-distribution}), and involve the solution of local problems and communication among nearest-neighbour subdomains. In order to achieve algorithmic scalability, i.e., a condition number that remains constant as the problem size and number of subdomains are scaled, they have to be equipped with a suitably defined coarse-grid correction.  The coarse-grid correction globally couples all subdomains and rapidly propagates the error correction information across the whole domain. However, it involves the solution of a global problem  whose size typically increases (at best) linearly with respect to the number of subdomains. If not tackled appropriately by the underlying parallel implementation\cite{art003,badia_multilevel_2016,zampini_PETSc_scalability}, this increase can jeopardize the practical scalability limits of this kind of preconditioners.

	Among the set of scalable \ac{dd} preconditioners available in the literature~\cite{Toselli2005, brenner_mathematical_2010}, \FEMPAR{} built-in preconditioning capabilities are grounded on the so-called \ac{bddc} preconditioning approach~\cite{dohrmann_preconditioner_2003,art003}, and its multi-level extension\cite{mandel_multispace_2008, badia_multilevel_2016} for extreme-scale computations. \ac{bddc} preconditioners belong to the family of non-overlapping \ac{dd} methods \cite{Toselli2005}. Computationally speaking, \ac{bddc} preconditioners require to solve a local Dirichlet and a local constrained Neumann problem at each subdomain, and a global coarse-grid problem~\cite{dohrmann_preconditioner_2003}. These methods rely on the definition of a \ac{fe} space, referred to as the \ac{bddc} space, with relaxed inter-subdomain continuity. The local constrained Neumann problems and the global coarse-grid problem are required in order to extract a correction of the solution from the \ac{bddc} space.
	Such space is defined by choosing some quantities to be continuous across subdomain interfaces, i.e., the \emph{coarse} or \emph{primal} \acp{dof}.
	The definition of the coarse \acp{dof} in turn relies on a geometrical partition of the mesh \acp{vef} laying on the subdomain interfaces into coarse objects, i.e., coarse \acp{vef}. Next, we associate to some (or all) of these objects a \emph{coarse} \ac{dof}.  Once a correction has been extracted from the \ac{bddc} space, the continuity of the solution at the interface between subdomains is restored with an averaging operator.

	The actual definition of the coarse \acp{dof} depends on the kind of \ac{fe} space being used for \ac{pde} discretization. For grad-conforming (i.e., $H^1$-conforming) \ac{fe} spaces, as those required for the discretization of the Poisson \ac{pde}, the coarse \acp{dof} of a \ac{fe} function $u_h$ are defined as the value of the function at vertices, or the mean values of the function on coarse edges/faces. These concepts have been generalized for div- and curl-conforming \ac{fe} spaces as well; see, e.g., \cite{Badia2019-electromagnetic-solvers}, and references therein, for the latter kind of spaces.

While \tutorial{03} uses a 2-level \ac{bddc} preconditioner suitable for the Poisson \ac{pde}, \FEMPAR{} actually goes much beyond than that by providing an abstract \ac{oo} framework for the implementation of widely applicable \ac{bddc}-like preconditioners. It is not the aim of this paper that the reader fully understands the complex details underlying this framework. However, it is at least convenient to have some familiarity with the data types on which the framework relies, and their basic roles in the construction of a \ac{bddc} preconditioner, as these are exposed in the code of \tutorial{03} in Sect.~\ref{sec:tutorial03_commented_code}. These are the following ones:
\begin{itemize}
 \item {\tt coarse\_triangulation\_t}.
 The construction of this object starts with the usual \ac{fe} discretization mesh distributed among parallel tasks; see Fig.~\ref{fig:mesh-distribution}. Each of these tasks {\em locally} classifies the mesh \acp{vef} lying on the interface among its local and ghost cells into coarse \acp{vef} (see discussion above).
 Then, these coarse \acp{vef} are glued together across parallel tasks
 by generating a global numbering of coarse \acp{vef} in parallel. Finally, all coarse cells (i.e., subdomains) and its coarse \acp{vef} are transferred from each parallel task to an specialized task (or set of tasks in the case it is distributed) that assembles them into a {\tt coarse\_triangulation\_t} object. This mesh-like container very much resembles the {\tt triangulation\_t} object (and indeed re-uses much of its code), except for the fact that the former does not discretize the geometry of any domain, as there is no domain to be discretized in order to build a \ac{bddc} coarse space.
 \item {\tt coarse\_fe\_space\_t}. This object very much resembles {\tt fe\_space\_t}. It handles a global numbering of the coarse \acp{dof} of the \ac{bddc} space. However, it does not provide data types for the evaluation of cell and facet integrals, as the cell matrices and vectors required to assemble the global coarse-grid problem are not actually computed as usual in \ac{fe} methods, but by  Galerkin projection of the sub-assembled discrete linear system using the basis functions of the coarse-grid space \cite{dohrmann_preconditioner_2003}. As {\tt coarse\_triangulation\_t}, {\tt coarse\_fe\_space\_t} is stored in a specialized parallel task (or set of tasks) that builds it by assembling the data provided by the tasks on which the \ac{fe} space is distributed.
 \item {\tt coarse\_fe\_handler\_t}. This is an abstract data type that very much resembles a local \ac{fe} space, but defined on a subdomain (i.e., a coarse cell). It defines the association among coarse \acp{dof} and coarse \acp{vef}, and provides mechanisms for the evaluation of the functionals associated to coarse \acp{dof} (i.e., the coarse \ac{dof} values), given the values of a \ac{fe} function. It also defines the so-called weighting operator as a basic customizable building block required to define the averaging operator required to restore continuity. Data type extensions of {\tt coarse\_fe\_handler\_t} suitably define these ingredients for the \ac{fe} space used for \ac{pde} discretization.
 \item {\tt mlbddc\_t}. This is the main data type of the framework. It orchestrates the previous objects in order to build, and later on apply the \ac{bddc} preconditioner at each iteration of a Krylov subspace solver. For example, using {\tt coarse\_fe\_handler\_t}, and the sub-assembled local Neumann problems (i.e., the local matrices that the user assembles on each local subdomain), it builds the local constrained Neumann problem required, among others, in order to compute the basis functions of the coarse-grid space, or to extract a correction from the \ac{bddc} space~\cite{dohrmann_preconditioner_2003}. It also builds the coarse cell matrices and vectors at each subdomain, and transfers them to the task (or set of tasks) that assembles the coarse-grid linear system. This task in turn uses {\tt coarse\_fe\_space\_t} in order to extract the local-to-global coarse \ac{dof} index map.
\end{itemize}

The scalability of the framework is boosted with the advanced parallel implementation approach discussed in detail in \cite{art003, badia_multilevel_2016}. This approach exploits a salient property of multilevel \ac{bddc}-like preconditioners, namely that there are computations at different levels that can be overlapped in time. To this end, the coarse-grid problem is not actually handled by any of the tasks on which the \ac{fe} mesh is distributed, but by an additional, specialized parallel task (set of tasks) that is (are) spawn in order to carry out such coarse-grid problem related duties. The {\tt environment\_t} \FEMPAR{} data type, which was already introduced in Sect.~\ref{sec:tutorial01_commented_code}, splits the full set of tasks into subgroups of tasks (i.e., levels), and defines communication mechanisms to transfer data among them. For example, for a 2-level \ac{bddc} preconditioner, one sets up an environment with 2 levels, and \FEMPAR{} devotes the tasks of the first
and second levels to fine-grid and coarse-grid related duties, resp., while achieving the desired overlapping effect among the computations at different levels. This will be illustrated in the next section.

\subsection{The commented code} \label{sec:tutorial03_commented_code}

The main program unit of \tutorial{03} is shown in Listing~\ref{lst:tutorial_03_program_unit}. For conciseness, we only show those lines of code of \tutorial{03}
which are different from those of \tutorial{02} (Listing~\ref{lst:tutorial_02_program_unit}).
The first worth noting difference is that \tutorial{03} uses its own tutorial-specific support modules in Lines~\ref{loc:tutorial_03_discrete_integration}-\ref{loc:tutorial_03_error_estimators} . While the one in Line~\ref{loc:tutorial_03_functions_names} is actually fully equivalent to its counterparts in
\tutorial{01} and {\tt \_02}, only minor adaptations were required in the other two.
Recall that these modules encompass an integration loop over the mesh cells. We aim to build a local sub-assembled Neumann problem at each subdomain in the former module (as required by non-overlapping \ac{dd} preconditioners; see \cite[Sect. 5.1]{Badia2019a}), and to only compute $e_K^2$ for the local cells in each parallel task in the latter (in order to avoid excess computation, and over-assembly of $e$ when these local quantities are reduced-sum in all parallel tasks in the {\tt compute\_error} helper subroutine). Following the discussion at the end of Sect.~\ref{sec:tutorial_03_fe_discretization}, the cell integration loops in these modules must be restricted to {\em local} cells. This is accomplished by embracing the body of the integration loop (see, e.g. Lines~\ref{loc:cg_di_fe_iterator_update_integration}-\ref{loc:cg_di_fe_iterator_next} of Listing~\ref{lst:tutorial_01_cg_discrete_integration}) by an {\tt if(fe\%is\_local())then...endif} statement. The {\tt is\_local()} \ac{tbp} of a \ac{fe} iterator returns {\tt .true.} if the iterator is positioned on a local cell,
and {\tt .false.} otherwise; see Fig~\ref{fig:mesh-distribution}.

\begin{lstlisting}[float=htbp,language={[03]Fortran},escapechar=@,caption=\Tutorial{03} program unit.,label={lst:tutorial_03_program_unit}]
program tutorial_03_poisson_sharp_circular_wave_parallel_amr
  use fempar_names
  use tutorial_03_discrete_integration_names @\label{loc:tutorial_03_discrete_integration}@
  use tutorial_03_functions_names            @\label{loc:tutorial_03_functions_names}@
  use tutorial_03_error_estimator_names      @\label{loc:tutorial_03_error_estimators}@
  ... ! Declaration of tutorial_03_... parameter constants  @\label{loc:tutorial_03_part2_start}@
  type(mpi_context_t)                         :: world_context @\label{loc:tutorial_03_world_context}@
  type(environment_t)                         :: environment
  type(p4est_par_triangulation_t)             :: triangulation
  type(p_l1_coarse_fe_handler_t), allocatable :: coarse_fe_handlers(:)
  type(h_adaptive_..._coarse_fe_handler_t)    :: coarse_fe_handler
  type(par_fe_space_t)                        :: fe_space
  ... ! strong_boundary_conditions, source_term, exact_solution
  ... ! discrete_solution, cg_discrete_integration
  type(iterative_linear_solver_t)             :: iterative_linear_solver
  type(parameterlist_t)                       :: mlbddc_parameters @\label{loc:tutorial_03_mlbddc_params}@
  type(mlbddc_t)                              :: mlbddc @\label{loc:tutorial_03_mlbddc}@
  ... ! output_handler, refinement_strategy
  type(std_vector_real_rp_t)                  :: my_rank_cell_array
  ...@\label{loc:tutorial_03_part3_start}@! Declaration of variables storing CLA values particular to tutorial_03_...
  ... ! Lines @\ref{loc:tutorial_02_part3_start}@-@\ref{loc:tutorial_02_part3_startup_loop_start}@ of Listing@~\ref{lst:tutorial_02_program_unit}@
  ... ! Lines @\ref{loc:tutorial_01_setup_triangulation}@-@\ref{loc:tutorial_01_setup_strong_bcs}@ of Listing@~\ref{lst:tutorial_01_program_unit}@
  call setup_coarse_fe_handler() @\label{loc:tutorial_03_setup_coarse_fe_handler}@
  call setup_fe_space()
  ... ! Lines @\ref{loc:tutorial_01_setup_discrete_solution}@-@\ref{loc:tutorial_01_assemble_operator}@ of Listing@~\ref{lst:tutorial_01_program_unit}@
  call setup_preconditioner() @\label{loc:tutorial_03_setup_preconditioner}@
  ... ! solve_system(), compute_error()
  ... ! setup_refinement_strategy(), output_handler_initialize()
  do     
    ... ! Lines @\ref{loc:tutorial_02_oh_write_current_step}@-@\ref{loc:tutorial_02_refinement_update}@ of Listing@~\ref{lst:tutorial_02_program_unit}@ 
    call setup_triangulation()
    call setup_fe_space()
    call redistribute_triangulation_and_fe_space() @\label{loc:tutorial_03_redistribute}@
    ... ! Lines @\ref{loc:tutorial_02_set_di}@-@\ref{loc:tutorial_02_compute_error}@ of Listing@~\ref{lst:tutorial_02_program_unit}@
 end do
 ... @\label{loc:tutorial_03_part3_stop}@ ! Lines @\ref{loc:tutorial_02_oh_finalize}@-@\ref{loc:tutorial_02_part3_stop}@ of Listing@~\ref{lst:tutorial_02_program_unit}@
contains
  ... ! Implementation of helper procedures
end program tutorial_03_poisson_sharp_circular_wave_parallel_amr
\end{lstlisting}

The reader may also observe subtle differences in the objects declared by \tutorial{03} (Lines~\ref{loc:tutorial_03_world_context}-\ref{loc:tutorial_03_mlbddc} of Listing~\ref{lst:tutorial_03_program_unit}) compared to those declared by \tutorial{02} (Lines-\ref{loc:tutorial_02_triangulation}-\ref{loc:tutorial_02_rest} of Listing~\ref{lst:tutorial_02_program_unit}). First, {\tt world\_context} is declared of type {\tt mpi\_context\_t}. This \FEMPAR{} data type represents a group of parallel tasks (as many as specified to the {\tt mpirun} script when the parallel program is launched) which uses \ac{mpi} as communication layer.  Second, the {\tt triangulation} and {\tt fe\_space} objects were declared to be of type {\tt p4est\_par\_triangulation\_t} and {\tt par\_fe\_space\_t}. These \FEMPAR{} data types  are the distributed-memory counterparts of the ones used by \tutorial{02}. The former is a data type extension of {\tt triangulation\_t} that follows the ideas in Sect.~\ref{sec:tutorial_03_fe_discretization}.
Finally, \tutorial{03} declares extra objects which are not necessary in \tutorial{02}. These are covered in the next paragraph.

First, \tutorial{03} declares an object of type {\tt h\_adaptive\_...\_coarse\_fe\_handler\_t}. This \FEMPAR{} data type is a type extension of {\tt coarse\_fe\_handler\_t} (see Sect.~\ref{sec:tutorial04_preconditioning}) suitable for grad-conforming \ac{fe} spaces on $h$-adaptive meshes \cite{Kus2017}.\footnote{We note that this data type is connected with the \ac{cli}. Using the corresponding \acp{cla}, one may select whether to associate coarse \acp{dof} to coarse vertices, and/or coarse edges, and/or coarse faces.} The {\tt coarse\_fe\_handlers(:)} array holds polymorphic pointers to data type extensions of {\tt coarse\_fe\_handler\_t}, as many as fields in the system of \acp{pde} at hand. As \tutorial{03} tackles a single-field \ac{pde}, this array is set up in the {\tt setup\_coarse\_fe\_handler} helper subroutine (Line~\ref{loc:tutorial_03_setup_coarse_fe_handler}) to be a size-one array pointing to {\tt coarse\_fe\_handler}. Second, \tutorial{03} declares {\tt iterative\_linear\_solver}. This object provides a rich suite of abstract implementations of Krylov subspace solvers (see Sect.~\ref{sec:tutorial04_preconditioning}). Third, it also declares the {\tt mlbdc} instance, in charge of building and applying the \ac{bddc} preconditioner at each iterative solver iteration (see Sect.~\ref{sec:tutorial04_preconditioning}). The configuration of this instance cannot be performed directly from {\tt parameter\_handler}, but by means of a rather involved parameter dictionary (declared in Line~\ref{loc:tutorial_03_mlbddc_params}) that lets one customize the solver parameters required for each of the subproblems solved by the \ac{bddc} preconditioner at each level. This parameter dictionary is set up in the {\tt setup\_preconditioner} helper subroutine (Line~\ref{loc:tutorial_03_setup_preconditioner}), which will be covered later in this section. Finally, \tutorial{03} declares {\tt my\_rank\_cell\_array} for post-processing purposes. This is a dynamic array which is adapted along with the triangulation at each \ac{amr} loop iteration. It has as many entries as local cells in each parallel task, and for all of these entries, it holds the parallel task identifier. It is written into output data files, so that the user may visualize how the adaptive mesh is distributed among the parallel tasks. For conciseness, it is left as an exercise to the reader to grasp how this array is handled by  \tutorial{03} in order to achieve this goal.

\Tutorial{03}'s main executable code spans Lines~\ref{loc:tutorial_03_part3_start}-\ref{loc:tutorial_03_part3_stop}. It very much resembles the one of \Tutorial{02}, despite
it is being executed in a significantly more complex, non-standard parallel execution environment. In particular,
all parallel tasks in {\tt world\_context} execute the bulk of code lines of \Tutorial{03}, despite these are split into two levels  by {\tt environment\_t} (Listing~\ref{lst:tutorial_03_environment}) and assigned different duties (and data) at different levels; see discussion at the end of Sect.~\ref{sec:tutorial04_preconditioning}. We have devoted daunting efforts in order to hide as much as possible this complex execution environment to library users. The vast majority of \acp{tbp} associated to the library data types can be called safely from any task in {\tt world\_context}. For example, one may call {\tt triangulation\%get\_num\_local\_cells()} from all tasks. In the case of L1 tasks (i.e., tasks belonging to the first environment level), this call returns the number of local cells in the mesh portion of the task, while it returns a degenerated value, i.e., zero, in the case of L2 tasks, as {\tt triangulation} is only distributed among the tasks in the first environment level. If, e.g., one allocates an array with the output of this call, this ends up with a zero-sized array in L2 tasks, which is perfectly fine with the Fortran standard. Another paradigmatic example are loops over the cells using iterators. L2 tasks do not enter the loop, as there are no cells to be traversed in this case. The only reasonable exception to this in \tutorial{03} is the {\tt compute\_error} helper subroutine, that the reader is encouraged to inspect at this point using the source code Git repository. This subroutine uses {\tt environment} in order to dispatch the path followed by  L1 and L2 tasks such that: (1) only a single L1 task logs into the screen the number of global cells and \acp{dof}, thus avoiding that the screen if flood with output messages coming from all L1 tasks; (2) only a single L2 task logs the number of coarse cells and \acp{dof} (for the same reason). The same task-level-dispatching mechanism is being used by many procedures within \FEMPAR{}, mostly those related with the \ac{bddc} preconditioner, although for a different purpose, namely to achieve the desired overlapping effect among computations at different levels.

\begin{lstlisting}[float=htbp,language={[03]Fortran},escapechar=@,caption=The \texttt{setup\_context\_and\_environment} procedure.,label={lst:tutorial_03_environment}]
subroutine setup_context_and_environment()
  !* Create a group of tasks (as many as specified to mpirun)
  call world_context%create()
  !* Force environment to split world_context into two subgroups of tasks (levels)
  !* composed by "world_context%get_num_tasks()-1" and a single task, resp.
  call parameter_handler%update(environment_num_levels_key, 2)
  call parameter_handler%update(environment_num_tasks_x_level_key, &
                                [world_context%get_num_tasks()-1,1])
  call environment%create(world_context, parameter_handler%get_values())
end subroutine setup_context_and_environment

  
\end{lstlisting}

\Tutorial{03} helper procedures very much resemble their counterparts in \tutorial{02}. A main difference is in the procedures that set up the triangulation and \ac{fe} space. When the value of {\tt current\_amr\_step} is zero, {\tt setup\_triangulation} calls the {\tt setup\_coarse\_triangulation} \ac{tbp} of {\tt triangulation}, right after the latter triangulation is built. This procedure triggers the process described in Sect.~\ref{sec:tutorial04_preconditioning} in order to build a {\tt coarse\_triangulation\_t} object. This object is kept inside {\tt triangulation}, although the user may have access to it via {\tt triangulation} getters. The same pattern is borrowed by {\tt setup\_fe\_space}, which calls the {\tt setup\_coarse\_fe\_space} \ac{tbp} of {\tt fe\_space}. For obvious reasons (see Sect.~\ref{sec:tutorial04_preconditioning}), this \ac{tbp} must be fed with the {\tt coarse\_fe\_handlers(:)} array. On the other hand, when {\tt current\_amr\_step} is not zero, it is not needed to explicitly call
these \acp{tbp}, as they are reconstructed automatically after mesh adaptation.

Apart from the aforementioned, the reader must note the call in Line~\ref{loc:tutorial_03_redistribute} of Listing~\ref{lst:tutorial_03_program_unit}, right after the triangulation and the \ac{fe} space are adapted within the current \ac{amr} loop iteration. The code of this helper subroutine is shown in Listing~\ref{lst:tutorial_03_redistribute}. This subroutine calls the {\tt redistribute} \ac{tbp} of {\tt triangulation}, which dynamically balances the computational load by redistributing the adaptive mesh among the parallel tasks. The default criteria is to balance the number of
cells in each task. Alternatively, the user might associate to each cell a partition weight. In this case, the primitive balances the sums of the cell partition weights among processors. The data that the user
might have attached to the mesh objects (i.e., cells and \acp{vef} set identifiers) is also migrated. On the other hand, Listing~\ref{lst:tutorial_03_redistribute} also calls the {\tt redistribute} \ac{tbp} of {\tt fe\_space}, which migrates the data that {\tt fe\_space} holds conformally to how the triangulation has been redistributed. Optionally, this \ac{tbp} can by supplied with a \ac{fe} function $u_h$ (or, more generally, an arbitrary number of them). In such a case, {\tt redistribute} migrates the \ac{dof} values of $u_h$ as well. This feature is required by numerical solvers of transient and/or non-linear \acp{pde}.
Finally, we note that  Listing~\ref{lst:tutorial_03_program_unit} redistributes the data structures at each iteration of the \ac{amr} loop, just as demonstrator of \FEMPAR{} capabilities. In an actual \ac{fe} application problem, one may tolerate load unbalance as long as it remains within reasonable margins. In general, the frequency of redistribution should be chosen in order to achieve an optimal trade-off among the overhead associated to migration, and the computational benefit that one obtains by dynamically balancing the computational load.

\begin{lstlisting}[float=htbp,language={[03]Fortran},escapechar=@,caption=The {\tt redistribute\_triangulation\_and\_fe\_space} helper subroutine.,label={lst:tutorial_03_redistribute}]
subroutine redistribute_triangulation_and_fe_space()
  call triangulation%redistribute()
  call fe_space%redistribute()
  ... ! Re-adjust the size and contents of my_rank_cell_array
  ... ! to reflect the current status of the triangulation
end subroutine redistribute_triangulation_and_fe_space
\end{lstlisting}

Listing~\ref{lst:tutorial_03_setup_preconditioner} shows the code of the {\tt setup\_preconditioner} procedure. It first builds the {\tt mlbddc\_parameters} parameter dictionary. To this end, it calls the {\tt setup\_mlbddc\_parameters\_...} subroutine, provided by \FEMPAR{}. This call takes the solver-related \ac{cla} values provided by {\tt parameter\_handler}, and populates {\tt mlbddc\_parameters} such that the same parameter values are used for the solvers of all subproblems that {\tt mlbddc} handles internally (e.g., the Dirichlet and constrained Neumann subproblems).\footnote{If one wants to use different solver parameters values for each of these subproblems, then  {\tt mlbddc\_parameters} has to be built manually.} The actual set up of the preconditioner occurs in Line~\ref{loc:setup_preconditioner_create} of Listing~\ref{lst:tutorial_03_setup_preconditioner}. We note that {\tt mlbddc} directly receives the \ac{fe} affine operator, instead of the coefficient matrix that it holds inside. This lets the \ac{bddc} framework to access to the application \ac{fe} discretization-related data, so that this information can be exploited when building an optimal preconditioner for the \ac{pde} problem at hand.

\begin{lstlisting}[float=htbp,language={[03]Fortran},escapechar=@,caption=The {\tt setup\_preconditioner} helper subroutine.,label={lst:tutorial_03_setup_preconditioner}]
subroutine setup_preconditioner()
  call setup_mlbddc_parameters_from_reference_parameters(&
       environment, &
       parameter_handler%get_values(), &
       mlbddc_parameters)
  call mlbddc%create( fe_affine_operator, mlbddc_parameters ) @\label{loc:setup_preconditioner_create}@
end subroutine setup_preconditioner
\end{lstlisting}

Finally, the {\tt solve\_system} subroutine is shown in Listing~\ref{lst:tutorial_03_solve_system}. In Line~\ref{loc:tutorial_03_solve_system_cg}, we force the Conjugate Gradients solver, as this is the most suitable iterative solver for the Poisson \ac{pde}. The rest of \ac{cla} values linked to {\tt iterative\_linear\_solver\_t} are not forced, so that the user may choose, e.g., among several convergence criteria and related solver tolerances, or whether to print on screen or not the convergence history of the solver. An iterative solver needs a matrix and a preconditioner to solve the system. These are provided to {\tt iterative\_linear\_solver} in Line~\ref{loc:tutorial_03_solve_system_set_operators}.

\begin{lstlisting}[float=htbp,language={[03]Fortran},escapechar=@,caption=The {\tt solve\_system} helper subroutine.,label={lst:tutorial_03_solve_system}]
subroutine solve_system()
  class(vector_t), pointer :: dof_values
  if ( current_amr_step == 0 ) then
     call iterative_linear_solver%create(environment)
     call parameter_handler%update(ils_type_key, cg_name) @\label{loc:tutorial_03_solve_system_cg}@
     call iterative_linear_solver%set_type_from_pl(parameter_handler%get_values())
     call iterative_linear_solver%set_parameters_from_pl(parameter_handler%get_values())
     call iterative_linear_solver%set_operators(fe_affine_operator%get_matrix(), mlbddc) @\label{loc:tutorial_03_solve_system_set_operators}@
  else 
     call iterative_linear_solver%reallocate_after_remesh()
  end if
  dof_values => discrete_solution%get_free_dof_values()
  call iterative_linear_solver%solve(fe_affine_operator%get_translation(),dof_values)
  call fe_space%update_hanging_dof_values(discrete_solution)
end subroutine solve_system
\end{lstlisting}

\subsection{Numerical results}

\begin{figure}[t!]
  \centering
  \includegraphics[width=0.45\textwidth]{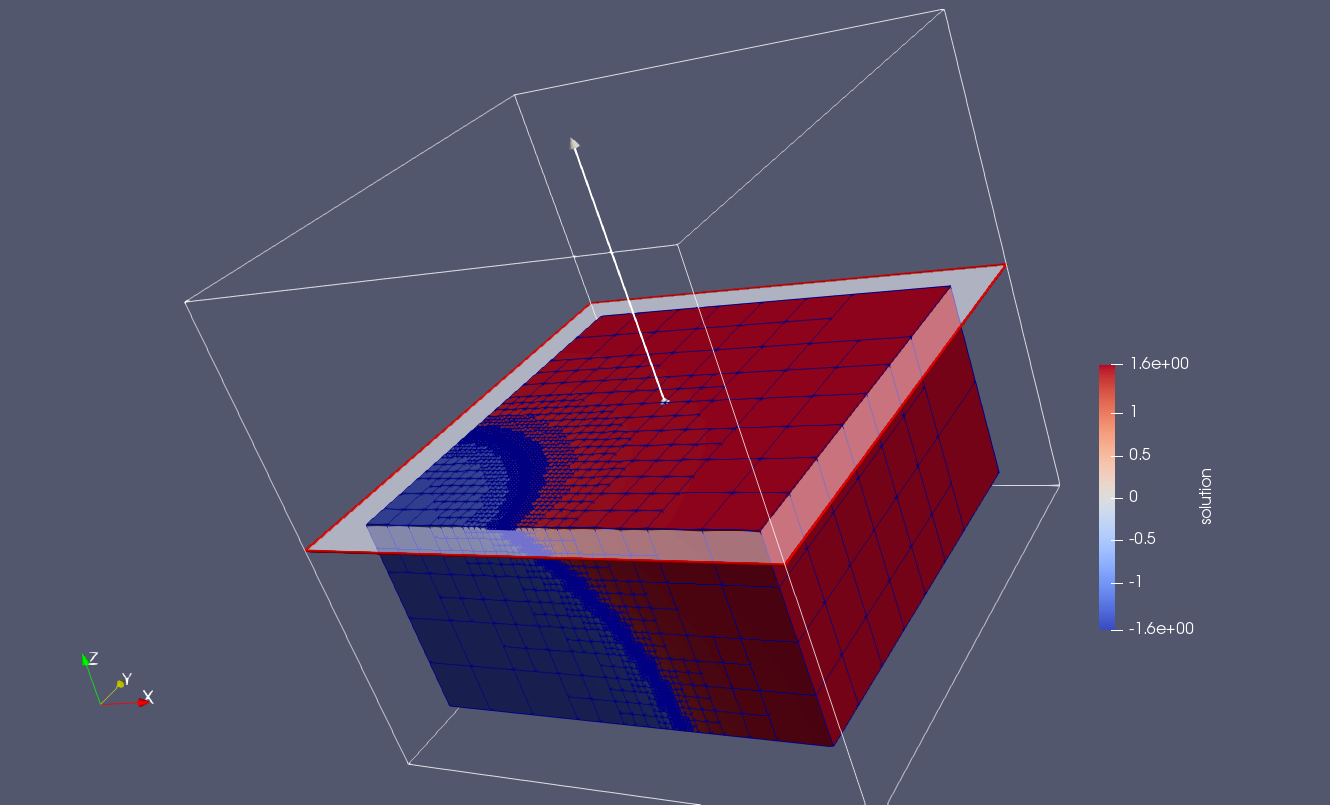}
  \includegraphics[width=0.45\textwidth]{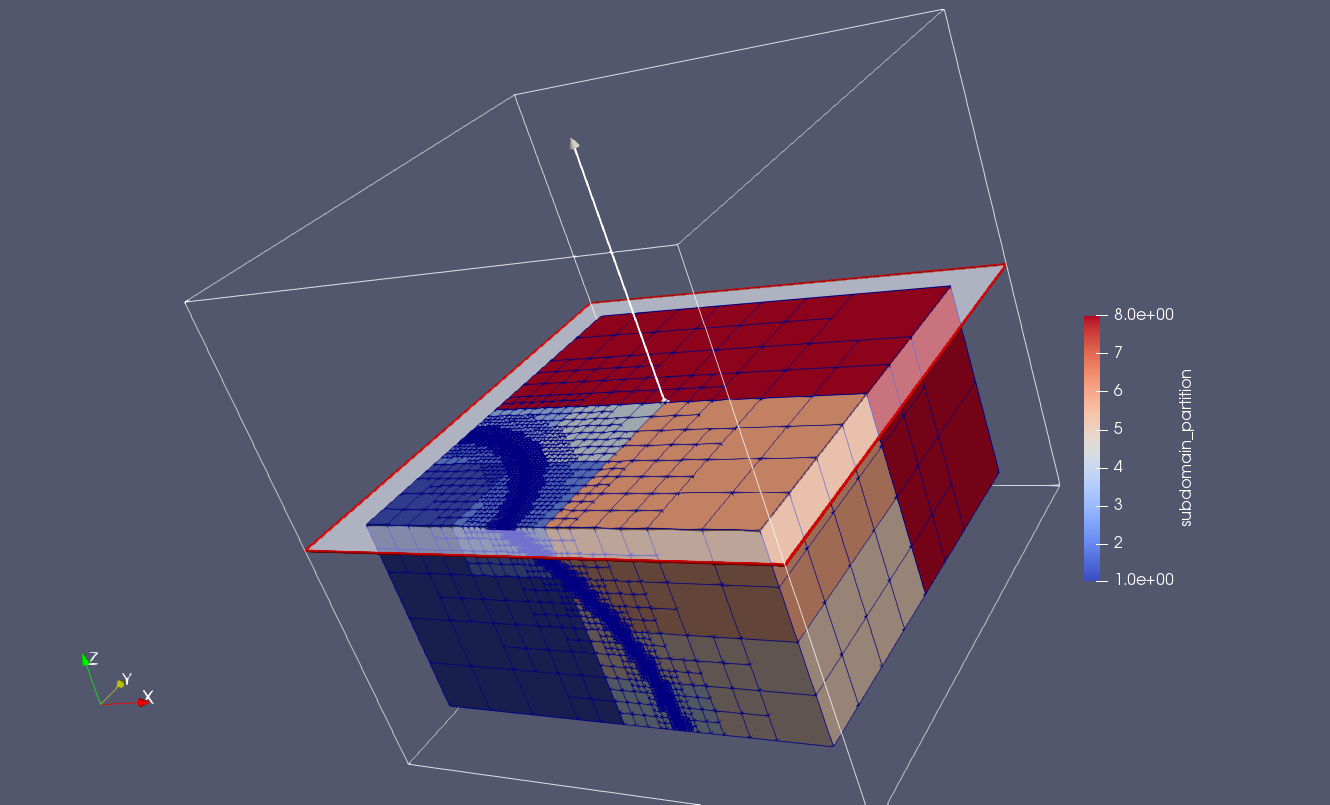}
\caption{Mesh and \ac{fe} solution (left) and its partition into 9 subdomains (right). 13 \ac{amr} steps; 218,590 cells; 153,760 (true) \acp{dof}.}
\label{fig:3d_benchmark_results}
\end{figure}

In Fig.~\ref{fig:3d_benchmark_results} we show the \ac{fe} solution computed by \tutorial{03} invoked with 10 parallel tasks, along with $\triang$ and its partition into 9 subdomains, for the 3D version of Problem~\eqref{sec:tutorial01_model_problem} discretized with an adapted mesh resulting from 13  \ac{amr} steps, resp., and trilinear Lagrangian \acp{fe}. The number of initial uniform refinement steps was set to 2, resulting in an initial conforming triangulation made of 8 hexahedral cells. The \ac{bddc} space was supplied with corner, edge, and face coarse \acp{dof}, resulting in a total of 77 coarse \acp{dof} for the subdomain partition in Fig.~\ref{fig:3d_benchmark_results}. The Preconditioned Conjugate Gradients solver converged to the solution in 14 iterations with a relative residual tolerance of $10^{-6}$.

\section{Conclusions}\label{sec-conclusions}

In this article we have provided three tutorials that cover some of the capabilities of the \FEMPAR{} library. The tutorials come with a comprehensive description of all the steps required in the simulation of \ac{pde}-based problems. They cover the numerical approximation of a linear \ac{pde}, structured and octree meshes with \ac{amr} strategies (both in serial and parallel environments), and the usage of parallel iterative solvers with scalable preconditioning techniques. This set of tutorials provides \FEMPAR{} users with a complete introduction to some key \FEMPAR{} tools. In any case, we refer to the tutorials folder in the \FEMPAR{} public repository \href{https://github.com/fempar/fempar}{\texttt{https://github.com/fempar/fempar}} for more advanced topics not covered here (e.g., nonlinear solvers, time integration, curl and div conforming \ac{fe} spaces, multi-field \ac{fe} spaces, or block preconditioning techniques).

\bibliographystyle{myelsarticle-num}
\bibliography{art039}

\end{document}